\documentclass[12pt]{iopart}

\usepackage{iopams}
\expandafter\let\csname equation*\endcsname\relax
\expandafter\let\csname endequation*\endcsname\relax
\usepackage{amsmath}
\usepackage{url}
\usepackage[]{units}
\usepackage{graphicx}
\usepackage{epstopdf}
\usepackage{cite}
\begin{document}
\title[]{Stochastic Gravitational Wave Backgrounds}

\author{Nelson~Christensen$^{1,2}$ \footnote{nelson.christensen@oca.eu}}
\address {$^{1}$ARTEMIS, Universit\'e C\^ote d'Azur, Observatoire C\^ote d'Azur, CNRS, 06304 Nice, France}
\address{$^{2}$Physics and Astronomy, Carleton College, Northfield, MN 55057, USA}

\begin{abstract}
A stochastic background of gravitational waves can be created by the superposition of a large number of independent sources. The physical processes occurring at the earliest moments of the universe certainly created a stochastic background that exists, at some level, today. This is analogous to the cosmic microwave background, which is an electromagnetic record of the early universe. The recent observations of gravitational waves by the Advanced LIGO and Advanced Virgo detectors imply that there is also a stochastic background that has been created by binary black hole and binary neutron star mergers over the history of the universe. 
Whether the stochastic background is observed directly, or upper limits placed on it in specific frequency bands, important astrophysical and cosmological statements about it can be made. This review will summarize the current state of research of the stochastic background, from the sources of these gravitational waves, to the current methods used to observe them.
\end{abstract}

%
%
\noindent Keywords: stochastic gravitational wave background, cosmology, gravitational waves

\section{Introduction}
Gravitational waves are a prediction of Albert Einstein from 1916\cite{Einstein:1916b,Einstein:1918}, a consequence of general relativity~\cite{Einstein:1916a}. Just as an accelerated electric charge will create electromagnetic waves (light), accelerating mass will create gravitational waves. And almost exactly a century after their prediction, gravitational waves were directly observed~\cite{PhysRevLett.116.061102} for the first time by Advanced LIGO~\cite{0264-9381-32-7-074001,0264-9381-27-8-084006}. The existence of gravitational waves had already been firmly established in 1982 through the observation of the orbital decay of a binary neutron star system~\cite{1982ApJ...253..908T}; as the two neutron stars orbited around one another, they were accelerating, so gravitational waves were emitted, carrying away energy, and causing the orbit to decay. Advanced LIGO has subsequently observed gravitational wave signals from merging binary black hole systems~\cite{PhysRevLett.116.241103,PhysRevX.6.041015,PhysRevLett.118.221101,GW170608}. Since then Advanced LIGO and Advanced Virgo~\cite{0264-9381-32-2-024001} have joined together to observe a binary black hole merger~\cite{PhysRevLett.119.141101} and a binary neutron star merger~\cite{PhysRevLett.119.161101}. The detection of gravitational waves from the binary neutron star merger, GW170817, provided the commencement of gravitational wave multi-messenger astronomy, with simultaneous observations of the event and its source across the electromagnetic spectrum~\cite{2041-8205-848-2-L12}. It can be argued that multi-messenger astronomy started with the joint electromagnetic and neutrino observations of SN 1987A~\cite{1989ARA&A..27..629A}.

A gravitational wave is a traveling gravitational field. An electromagnetic wave is a traveling electric field and a magnetic field, both transverse to the direction of propagation. Similarly, the effects of a gravitational wave are transverse to the direction of propagation. The effects of a gravitational wave are similar to a tidal gravitational field. In terms of general relativity, a gravitational wave will stretch one dimension of space while contracting the other. Just like electromagnetic waves, gravitational waves carry energy and momentum with them.

Gravitational waves are far too weak to be created by some process on the Earth and then subsequently detected. Energetic astrophysical events will be the source of observable gravitational wave signals. The events could be the inspiral of binary systems involving black holes or neutron stars. Core collapse supernovae could produce a detectable signal if they occurred in our galaxy, or perhaps in nearby galaxies. A spinning neutron star would produce a periodic gravitational wave signal if the neutron star had an asymmetry that made it nonaxisymmetric. Finally, there could be a stochastic background of gravitational waves made by the superposition of numerous incoherent sources. The recent detection by LIGO and Virgo of gravitational waves from the coalescence of binary black hole and binary neutron star systems implies that there is a stochastic background created by these sorts of events happening throughout the history of the universe~\cite{PhysRevLett.116.131102,PhysRevX.6.041015,Abbott:2017xzg}. Because of the recent LIGO-Virgo results there will be an emphasis in this report on the stochastic background that LIGO-Virgo may soon observe, however the searches via other methods will also be addressed. Certainly different processes in the early universe have created gravitational waves. For example, quantum fluctuations during inflation~\cite{PhysRevD.23.347}, the speculated period of exponential growth of the universe at its earliest moments, have created gravitational waves that would be observed as a stochastic background today~\cite{PhysRevD.48.3513}.

This report will give an overview of the stochastic gravitational wave background (or more simply in this report, the stochastic background). Presented will be a summary of the various means by which a stochastic background could be created. Also, the different ways that a stochastic background could be detected will be presented, along with the information that can be extracted from its observation, or even the absence of its observation.

\subsection{Gravitational Waves}
Given here is a brief review of gravitational waves. For a comprehensive summary of gravitational wave physics, sources, and
detection methods, see~\cite{maggiore2007gravitational,creighton2012gravitational,Saulson:2013dga,Romano2017}. Working with linearized general relativity, the gravitational wave is assumed to make only a slight modification to flat space, 
\begin{equation}
g_{\mu \nu} \approx \eta_{\mu \nu} + h_{\mu \nu} ~ ,
\end{equation}
where $g_{\mu \nu}$ is the spacetime metric, $\eta_{\mu \nu}$ in the Minkowski metric (representing flat spacetime), and $h_{\mu \nu}$ is the metric perturbation. The generation of gravitational waves is a consequence of general relativity, and can be predicted via the Einstein equation. To first order in the metric perturbations, gravitational waves are created when the mass quadrupole moment is accelerating, namely that it has a non-zero second derivative with respect to time. Gravitational waves also carry energy and momentum. When a system emits gravitational waves, it loses energy.
The existence of gravitational waves was first confirmed through the observation of the orbital decay of the binary pulsar PSR 1913+16~\cite{1982ApJ...253..908T,0004-637X-829-1-55}; the rate at which the orbit for this system is decaying exactly matches the prediction from general relativity for the loss of energy through gravitational wave emission. This is also the reason for the coalescence of the binary black holes and the binary neutron stars observed by Advanced LIGO and Advanced Virgo, such as GW150914~\cite{PhysRevLett.116.061102}, GW151226~\cite{PhysRevLett.116.241103}, GW170104~\cite{PhysRevLett.118.221101}, GW170608~\cite{GW170608}, GW170814~\cite{PhysRevLett.119.141101} and GW170817~\cite{PhysRevLett.119.161101}.

After emission, a gravitational wave essentially travels as a plane wave. Imagine a wave traveling in the 
{\bf z}-direction. Just like electromagnetic radiation, there are two possible polarizations, and the physical effects are transverse to the direction of propagation. We can arbitrarily choose our {\bf x} and {\bf y} axes. One polarization (which we will call the $+$ polarization will cause space to be expanded and contracted along these {\bf x} and {\bf y} axes. The other polarization, which we will call the $\times$ polarization, will cause space to be expanded and contracted along the {\bf x$^{\prime}$} and {\bf y$^{\prime}$} axes, where these axes are rotated by $45^o$ from the other axes.

Let us look in detail at the effect of the $+$ polarization. Consider the plane wave moving in the {\bf z}-direction
\begin{equation}
h_{i j}(z,t) = h_{+} \left( \begin{tabular}{ccc}
  1 & 0 & 0 \\
  0 & -1 & 0 \\
  0 & 0 & 0 
  \end{tabular} \right)_{i j}  e^{i(kz - \omega t)} ~ . 
\end{equation}
Spacetime is stretched due to the
strain created by the gravitational wave. Starting with a length $L_{0}$ along the {\bf x}-axis, the
gravitational wave causes the length to oscillate like 
\begin{equation}
L(t) = L_{0} + \frac{h_{+} L_{0}}{2} \cos(\omega t) ~ .
\end{equation}
There is a change in its length of
\begin{equation}
\Delta L_{x} = \frac{h_{+} L_{0}}{2} \cos(\omega t) ~.
\end{equation}
Along the {\bf y}-axis, a similar length $L_{0}$ subjected to the same gravitational wave oscillates like 
\begin{equation}
\Delta L_{y} = - \frac{h_{+} L_{0}}{2} \cos(\omega t) ~.
\end{equation}
In this example, the {\bf x}-axis stretches while the {\bf y}-axis contracts, and then vice versa as the wave propagates through the region of space.
In terms of the relative change of the lengths of the two arms (at $t=0$),
\begin{equation}
\Delta L = \Delta L_{x} - \Delta L_{y} = h_{+} L_{0} \cos(\omega t) ~,
\end{equation}
or
\begin{equation}
 h_{+} = \frac{\Delta L}{L_{0}} ~.
\end{equation}
The amplitude of a gravitational wave, $h_{+}$, is the amount of strain that it produces on spacetime.
The other gravitational wave polarization ($h_{\times}$) produces a similar strain on axes $45^o$ from ({\bf x},{\bf y}). The stretching and contracting of space is the physical effect of a gravitational wave, and detectors of gravitational waves are designed to measure this strain on space.

\subsection{Sources of Gravitational Waves}
When searching for gravitational waves the signals are roughly divided into four categories: coalescing binaries, unmodeled bursts (for example from core collapse supernova), continuous waves (for example from pulsars), and stochastic. The signal search techniques are then optimized for these particular signals.

Compact binary coalescence will produce a typical chirp-like signal. In the LIGO-Virgo observational band, from \unit[10]{Hz} up to a few kHz, these signals will be made from binary systems consisting of neutron stars (with masses \unit[$\sim$ 1.4]{$M_\odot$}) and black holes (with masses up to \unit[$\sim$ 100]{$M_\odot$}). As the binary system's orbit decays via energy loss by gravitational wave emission, the two objects spiral into one another. The orbital frequency increases, and consequently the gravitational wave frequency and amplitude also increase. In addition to the inspiral (chirp) signal, there will also be a signal associated with the merger of the two objects, and if a black hole is created, the ringdown signal as the black hole approaches a axisymmetric form. Since the binary inspiral signal is relatively straightforward to calculate, the LIGO-Virgo signal search is based on comparing the data with templates. As the ability to predict the form of the signal has improved, these templates now account for spin of the masses~\cite{0264-9381-33-21-215004,PhysRevD.95.042001,PhysRevD.89.024003}. Once the signals are detected, Bayesian parameter estimation routines are used to extract the physical parameters of the system. These methods now incorporate the full extent of the waveform: inspiral, the merger of the two masses, and the black hole's ringdown to a axisymmetric form~\cite{PhysRevX.6.041014}. It is interesting to note that stellar mass binary black hole systems, similar to GW150914~\cite{PhysRevLett.116.061102}, will also be visible in the proposed space-based gravitational wave detector~\cite{PhysRevLett.116.231102}, the Laser Interferometer Space Antenna (LISA)~\cite{eLISA:2014,2017arXiv170200786A}. LISA will be able to observe these systems weeks to years before they coalesce in the LIGO-Virgo band. LISA will observe gravitational waves with frequencies between \unit[0.1]{mHz} to \unit[100]{mHz}. In this band LISA will also observe binary black hole systems with masses up to \unit[$\sim 10^{7}$]{$M_\odot$}.
Pulsar timing methods (whereby the regular radio signals from pulsars are used like clocks in the sky, and the presence of a gravitational wave would vary the arrival time of the pulses) will search for supermassive binary black holes systems, with masses from \unit[$3 \times 10^{7}$]{$M_\odot$} to \unit[$3 \times 10^{9}$]{$M_\odot$} (gravitational wave periods of order years)~\cite{0264-9381-30-22-224005}.

There are several possible sources of unmodeled bursts of gravitational waves. Core collapse supernovae are one of the most exciting possibilies. The gravitational wave emission from these sorts of events are extremely difficult to predict~\cite{Fryer2011}. Other burst signals could come, for example, from pulsar glitches, or the transition of a neutron star to a black hole. These types of signals are typically searched via excess power in the data. LIGO and Virgo have recently searched for signals of thess types with durations from a few milliseconds up to \unit[10]{s}~\cite{PhysRevD.95.042003}. The inspiral and merger of a very massive binary black hole pair will be of short duration in the LIGO-Virgo observing band, so these excess power detection methods will be the most effective means of observing them. There are also different mechanisms by which there could be gravitational wave transients of significant amplitude for extended periods; LIGO and Virgo are currently looking for burst events lasting up to \unit[1000]{s}~\cite{PhysRevD.93.042005,0264-9381-35-6-065009}. 
In addition to excess power types of searches, it is also possible to search for cosmic string signals via a dedicated template based search~\cite{Aasi:2013vna}.
Cosmic strings are theorized to be one-dimensional topological defects created after a spontaneous symmetry phase transition~\cite{0305-4470-9-8-029,vilenkin:1994} as predicted in a range of field theories.
While cosmic string kinks and cusps will produce short duration transient gravitational wave signals, the forms of these signals are technically predictable.

Neutron stars are extremely dense, and often spinning at incredible rates. It is suspected that neutron stars typically have masses around \unit[1.4]{$M_{\odot}$}, with a radius around \unit[12]{km}. Neutron stars can have significant angular velocities; there is  evidence of a pulsar with a rotation rate of \unit[714]{Hz}~\cite{Hessels1901}. A rotating sphere will not emit gravitational waves (due to conservation of mass); more generally, a rotating axisymmetric object will not emit gravitational waves (due to conservation of angular momentum). However, if there is some asymmetry in the shape of the rotating neutron star, then it can emit gravitational waves. These gravitational waves would be periodic, but due to other factors (loss of energy from gravitational wave emission, or accretion from a companion in a binary), there can be a frequency derivative. The Doppler shift between the source and the detector must also be considered. With these factors in mind, LIGO and Virgo are currently searching for gravitational waves from rapidly rotating neutron stars~\cite{PhysRevD.94.102002,PhysRevD.85.042003}

The incoherent sum of numerous unresolved gravitational wave signals will result in a stochastic background of gravitational waves. This is the main topic of this report, and much more information on this background is presented below. 

The magnitude of the stochastic gravitational wave background is usually reported in terms of its energy density per logarithmic frequency interval with respect to the closure density of the universe ($\rho_{c} = \frac{3 c^2 H_0^2}{8 \pi G} \approx 7.6 \times 10^{-9}$ erg/cm$^3$ with $H_0 = 67.74$ km/s/Mpc, $h = 0.6774$~\cite{Planck2015}, where $c$ is the speed of light and $G$ is Newton's constant), or specifically
\begin{equation}
\label{eq:Omega1}
\Omega_{GW}(f) = \frac{f}{\rho_{c}} \frac{d \rho_{GW}}{df} ~ .
\end{equation}
One can also consider the energy density of gravitational waves over a particular frequency band, namely $\Omega_{GW} = \int d \ln f ~ \Omega_{GW}(f)$~\cite{PhysRevX.6.011035}. 
The stochastic gravitational wave background could come from cosmological sources: the inflationary epoch, phase transitions in the early universe, alternative cosmologies, or cosmic strings. Alternatively, there could be an astrophysically-produced cosmological background. This could be produced from supernovae, magnetars, or the inspiral and merger of compact objects (neutron stars or black holes) over the history of the universe. Because of the recent observation of stellar mass binary black hole and binary neutron star mergers by Advanced LIGO and Advanced Virgo, it is likely that the stochstic background in the LIGO-Virgo observing band will be dominated by this source, with $\Omega_{GW}(f) \approx 10^{-9}$ at \unit[25]{Hz}~\cite{PhysRevLett.116.131102,Abbott:2017xzg}.

\subsection{Summary of recent gravitational wave detections}
\label{sec:detection_summary}
In the first observing run of Advanced LIGO (O1, September 12, 2015 to January 19, 2016) three gravitational wave signals were observed. GW150914 was reported as a definitive gravitational wave observation, with the signal created by the merger of a binary black hole pair with masses \unit[36]{$M_\odot$} and \unit[29]{$M_\odot$}, at a distance of \unit[410]{Mpc}. The total energy emitted in gravitational waves was \unit[3]{$M_\odot c^2$}~\cite{PhysRevLett.116.061102}. The second definitive gravitational wave observation was GW151226. This event was the result of the merger of two black holes with masses of \unit[14]{$M_\odot$} and \unit[7.5]{$M_\odot$}, at a distance of \unit[440]{Mpc}. A total of \unit[1]{$M_\odot c^2$} of energy was released as gravitational waves~\cite{PhysRevLett.116.241103}.
Finally, event LVT151012 was almost certainly a gravitational wave event, but because of the long distance to the source, \unit[1000]{Mpc}, it had a reduced gravitational wave amplitude and signal to noise ratio, and hence a lower statistical significance. The masses for this system were \unit[23]{$M_\odot$} and \unit[13]{$M_\odot$}. The energy released in gravitational waves was \unit[1.5]{$M_\odot c^2$}~\cite{PhysRevX.6.041015}.

The second observing run (O2, November 30, 2016 to August 25, 2017) of Advanced LIGO and Advanced Virgo has provided more events. Advanced Virgo joined O2 on August 1, 2017. Advanced LIGO observed gravitational waves from binary black hole mergers GW170104 (with masses of \unit[19.4]{$M_\odot$} and \unit[31.2]{$M_\odot$} at a distance of \unit[880]{Mpc})~\cite{PhysRevLett.118.221101} and GW170608 (with masses of \unit[12]{$M_\odot$} and \unit[7]{$M_\odot$}, the lightest binary black hole system observed to date, at a distance of \unit[340]{Mpc})~\cite{GW170608}. The first three-detector observation of gravitational waves between Advanced LIGO and Advanced Virgo was the detection of GW170814, another binary black hole system (with masses of \unit[25.3]{$M_\odot$} and \unit[30.5]{$M_\odot$} at a distance of \unit[540]{Mpc})~\cite{PhysRevLett.119.141101}. The Advanced LIGO and Advanced Virgo network then detected gravitational waves from a binary neutron star inspiral, GW170817~\cite{PhysRevLett.119.161101}; a gamma ray burst was detected 1.7 s after the merger~\cite{2041-8205-848-2-L13,2041-8205-848-2-L14,2041-8205-848-2-L15}, and the source was identified across the electromagnetic spectrum~\cite{2041-8205-848-2-L12}, thus beginning the era of gravitational wave multi-messenger astronomy.

\subsection{What is a stochastic gravitational wave background?}
\label{sec:sgwb}
A stochastic background of gravitational waves is very different from transient gravitational waves (binary inspirals, or burst events) or continuous periodic gravitational waves (coming from pulsars). These other sources are sending gravitational waves from specific locations in the sky. A stochastic background will be coming from all directions. To first approximation, the stochastic background is assumed to be isotropic; one could determine its statistical properties by observing any part of the sky~\cite{NLCthesis,Christensen:1992wi}. Searches for the stochastic background typically proceed with the hypothesis that it is uniform across the sky~\cite{PhysRevLett.118.121101}.
This is analogous to the cosmic microwave background, which is essentially isotropic, but, in fact it is ultimately anisotropic (with temperature anisotropies at the level of $10^{-5}$)~\cite{0067-0049-208-2-20,0067-0049-208-2-19}. Similarly, there are signal searches that attempt to measure an anisotropic stochastic gravitational wave background~\cite{PhysRevLett.118.121102}.

Unlike the other gravitational wave signals, a stochastic-background would just appear as \emph{noise} in a single gravitational wave detector. 
For example, consider some detector attempting to measure gravitational waves. The signal $s(t)$ from that detector would be the sum of the gravitational wave, $h(t)$, and noise, $n(t)$, or specifically, 
\begin{equation}
\label{eq:define}
s(t) = n(t) + h(t) ~ .
\end{equation}
However, the magnitude of the stochastic background will always be much smaller than the noise in the detector, $n(t) >> h(t)$. The only way to detect the stochastic background will be to take the correlation between two detector outputs,
\begin{equation}
\label{eq:coherence}
\begin{aligned}
\left \langle s_1(t) ~ s_2(t) \right \rangle ~ &= ~ \left \langle \left( n_1(t) + h(t) \right) \left( n_2(t) + h(t) \right) \right \rangle  \\
&= ~ \left \langle n_1(t) ~ n_2(t) \right \rangle + \left \langle n_1(t) ~ h(t) \right \rangle + \left \langle h(t) ~ n_2(t) \right \rangle + \left \langle h(t) ~ h(t) \right \rangle \\
&\approx ~ \left \langle h(t) ~ h(t) \right \rangle ~ ,
\end{aligned}
\end{equation}
(where the $\left \langle ~ \right \rangle$ represents the time average)
since it is assumed that the noise in each detector is statistically independent from one another, 
and also from the stochastic background.

In reality, the two detectors will be displaced from one another, so the detected signal will not be quite the same; the consequences of this will be articulated below.
Also, having two co-located detectors typically leads to common noise, as was the case for initial LIGO when it used two co-located detectors to attempt to measure the stochastic background~\cite{PhysRevD.91.022003}; Advanced LIGO does not have co-located detectors. As a consequence, LIGO and Virgo are attempting to measure the stochastic background through the correlation of the output of detectors displaced thousands of kilometers from one another. The assumption was that there would be no common noise; but even this assumption cannot be sustained~\cite{Thrane:2013npa,0264-9381-34-7-074002,0264-9381-33-22-224003}.

As will be described below, there will be numerous different methods used to try to measure a stochastic background in different frequency regimes. In all likelihood, the stochastic background's energy level will change very little over the observational band of the detector. There will not be large variations in the background when looking at it in the frequency domain, nor the time domain. The stochastic background would essentially be impossible to detect in a single detector. But through the correlation of data from different detectors, one could possibly extract the signal. In terms of formal statistical definitions, it is assumed that the background is stochastic, stationary, and ergodic~\cite{1959ZNatA..14..767B}.

There is certainly a stochastic gravitational wave background at some level. From all of the activity over the history of the universe space-time is constantly oscillating. Using the stochastic background to probe the earliest moments of the universe, for example from inflation~\cite{PhysRevD.23.347}, would provide an unprecedented window to the physics of the early universe~\cite{PhysRevD.48.3513,PhysRevLett.102.231301}. The gravitational waves produced in the early universe will have frequencies today that extend from $1/T_{\rm{Hubble}}$ to at least \unit[$10^{14}$]{Hz}, if not higher~\cite{NLCthesis,Christensen:1992wi}. However, for LIGO and Virgo, their observational band (from \unit[$10$]{Hz} to a few kHz) is likely to be dominated by a stochastic background produced by the merger of binary black holes and binary neutron stars over the history of the universe~\cite{PhysRevLett.116.131102,Abbott:2017xzg}.

A properly calibrated gravitational wave detector will produce an output of the measured gravitational wave strain, $h(t)$ (which is dimensionless). From the correlation of the output of two detectors one can measure the root mean square (rms) of the strain, $h_{rms}^2$, or the spectral density $S_{h}(f)$,
\begin{equation}
\label{Eq:specden}
h^2_{rms} = \left \langle \sum_{i,j} h_{ij} h_{ij} \right \rangle = \int_{0}^{\infty} df S_{h}(f) ~ .
\end{equation}
The energy density of the gravitational waves can be related to the spectral density, namely
\begin{equation}
\rho_{GW} = \int_{0}^{\infty} df \rho_{GW}(f) = \int_{0}^{\infty} df  S_{h}(f) \frac{\pi c^2 f^2}{8 G} ~ ,
\end{equation}
with
\begin{equation}
\frac{d \rho_{GW}}{df} = \rho_{GW}(f) ~ .
\end{equation}
In this case, Eq.~\ref{eq:Omega1} can be written as
\begin{equation}
\label{eq:Omega2}
\Omega_{GW}(f) = \frac{f \rho_{GW}(f)}{\rho_c} ~ .
\end{equation}

\subsection{The importance of observing a stochastic gravitational wave background}
\label{sec:importance}
Whether produced by cosmological or astrophysical sources, an observed stochastic gravitational wave background would provide a wealth of information about this universe. This is analogous to the cosmic microwave background; the observation of it and its anisotropies has revolutionized our understanding about the universe~\cite{0067-0049-208-2-20,0067-0049-208-2-19,refId0,Planck2015}. An even deeper view of the universe could come from the stochastic gravitational wave background. Gravitational waves from inflation would help to describe the universe at its earliest moments~\cite{PhysRevD.50.1157,Starobinski:1979aa,PhysRevLett.99.221301,PhysRevD.85.023525,PhysRevD.85.023534,1475-7516-2015-01-037,PhysRevD.55.R435,2006JCAP...04..010E}.
There is also the possibility that the initial state of the universe was perturbed via string cosmology. 
With string cosmology there could be a phase of accelerated evolution in advance of the Big Bang.
This would also create a disctinctive background of gravitational waves~\cite{GASPERINI1993317,doi:10.1142/S0217732393003433,PhysRevD.73.063008,1475-7516-2016-12-010}.  
These pre-Big Bang cosmologies might produce gravitational waves that could be observed in the LIGO-Virgo observational band~\cite{PhysRevD.73.063008,GASPERINI1993317,doi:10.1142/S0217732393003433,1475-7516-2016-12-010}. 
Cosmic strings, theorized topological defects produced by phase transitions in the early universe, vibrate and lose energy via gravitational wave emission over the history of the universe~\cite{0305-4470-9-8-029,Sarangi2002185,PhysRevLett.98.111101,Damour:2004kw}. If cosmic strings exist, they will create a stochastic background of gravitational waves, the observation of which would bring confirmation of physics beyond the Standard Model~\cite{PhysRevD.86.023503}.
A first order phase transition in the early universe would see the production of bubbles of different phases. The growth of sperical bubbles would not create gravitational waves, but the collision of bubbles would. The observation of a stochastic background produced by first order phase transitions would certainly provide significant information on cosmology and high-energy physics~\cite{PhysRevLett.69.2026,PhysRevD.49.2837,PhysRevD.90.107502}. 

An astrophysically produced stochastic gravitational wave background certainly exists at some level. The recent observations by Advanced LIGO and Advanced Virgo of binary black hole and binary neutron star mergers~\cite{PhysRevLett.116.061102,PhysRevLett.116.241103,PhysRevX.6.041015,PhysRevLett.118.221101,GW170608,PhysRevLett.119.141101,PhysRevLett.119.161101} imply that a stochastic background will be produced by these events happenning over the full history of the universe~\cite{PhysRevLett.116.131102,PhysRevLett.118.121101,Abbott:2017xzg}. A stochastic background produced by binary black hole mergers is likely to be the loudest background in the LIGO-Virgo band, and one that may ultimately be observable by them~\cite{Abbott:2017xzg}. The merger of binary neutron star systems over the course of the universe will also contribute significantly to the stochastic background~\cite{doi:10.1093/mnras/stt207,PhysRevD.85.104024,PhysRevD.84.084004,Abbott:2017xzg}. An astrophysically produced stochastic background would have contributions from core collapse supernovae~\cite{PhysRevD.72.084001,Zhu:2010af}, rotating neutron stars~\cite{PhysRevD.86.104007}, differentially rotating neutron stars~\cite{PhysRevD.87.063004}, and magnetars~\cite{PhysRevD.95.083003}  throughout the universe. Any information derived from an astrophysically-produced stochastic background would provide significant information about astrophysical processes over the history of the universe.
Clearly the differentiation between the different sources of a stochastic background will be difficult to observe and will ultimately require the observation of the frequency dependence of the stochastic background over an extended frequency band.

\subsection{Methods used to measure a stochastic background}
\label{subsec:methods}
There are many methods that are currently being used today to try to observe a stochastic background of gravitational waves. 
A number of techniques have been proposed for future attempts to observe the stochastic background. 
These methods will be reviewed below. However, a recent review provides an extremely comprehensive explanation of all of the methods used and proposed to observe the stochastic background, and the interested reader is encouraged to consult that summary~\cite{Romano2017}. In addition the article~\cite{PhysRevX.6.011035} provides an excellent overview on observational limits on the stochastic background over 29 decades in frequency.

LIGO and Virgo have used correlation methods between two or more interferometric detectors to attempt to measure the stochastic background~\cite{NLCthesis,Christensen:1992wi,PhysRevD.59.102001}. While no signal was detected, upper limits have been placed on the energy density of the background from \unit[$20$]{Hz} to \unit[$1000$]{Hz}~\cite{PhysRevD.69.122004,PhysRevLett.95.221101,0004-637X-659-2-918,Abadie:2011fx,Abbott:2011rs,PhysRevLett.113.231101,PhysRevD.91.022003,PhysRevLett.118.121101}. Pulsar timing has been used to try to detect a stochastic background in the \unit[$10^{-9}$]{Hz} to \unit[$10^{-8}$]{Hz} band~\cite{PhysRevX.6.011035}. The temperature and polarization anisotropies of the cosmic microwave background can be used to constrain the energy density of the stochastic gravitational wave background in the \unit[$10^{-20}$]{Hz} to \unit[$10^{-16}$]{Hz} band~\cite{PhysRevX.6.011035,Pagano2016823}. The normal modes of oscillation of the Earth can even be used to constrain the stochastic background energy density in the \unit[$0.3$]{mHz} to \unit[$5$]{mHz} band~\cite{PhysRevD.90.042005}.

In the future (probable launch in the 2030s), the space based gravitational wave detector LISA~\cite{eLISA:2014} will search for a stochastic background in the \unit[0.1]{mHz} to \unit[100]{mHz} band. Earth based atomic interferometers are being proposed to search for gravitational waves, including a stochastic background, in the \unit[0.3]{Hz} to \unit[3]{Hz} band~\cite{PhysRevD.93.021101}. A detector such as this would occupy an important location in the frequency spectrum between LISA and LIGO-Virgo. The proposed, space-based DECi-hertz Interferometer Gravitational wave Observatory (DECIGO) would attempt to observe gravitational waves from 0.1 Hz to 10 Hz~\cite{1742-6596-122-1-012006,1742-6596-840-1-012010}.

Presented in Sec.~\ref{sec:detection} will be a more detailed description of the methods to observe the stochastic background, what their senistivities are at present, and what their sensitivities are expected to be in the future.

\section{Summary of sources of a possibly observable stochastic gravitational wave background}
\label{sec:sources}
There are a number of sources of a stochastic background. Below we summarize the most probable backgrounds produced via cosmological or astrophysical phenomena. An excellent review of astrophyically-produced stochastic backgrounds can be found in~\cite{2011RAA....11..369R}, however the implications of the observations by Advanced LIGO and Advanced Virgo of gravitational waves from binary black holes and binary neutron stars has significantly increased the probability that an astrophysically-produced stochastic background will be observed in the near future~\cite{PhysRevLett.116.131102,Abbott:2017xzg}.

\subsection{Inflation}
\label{subsec:inflation}
The electromagnetic analog to the stochastic background is the cosmic microwave background (CMB). In the early universe the fundamental particles and photons were in thermal equilibrium. Up until about 400,000 years after the Big Bang, protons, electrons and photons formed a cosmic soup, and continuously bounced off one another. However, due to the expanding universe the temperature of the universe dropped, and neutral hydrogen was eventually formed. This event is referred to as {\it recombination}, although it is the first time in which electrons and protons combined to form neutral hydrogen. At this moment the photons were free to propagate away, and essentially did not interact anymore with matter. 

The CMB was observed for the first time, albeit accidentally, in 1964 (when the age of the universe was 13.8 billion years~\cite{1538-4357-563-2-L95,Planck2015,0067-0049-208-2-19}) by Arno Penzias and Robert Wilson of Bell Laboratories, in New Jersey, USA~\cite{1965ApJ...142..419P}. The explanation of the cosmological origin of the observation was published simultaneously~\cite{1965ApJ...142..414D}, although the existence of the CMB had been predicted before~\cite{1948Natur.162..774A}. The CMB is observed today to have a perfect black body temperature distribution corresponding to 2.726 K~\cite{1994ApJ...420..439M,0004-637X-707-2-916}. There are slight temperature anisotropies across the sky of order 30 $\mu$K RMS~\cite{1992ApJ...396L...1S}. From these temperature fluctuations, specifically how they vary as a function of angular scale, it is possible to estimate the cosmological parameters that describe our universe~\cite{0264-9381-18-14-306,1538-4357-563-2-L95,0067-0049-208-2-19,Planck2015}.

While the cosmological information provided by the CMB is astounding, specific features of the CMB raise a number of questions. For example, any two points on the sky separated by more than 2$^o$ were causally disconnected at the time of recombination. This then begs the question: how is it possible that the temperature of two points on opposite sides of the sky have the same temperature (to a part in $10^{5}$) if they have not been in thermal equilibrium with each other? This is what is known as the {\it Horizon Problem}. The temperature fluctuations of the CMB as a function of angular scale on the sky can be used as input for Bayesian parameter estimation methods~\cite{0264-9381-18-14-306,1538-4357-563-2-L95} that then allow for the estimation of cosmological parameters~\cite{0067-0049-208-2-19,Planck2015}. From this, as well as other methods, it is apparent that the present energy density of the universe (considering radiation, baryonic matter, dark matter, dark energy) seems to be equal, or nearly equal, to the closure density of the universe
\begin{equation}
\rho_{c} = \frac{3 H_{0}^{2}}{8 \pi G} = 7.8 \times 10^{-9} \rm{ergs/cm}^{3}
\end{equation}
with a Hubble constant of $H_{0} = 67.74 ~ \rm{km/s/Mpc}$~\cite{Planck2015}. If the current energy density of the universe is equal to the critical energy density then the curvature of the universe is zero, namely the universe is flat. The question then becomes, how is it possible that we find ourselves in such a special state of curvature? And if we are just close to a curvature of zero now, then earlier in the universe the curvature must have been even closer to zero. This is what is known as the {\it Flatness Problem}.

The theory of {\it inflation} solves these problems~\cite{PhysRevD.23.347,Linde82}. It is assumed that in the very earliest moments the universe went through a period where its size grew exponentially, namely $a(t) \propto e^{H_{vac} t}$, where $a(t)$ is the scale parameter of the universe, and $H_{vac}$ is the Hubble parameter at that time~\cite{Binetruy_2015}. This expansion could be caused by the presence of some scalar field, let us call it $\phi$, which would give the space at that time some energy density, $\rho_{vac}$, which would then be related to the square of the Hubble parameter by $H_{vac}^2 \propto \rho_{vac}$~\cite{Binetruy_2015}. Eventually the decay of the scalar field to our present vacuum would put an end to the exponential inflation at that time, and provide the energy for the production of the fundamental particles that we are aware of today.

This rapid expansion of the universe has the effect of driving the curvature of the universe to zero, thus solving the Flatness Problem. It also means that our entire observable universe occupied a region which was in casual contact, and presumably thermal equilibrium, before the effect of the exponental expansion drove the regions apart from one another.

At this early period in the universe quantum mechanics would have played an important role in the evolution of the universe. All quantum fields have vacuum fluctuations associated with them. This would have been true for the inflationary field $\phi$ as well. Scalar fluctuations in the field could have served as the initial seeds for the distributions of matter that we see in the universe today. However, there would also have been tensor fluctuations, and these would have produced gravitational waves~\cite{starobinsky1979jetp,starobinksii,kolbturner,bar-kana,PhysRevD.85.023525,PhysRevD.85.023534,1475-7516-2015-01-037,PhysRevD.55.R435}. Gravitational waves could also be produced at the end of inflation, during the period of {\it pre-heating}, when the scalar filed is descaying into the material that makes up the present day universe~\cite{PhysRevLett.99.221301,2006JCAP...04..010E}. These primordial gravitational waves, if observed, could provide information about the universe in this inflationary era.

The gravitational waves produced during inflation would exist today over wavelengths corresponding to the size of the observable universe, down to sub-atomic distances. 
For frequencies above $10^{-17}$ Hz the predicted background is around $\Omega_{GW} \approx 10^{-15}$, a level that will likely be difficult to observe by any technique at any wavelength. Note that for lower frequencies ($10^{-17}$ Hz corresponds to a period of 23\% of the age of the universe) there is an increase in the predicted energy density of the stochastic background as perturbations from the early universe that were {\it frozen out} (being larger than observable size of the universe) re-enter and propagate again as gravitational waves.
Of course, alternative inflationary scenarios could produce a stochastic background at different levels. 

\subsection{Cosmic Strings}
\label{subsec:cosmic_strings}
Cosmic strings are a unique possibility for new physics that could be observed via gravitational waves. 
These would be one-dimensional topological defects, or false vacuum remnants, produced after a spontaneous symmetry phase transition~\cite{0305-4470-9-8-029,vilenkin:1994} from a 
broad variety of field theories, for example, Grand Unified 
Theories applied in the early universe~\cite{PhysRevD.68.103514}. Their formation happens at the end of inflation~\cite{SAKELLARIADOU200968}.

Cosmic strings are classical objects. Cosmic superstrings are other theorized objects; these would be quantum objects, even though they would extend to cosmological distances~\cite{Sakellariadou2881}. The formation of cosmic superstrings would occur at the end of {\it brane} inflation, when D-branes annihilate, or via brane collisions~\cite{Sakellariadou2881}.

When cosmic strings intersect they always swap partners, or when a single string folds upon itself, the connection interchange creates a cosmic string loop~\cite{SHELLARD1987624,PhysRevD.41.1751}. On the other hand, when cosmic superstrings intersect the probability of swapping partners is less than one~\cite{Sakellariadou2881}, even much less than one~\cite{1475-7516-2005-04-003}. This can lead to an excess in the density of cosmic superstrings~\cite{1475-7516-2005-04-003}. The intercommutation probability, $p$, is a very important parameter concerning the production of gravitational waves in the universe. 
There are predictions that the intercommutation probability $p$ should be in the range of $10^{-1}$ to 1 for D-strings, or $10^{-3}$ to $1$ for F-strings~\cite{Jackson:2004zg}.
Cosmic strings and cosmic superstrings create gravitational waves~\cite{Damour:2004kw}. 
When cosmic strings intersect, cusps and kinks will be formed. 
Cosmic string kinks~\cite{PhysRevD.50.2496,PhysRevD.43.3173,PhysRevD.95.023519} are discontinuities on the tangent vector of a string, while cusps are points where the string instantaneously reaches the speed of light~\cite{Damour:2001bk,Damour:2004kw,PhysRevLett.98.111101}.
These cusps and kinks will create bursts of gravitational waves, whose waveforms can be predicted~\cite{Damour:2000wa,Damour:2001bk,Damour:2004kw}. The superposition of these gravitational waves from cosmic strings produced over the history of the universe will create a stochastic background of gravitational waves~\cite{Damour:2004kw,PhysRevLett.98.111101}.

Cosmic strings are characterized by the dimensionless tension of the string, $G \mu$ (assuming $c = 1$), where $\mu$ is the mass per unit length and $G$ is Newton's constant. The product $G\mu$ is thus an unknown parameter that will affect the production of gravitational waves, and can be constrained by searches for gravitational waves (even null results)~\cite{Abbott:2011rs,Aasi:2013vna,0264-9381-32-4-045003,0004-637X-821-1-13}. 

Assuming that the magnitude of loops is defined by the gravitational backreaction scale (namely, the effect of the emitted gravitational waves changing the state of cosmic string that created them), the null search results from initial LIGO place upper limits on the
string tension of $G \mu < 10^{-8}$ for particular regions of the cosmic string parameter space~\cite{Aasi:2013vna}. 
The string tension has also been constrained through observations of the CMB to be less than $10^{-7}$~\cite{Ade:2013xla,Lizarraga:2016onn,0264-9381-32-4-045003,Lazanu:2014eya}. Cosmic string loops will oscillate, producing gravitational waves~\cite{PhysRevD.42.354,Hogan87}. 
Combining gravitational wave observations~\cite{Aasi:2013vna} and cosmological data (CMB~\cite{refId0A15,2041-8205-749-1-L9,1475-7516-2014-04-014}, baryon acoustic oscillations~\cite{doi:10.1111/j.1365-2966.2011.19250.x,doi:10.1111/j.1365-2966.2012.21888.x,doi:10.1111/j.1365-2966.2012.22066.x}, gravitational lensing data~\cite{refId0A17}), and again assuming that the size of the loops is determined by the gravitational backreaction scale, string tension values greater than $4 \times 10^{-9}$
are excluded for an intercommutation probability of $p = 10^{-3}$~\cite{0264-9381-32-4-045003}. 

The data from Advanced LIGO and Advanced Virgo are now being used to search for cosmic string gravitational wave signals.
The anlaysis of the Advanced LIGO data from the first observing run, O1, has recently been published~\cite{Abbott:2017mem}. No gravitational wave signals from cosmic strings were observed.
That fact, along with the upper limits set on the energy density of the stochastic background from the Advanced LIGO O1 data,  $\Omega_{GW} < 1.7 \times 10^{-7}$ for 20 - 86 Hz~\cite{PhysRevLett.118.121101}, were used to constrain three cosmic string models. One model (M1) assumes that all cosmic string loops were formed with roughly the same size, and the loops do not self-interact after they were created~\cite{vilenkin:1994,vilenkin2000cosmic,PhysRevD.73.105001}. The next model (M2) uses numerical calculations to predict the size of the cosmic string loops when they were created, as well as the creation rate as a function of time~\cite{PhysRevD.89.023512}. The third model (M3) differs from M2 in that it considers the distribution (as a function of time) of loops that do not self interact; it also considers the back-reaction on the loops when gravitational waves are emitted~\cite{1475-7516-2007-02-023,1475-7516-2010-10-003}. The lack of detection of such gravitational wave bursts in the Advanced LIGO O1 data constrains M3, assuming an intercommutation probability $p = 1$, to have a string tension $G \mu < 1 \times 10^{-9}$; the O1 burst search does not significantly constrain M1 and M2. The results of the Advanced LIGO O1 upper limits for the energy density of the stochastic background essentially exclude M3. For M1, the O1 stochastic search result constrains the string tension, assuming $p=1$, to be $G \mu < 5 \times 10^{-8}$; for M2 the constraints are weaker, and with a reduction of intercommutation probability to $p = 0.1$, a tension constraint of $G \mu < 5 \times 10^{-8}$ can also be set. See ~\cite{Abbott:2017mem} for the complete details of this study.

\subsection{First Order Phase Transitions}
\label{subsec:phase_transition}
In the physics world there are many types of phase transitions. In our day to day lives we see transitions from solid, liquid and gaseous matter. A particular medium in thermal equilibrium will have uniform characterists pertaining to its physical qualities. But when a phase transition occurs,
some of these physical characteristics will change. Some of the changes can even happen discontinuously~\cite{Papon:1339609}. 

A first order phase transition has a discontinuity in the first derivative of the free energy with respect to a thermodynamic parameter. Consider the Gibbs free energy
\begin{equation}
G(p,T) = U + pV - TS ~ ,
\end{equation}
where $p$ is the pressure, $T$ is the temperature, $U$ is the internal energy of the system, $V$ is the volume, and $S$ is the entropy.

From the Maxwell relations we have $S = -\frac{\partial G}{\partial T}_{p}$ and $V = \frac{\partial G}{\partial p}_{T}$. If these quantities were discontinuous, then we would have a first order phase transition. As a simple example, consider water changing from a liquid to a gas, namely the water is boiling. Both the entropy, $S$, and the volume, $V$, change abruptly when going from one phase to the other. In fact, the change in entropy can be related to the latent heat of the process, $L = T \Delta S$.

Second order phase transitions have a discontinuity in the second derivative (with respect to thermodynamic parameters) of the free energy, while the first derivatives remain continuous. As an example, second order phase transitions are observed in superconductors, or the ferromagnetic phase transition in iron.

First order phase transitions in the early universe could produce a significant stochastic background of gravitational waves. The boiling water analogy can be made, but now one can imagine bubbles of a different phase of the universe forming from within another older phase. The early universe certainly experienced a number of phase transitions. If one considers the Standard Model, there was presumably a grand unification period when the electromagnetic, weak and strong forces were all unified. As the universe cooled there would have been a transition to phase with the electroweak force and the strong force separated. Eventually an electroweak phase transition would see the separation of the electromagnetic force from the weak force. The standard electroweak phase transition is not a first order phase transition, but slight modifications to the Standard Model could produce a first order electroweak phase transition~\cite{PhysRevLett.119.141301}. It is estimated that a {\it cross-over} between the unified electroweak phase and the subsequent broken phase would have happened at a temperature of $T_{c} = 159.5 \pm 1.5$ GeV~\cite{PhysRevD.93.025003}. However, if some modification to the Standard Model would have produced a first order phase transition at this energy scale then there would be a stochastic background of gravitational waves peaking at a frequency of about 260 mHz~\cite{PhysRevD.49.2837}. What makes this so exciting is that this is within the observing band of LISA~\cite{eLISA:2014,2017arXiv170200786A}. This is one of the reasons why outside of the LHC experiments at CERN, LISA may offer the best prospects for acquiring high energy physics information, and especially possible extensions to the Standard Model.

The Standard Model extensions to the electroweak phase transition, if they existed, would have important physical consequences. Electroweak baryogenesis could help to explain cosmic baryon asymmetry~\cite{1367-2630-14-12-125003}. 
Electroweak baryogenesis pertains to mechanisms that would produce an asymmetry in baryon density during the electroweak phase transition, and could then possibly explain the observed abundance of matter over anti-matter (baryon asymmetry) in the universe. Electroweak baryogenesis also satifies the famous Sakharov conditions~\cite{0038-5670-34-5-A08}: the interactions occur out of thermal equilibrium; charge (C) and charge-parity (CP) symmetries are violated; there is a violation of baryon number. Electroweak baryogenesis provides an example of a first order phase transition that could address baryon asymmetry and also produce gravitational waves in the early universe. In this modification to the electroweak theory bubbles (of a new vacuum phase) would be created when the Higgs field transitions into the vacuum state where the electroweak symmetry is spontaneously broken. These bubbles would then expand. The C and CP violation would occur when particles present scatter off of the front of the expanding bubble walls. The C and CP asymmetries occurring in front of the expanding bubble wall would produce baryon number violation, giving more baryons (matter) than antibaryons (antimatter)~\cite{1367-2630-14-12-125003,1705.01783}.

In addition to the possible explanation for one of the great mysteries of the universe - why we have a suplus of matter over antimatter - we also have a mechanism that can create a significant background of gravitational waves. The characteristics of the gravitational waves produced by a first-order phase transition depend on the expansion speed of the bubble walls, the latent heat of the transition, and the rate at which bubbles of the new phase are created~\cite{1475-7516-2016-04-001,1705.01783}.

With first order phase transitions, gravitational waves are created via different physical meachanisms. An expanding bubble will be spherical, so will not produce gravitational waves; however, when bubble walls collide, there will be gravitational wave production. The plasma that is present can also experience shocks, and these discontinuities between regions of different plasma properties could also generate gravitational waves~\cite{PhysRevD.45.4514,PhysRevD.77.124015,1475-7516-2008-09-022,1475-7516-2016-04-001}. After the bubble collisions there will be sound waves in the plasma; these can create gravitational waves~\cite{PhysRevLett.112.041301,PhysRevD.92.123009,1475-7516-2016-04-001}. Because of the very large Reynolds number that would exist for this fluid, turbulent motion results; a large magnetic Reynolds leads to an amplification of the magnetic fields created by the movement of charges during the phase transition~\cite{1475-7516-2009-12-024}. Finally, magnetohydrodynamic turbulence can produce gravitational waves; the magnetic fields and turbulent motions can create stresses that are anisotropic. This can ultimately be an efficient way to convert magnetic energy to gravitational wave energy~\cite{PhysRevD.74.063521,PhysRevD.92.043006,1475-7516-2016-04-001}. All of these processes would typically be present after a first order phase transition. The amount of gravitational waves produced by these different effects would depend on the dynamics of the first order phase transition. The sensitivity of LISA for detecting a stochastic background will be of order $\Omega_{GW} \sim 5 \times 10^{-13}$ at $10^{-3}$ Hz~\cite{1475-7516-2016-12-026}. Many of the modifications to the electroweak phase transition, making it first order, would create a stochastic gravitational wave background that could be detectable by LISA~\cite{1475-7516-2016-04-001}. 
The possibility to detect a stochastic background created by a first order phase transition in the early universe is an amazing opportunity to observe new physics outside of the standard model.

\subsection{Pre Big Bang Models}
\label{subsec:pre_BB}
Some pre Big Bang models are an extension of the standard inflationary cosmology. The theories consider the consequences for cosmology when some version of superstring theory is applied. As noted above, the stochastic background of gravitational waves generated via quantum fluctuations during inflation would result in an energy density that is essentially flat in frequency, and at a very small level currently, $\Omega_{GW} \sim 10^{-15}$. In pre Big Bang models the universe would begin with a string perturbative vacuum scenario~\cite{GASPERINI1993317,doi:10.1142/S0217732393003433,GASPERINI20031,1475-7516-2016-12-010}.
The universe materializes via a highly perturbative initial state before the Big Bang. In the standard inflationary scenario there would have been an initial singularity~\cite{PhysRevLett.72.3305,PhysRevLett.90.151301}. Superstring theory allows for the assumption that there is no singularity associated with the Big Bang, and hence it is logical to extend time to before the Big Bang.

With string cosmology (namely the pre Big Bang scenario) there will be a different behaviour for the curvature scale of the universe, as opposed to that in the standard inflationary cosmology. Standard inflation has a constant curvature scale before reaching the radiation dominated (standard Big Bang) era. However, with string cosmology there would be a growth in the curvature scale, going from a low curvature scale to some maximum curvature scale that would be defined by the string scale. This is the so called string inflation.
The curvature scale of the universe would then diminsh, and the radiation dominated era of the standard cosmology would ensue. The universe would not have experienced a singularity with an infinite curvature scale, but instead would be finite through the effects of the stringy phase~\cite{GASPERINI20031}.

This dynamical process in the early universe would be a source of gravitational waves, and would create a stochastic background that would be present today~\cite{Buonanno:1996xc,GASPERINI20031,PhysRevD.73.063008}. Initially the universe would be in a low energy, low curvature-scale, \textit{dilaton} phase. The dilaton is the assumed fundamental scalar for the string theory. An inflationary evolution would occur due to the kinetic energy of dilaton field. As the curvature scale increases the universe is described by a high energy string phase. Eventually the curvature scale approaches the string scale and higher order corrections become important in the string action. This is when the universe transitions to the radiation dominated era described by the standard cosmology. These transitions from different expansion rates for the universe will create gravitational waves~\cite{Buonanno:1996xc,GASPERINI20031,PhysRevD.73.063008}. This process can then create a background that can peak at higher frequencies, possibly within the observation band of LIGO-Virgo, or LISA. The parameters pertaining to the string phase will affect the frequency dependence of the stochastic background~\cite{Buonanno:1996xc,PhysRevD.73.063008}. Whether or not LIGO and Virgo will be able to to observe a stochastic background from a pre Big Big cosmology has been the source of active investigation~\cite{PhysRevD.73.063008,1475-7516-2016-12-010}.

Various observations are already constraining pre Big Bang models. No stochastic background has been detected at this point, so the upper limits on the energy density of the stochastic background in various frequency bands can make some restrictions on pre Big Bang theories.
Specifically, observations of the cosmic microwave background (CMB), and stochastic background energy limits set by Advanced LIGO and pulsar timing are able to currently constrain pre Big Bang parameters~\cite{PhysRevX.6.011035,0264-9381-32-4-045003,1475-7516-2016-12-010,1475-7516-2017-06-017}.

This string cosmology would produce both scalar and tensor perturbations to the metric of the universe. Observations of the CMB, for example from Planck, estimate cosmological parameters such that it appears that scalar perturbations are creating a stochastic background that is decreasing with frequency, in contrast to pre Big Bang predictions~\cite{Planck2015}. This constrains the parameters responsible for the very low frequency gravitational waves produced in the pre Big Bang evolution~\cite{1475-7516-2016-12-010}. 

Pulsar timing arrays provide another important limit on the stochastic background that constrains pre Big Bang models~\cite{PhysRevX.6.011035}.
For example, the Parkes Pulsar Timing Array placed a limit on the energy density of the stochastic background to be $\Omega_{GW}(f) < 2.3 \times 10^{-10}$ at $f = 1/\rm{year}$~\cite{Shannon1522,PhysRevX.6.011035}. Finally, the recent upper limit by Advanced LIGO, $\Omega_{GW}(f) < 1.7 \times 10^{-7}$ from  20 -- 86 Hz further constrains pre Big Bang models~\cite{PhysRevLett.118.121101}. In order for the pre Big Bang models to exist within these observational constraints fine tuning must be done on the string parameters. That said, it has still been demonstrated that pre Big Bang models could produce a stochastic background that peaks within the Advanced LIGO - Advanced Virgo observational band, or the LISA observational band~\cite{1475-7516-2016-12-010}.

\subsection{Binary Black Holes}
\label{subsec:bbh}
A stochastic background produced by binary black holes is highly probable. After Advanced LIGO's observations of two significant events, and another probable event, in its first observing run (O1) it became clear that there will likely be a stochastic background produced by all binary black hole mergers over the history of the universe~\cite{PhysRevLett.116.061102}. More binary black hole inspiral gravitational wave events were subsequently observed by Advanced LIGO and Advanced Virgo in the second observing run (O2)~\cite{PhysRevLett.118.221101,GW170608,PhysRevLett.119.141101}.
This astrophysically produced background will likely be the loudest stochastic background in the observing band of LIGO and Virgo, from 10 Hz up to 1000 Hz.

\begin{figure}
\begin{center}
\includegraphics[width=5.5in]{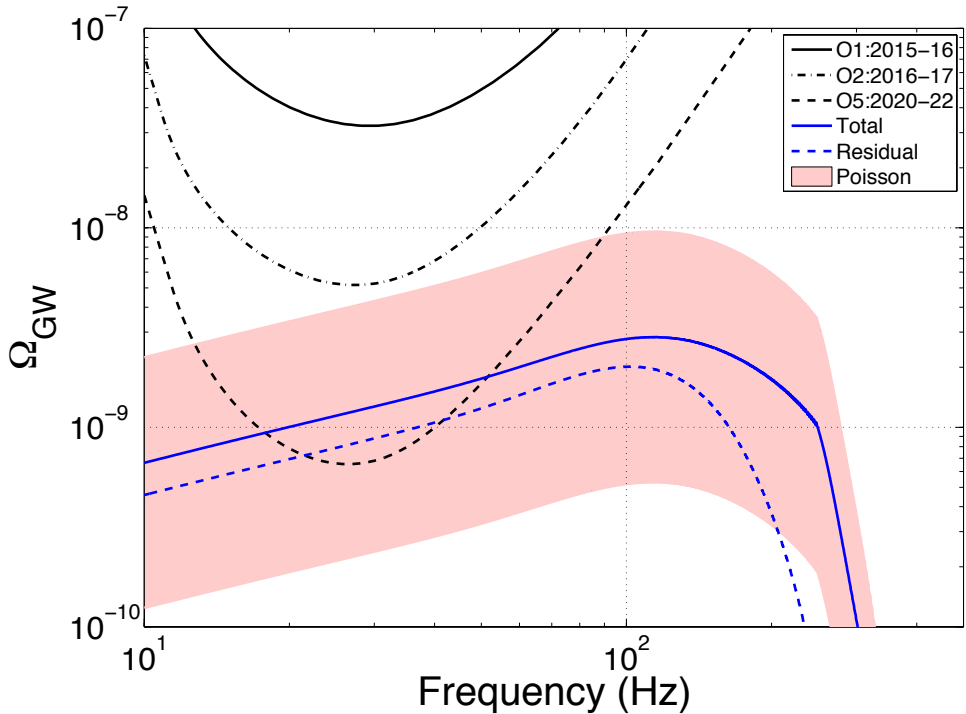}
\caption{As presented in \protect \cite{PhysRevLett.116.131102}, the predicted Advanced LIGO - Advanced Virgo network sensitivity to a stochastic background produced by binary black holes that were formed through binary stellar evolution. Displayed are the energy
density spectra (solid for the total background; dashed for the residual background, excluding resolved sources,
assuming final Advanced LIGO and Advanced Virgo
sensitivity).  The pink region represents the uncertainty in the estimation. The black curves (O1, O2 and O5) display the 1$\sigma$ sensitivity of the Advanced LIGO - Advanced Virgo network expected (at the time of the publication of \protect \cite{PhysRevLett.116.131102}) for the observing runs O1 and O2, and at the design sensitivity (2 years of observation in O5). Figure from \protect \cite{PhysRevLett.116.131102}.}
\label{fig:BBH_background1a}
\end{center}
\end{figure}

Immediately after the observation of GW150914~\cite{PhysRevLett.116.061102} LIGO and Virgo reported on the implications that the observation of a stellar mass binary black hole merger would have on the stochastic background~\cite{PhysRevLett.116.131102}. The detection made clear that the universe contains a population of stellar mass binary black holes. Consequently the binary black hole produced stochastic background should be larger than what was expected previously. This stochastic background would be created from all of the binary black hole mergers in the observable universe over its 13.8 billion year history. Using various scenarios and parameters for the formation of stellar mass binary black hole systems, LIGO and Virgo used the observation of GW150914 to predict that around 25 Hz (where Advanced LIGO and Advanced Virgo will have the best sensitivity for detecting a stochastic background) the estimated energy of the binary black hole produced stochastic background will be $\Omega_{GW}(f = 25 ~ \rm{Hz}) = 1.1^{+ 2.7}_{-0.9} \times 10^{-9}$~\cite{PhysRevLett.116.131102}. See Fig.~\ref{fig:BBH_background1a}.

LIGO and Virgo have now observed several binary black hole mergers. In O1 there were GW150914~\cite{PhysRevLett.116.061102}, GW151226~\cite{PhysRevLett.116.241103} and the probably (but not definitive) LVT151012~\cite{PhysRevX.6.041015}. At the time of this writing, LIGO and Virgo have announced the detection of three binary black hole mergers observed in their second observing run, O2: GW170104~\cite{PhysRevLett.118.221101}, GW170608~\cite{GW170608}, and GW170814~\cite{PhysRevLett.119.141101}. Using these observations, LIGO and Virgo now estimate the energy of the binary black hole produced stochastic background will be $\Omega_{GW}(f = 25 ~ \rm{Hz}) = 1.1^{+ 1.2}_{-0.7} \times 10^{-9}$ ~\cite{Abbott:2017xzg}. The level is the same as that of the initial observation~\cite{PhysRevLett.116.131102}, but the error has narrowed.

In order to estimate the stochastic background from binary black hole mergers one must take into account many factors. For example, it is necessary to understand the mechanism by which these binaries are formed, which would then help to explain how often these sorts of mergers occur in the universe. The formation rate will depend on when this happens in the age of the universe, and the metallicity of the formation environment. The merger rate, as a function of redshift, will also be required. 

A comprehensive explanation of how to calculate the contribution of binary black hole mergers to the stochastic background is given in~\cite{PhysRevLett.116.131102}, and presented here is a summary of that demonstration. 
Some set of intrinsic source parameters $\theta$ will describe the ensemble of binary black holes. These source parameters could be things like the masses and spins of the black holes.
The distribution of these parameters are essentially unknown at present. 
However, the recent observations of binary black hole mergers by LIGO and Virgo~\cite{2041-8205-833-1-L1,0067-0049-227-2-14,2041-8205-832-2-L21,PhysRevLett.118.221101,GW170608,PhysRevLett.119.141101} and previous assumptions~\cite{Kim2003} allow for dividing this ensemble into different subsets. Consider a subset of binary black holes $k$ described by parameters $\theta_k$ (for example, the mass and spin values). Call $R_m(z;\theta_k)$ the merger rate per comoving volume per unit source time; this depends on the formation rate of black hole binaries as a function of redshift and also the distribution of the time delays between binary black hole formation and merger~\cite{Nakar:2007,PhysRevLett.116.131102}.
Then the total gravitational wave energy density spectrum for this particular class is (see, e.g.~\cite{2011RAA....11..369R,2011ApJ...739...86Z,PhysRevD.84.084004,2011PhRvD..84l4037M,PhysRevD.85.104024,2013PhRvD..87d2002W,doi:10.1093/mnras/stt207,2015A&A...574A..58K}):
\begin{equation}
\label{eq:Omega4}
\Omega_{GW}(f;\theta_{k}) = \frac{f}{\rho_{c} H_{0}} \int_{0}^{z_{max}} dz \frac{R_{m}(z,\theta_{k}) \frac{d E_{GW}}{d f_{s}}(f_{s},\theta_{k})}{(1 + z) E(\Omega_{M},\Omega_{\Lambda},z)} ~.
\end{equation}
Note the term that accounts for cosmology, namely the dependence of how the comoving volume depends on redshift appears through $E(\Omega_{M},\Omega_{\Lambda},z) = \sqrt{\Omega_{M}(1 + z)^{3} + \Omega_{\Lambda}}$. The spectral density of the energy of gravitational waves emitted at the source is $\frac{d E_{GW}}{d f_{s}}(f_{s},\theta_{k})$.
Then to calculate the total energy density a sum is done over all source classes $k$~\cite{PhysRevLett.116.131102}.

The formation scenarios for binary black hole systems are important for predicting the expected rate of mergers over the history of the universe~\cite{2041-8205-818-2-L22}. This would affect the predicted level of the subsequently produced stochastic background.
In one scenario, the binary black holes are created as isolated binaries of massive stars
in galactic fields~\cite{Belczynski:2016obo,lrr-2006-6,2041-8205-818-2-L22}.
An important observation for forming black holes similar to those observed in GW150914 is that there is a need for low metallicity, typcially smaller than 10\% of the solar metallicity; the initial stars would have masses in the range of $40 - 100 M_{\odot}$~\cite{Belczynski:2016obo}. 
The other formation channel for binary black holes is through dynamical interactions in
dense stellar environments, for example, as one might find in globular clusters~\cite{Rodriguez:2016kxx,2013LRR....16....4B,2041-8205-818-2-L22}.
Studies indicate that globular clusters can produce a significant population of massive black hole binaries that merge in the local Universe, with most of the resulting binary black hole systems having total masses from $32 M_{\odot}$ to $64 M_{\odot}$~\cite{Rodriguez:2016kxx}.
The formation rate as a function of redshift will ultimately affect the production of a stochastic background, and that will depend on how the binary black hole systems are formed. Clues as to the dominant formation channel may come through the observations of the spins and orbital eccentricities of a large number of gravitational wave events from binary black hole mergers~\cite{2041-8205-818-2-L22}.

Black holes that have been proposed to have been produced in the early universe are referred to as {\it primordial} black holes~\cite{1967SvA....10..602Z,1971MNRAS.152...75H,1974MNRAS.168..399C,1974A&A....37..225M,1975Natur.253..251C,PhysRevD.94.083504}. Primordial black holes have now also been suggested as the source of binary black hole systems in the universe. The possibility that dark matter could consist of primordial black holes has been raised after Advanced LIGO's and Advanced Virgo's observation of gravitational waves from binary black hole mergers. The observed masses for the binary black holes systems have been relatively large. There are claims that the mass window of  $20 - 100 M_{\odot}$ cannot be excluded as the source of dark matter and could be the source of the LIGO-Virgo observations~\cite{PhysRevLett.116.201301}.
Given the presumed existence of primordial black holes, their implications for contributing to a binary black hole produced stochastic background has been investigated. One conclusion has been that the magnitude of the energy density from primordial black holes is much lower than that arising from the stellar produced binary black hole mergers~\cite{PhysRevLett.117.201102}. Other work has suggested that primordial black hole formation could be responsible for supermassive binary black hole mergers creating a stochastic background at the limit of what could be detected by pulsar timing experiments today~\cite{PhysRevD.95.043511}.

Predictions suggest that there will be a binary black hole merger once every few tens of minutes in the observable universe~\cite{Abbott:2017xzg}. The binary black hole merger signals will only appear within the LIGO-Virgo observing band for of order a second. As such, the binary black hole mergers form a non-Gaussian background of {\it popcorn} noise~\cite{PhysRevD.92.063002}. Through a mock data challenge, it has been verfied that the standard stochastic search pipeline used by LIGO-Virgo is capabable of efficiently detecting such a background~\cite{PhysRevD.92.063002}, even if there are likely more efficient ways to do so. This is an on-going field of research.

A stochastic background produced by binary black hole mergers will mask a cosmologically produced background. While an astrophysically produced stochastic would provide a wealth of information, the observation of gravitational waves from the Big Bang is the {\it Holy Grail} of gravitational wave astronomy. For second generation gravitational wave detectors, such as Advanced LIGO, Advanced Virgo and KAGRA, it will be impossible to directly detect the majority of the binary black hole mergers over the history of the universe. However, the proposed third generation detectors, such as the Einstein Telescope~\cite{0264-9381-27-19-194002} or the Cosmic Explorer~\cite{0264-9381-34-4-044001}, should be able to directly observe almost every stellar mass binary black hole merger in the observable universe. And whereas Advanced LIGO and Advanced Virgo should be observing a binary black hole stochastic background at the $\Omega_{GW} \sim 10^{-9}$ level, by removing this binary black hole foreground the third generation detection detectors could be sensitive to a cosmologically produced background at the  $\Omega_{GW} \sim 10^{-13}$ level with 5 years of observations ~\cite{PhysRevLett.118.151105}. With this sensitivity the third generation detectors will get into the realm where important cosmological observation can potentially be made~\cite{Caprini:2018mtu}.

With the large number of signals present, it is interesting to consider the required data analysis challenges that will be faced by the third generation gravitational wave detectors~\cite{0264-9381-27-19-194002,0264-9381-34-4-044001}. It is certainly probable that there will be overlapping signals.  The Advanced LIGO - Advanced Virgo study describing the implications for a stochastic background given the observations of gravitational waves from binary black hole and binary neutron star mergers directly addresses the possibility of overlapping signals~\cite{Abbott:2017xzg}. Given the expectations for the stochastic background produced by these compact objects a simulated time series of the signals was produced. Because of the long time that binary neutron star gravitational wave signals occupy the observation band, these type of signals (from sources throughout the observable universe) overlap, whereas the binary black hole produced gravitational wave signals are in the observation band for shorter periods, and form a popcorn type of signal~\cite{Abbott:2017xzg}. The predicted time between binary neutron star mergers in the observable universe is $13^{+49}_{-9}$ s, and assuming frequencies above 10 Hz, the number of overlapping signals at a given time is expected to be $15^{+30}_{-12}$. For binary black hole mergers the predicted time between these events in the observable universe is $223^{+352}_{-115}$ s, while the number of overlapping signals at a given time is predicted to be $0.06^{+0.06}_{-0.04}$~\cite{Abbott:2017xzg}.

The third generation gravitational wave detectors will have a lower frequency cutoff, probably 5 Hz. This means that the probability of signal overlap will be higher than for Advanced LIGO - Advanced Virgo since the signal will spend even more time in the detector. This is similar to the situation faced by LISA~\cite{2017arXiv170200786A}, which will need to deal with a very large number of overlapping signals (since, at these low frequencies, source behavior is more like a continuous signal than a transient one); the same is true for other space based detectors~\cite{1742-6596-122-1-012006,1742-6596-840-1-012010,PhysRevD.80.104009}. Many methods have been developed to detect and characterize numerous overlapping gravitational wave signals with these space based gravitational wave detectors~\cite{PhysRevD.72.022001,0264-9381-22-18-S04,PhysRevD.89.022001,PhysRevD.73.042001,PhysRevD.77.123010}; these types of methods to identify and them remove the compact binary merger gravitational wave signals will help the get the third generation gravitational wave detectors (and the space based detectors too) closer to measuring a cosmologically produced stochastic background~\cite{PhysRevLett.118.151105}.

\subsection{Binary Neutron Stars}
\label{subsec:bns}
A stochastic background produced by binary neutron stars will definitely exist at some level. The dramatic observation by Advanced LIGO and Advanced Virgo of the binary neutron star inspiral GW170817~\cite{PhysRevLett.119.161101} has led to numerous important astrophysical observations. The associated short gamma ray burst, GRB 170817a, implies that binary neutron star mergers are the source of short gamma ray bursts, in general~\cite{2041-8205-848-2-L14,2041-8205-848-2-L13}. The observation of the kilonova following the merger seems to confirm many predictions, including how the heaviest elements are created in the universe~\cite{2041-8205-848-2-L12}. From the gravitational wave signal one can infer the luminosity distance to the source; then using the measured redshift of the host galaxy, a measurement of the Hubble constant could be made, independent of the cosmic distance ladder~\cite{Abbott:2017xzu}.

The observation of this binary neutron star merger also has important implications for the production of a stochastic gravitational wave background, and the ability of Advanced LIGO and Advanced Virgo to observe it~\cite{Abbott:2017xzg}. This background would be from every binary neutron star merger throughout the observable universe; most of these are too small to be observed directly by LIGO and Virgo, but the background that they create may be detected. Using the observation of GW170817 and the total observing time by Advanced LIGO, the prediction for the energy density of a binary neutron star
produced stochastic background will be $\Omega_{GW}(f = 25 ~ \rm{Hz}) = 0.7^{+ 1.5}_{-0.6} \times 10^{-9}$. This can be compared with the predicted level of the binary black hole produced stochastic background of $\Omega_{GW}(f = 25 ~ \rm{Hz}) = 1.1^{+ 1.2}_{-0.7} \times 10^{-9}$. The combination of the two gives the total astrophysically produced stochastic background, as predicted by the LIGO and Virgo observations of $\Omega_{GW}(f = 25 ~ \rm{Hz}) = 1.8^{+ 2.7}_{-1.3} \times 10^{-9}$~\cite{Abbott:2017xzg}.

Then assuming the expected evolution of the sensitivity for Advanced LIGO and Advanced Virgo (as the detectors approach their design sensitivities)~\cite{Aasi:2013wya}, it is estimated that the LIGO-Virgo network could observe this background with a signal to noise ratio of 3 after a total of approximately 40 months of observing in the Advanced LIGO - Advanced Virgo era (with observations starting with the first observing run, O1)~\cite{Abbott:2017xzg}. Considering the uncertainties in the estimation of the background, and then taking the most optimistic assumptions, the astrophysical background might be observed at the
$3~\sigma$ level after 18 months of Advanced LIGO - Advanced Virgo era observations; this could then come during O3, the third observation run, scheduled to begin in the fall of 2018~\cite{Abbott:2017xzg}. The eventual detection of the astrophysically produced stochastic background by the LIGO-Virgo network is considered to be likely.

It is interesting to consider the nature of these two types of stochastic signals. When Advanced LIGO and Advanced Virgo reach their design sensitivities the low-frequency cutoff for observations will be 10 Hz. For the binary black hole produced stochastic background, the events come individually, once every $223^{+352}_{-115}$ s. The average duration of a signal in the interferometers' observing band is approximately 14 s. The probability of two signals overlaping is therefore quite small. The average number of overlapping binary black hole gravitational wave signals is $0.06^{+0.06}_{-0.04}$. The situation is quite different for the binary neutron star produced stochastic background. For these signals the average length of time that they are in the observing frequency band is 190 s. These events arrive every $13^{+49}_{-9}$ s. Consequently, the average number of overlapping binary neutron star gravitational wave signals is $15^{+30}_{-12}$~\cite{Abbott:2017xzg}.
A continuous background is created by the binary neutron star inspirals. But whether created by binary black holes or binary neutron stars, this astrophysically produced stochastic background is likely to be detected by the LIGO-Virgo network in the coming years.

\subsection{Close Compact Binary Stars}
\label{subsec:wdb}
While systems like binary black holes and binary neutron stars are the sources of interesting gravitational wave signals, other binary star systems will also produce gravitational waves. Close compact binary stars, most of which are white dwarf binaries, will produce thousands of signals that will be resolvable by LISA in the frequency band around a few $10^{-4}$  Hz to a few $10^{-2}$ Hz.
In addition to binaries containing white dwarfs, there will be neutron stars and stellar-origin black holes in different combinations~\cite{2017arXiv170200786A}. 
A background of gravitational waves will be formed by all of the 
unresolvable galactic~\cite{Nelemans:2001hp} and extragalactic~\cite{Farmer:2003pa} binaries; the sum of all of the gravitational waves that are not individually resolvable will form a stochastic background which could make the observation of a cosmologically produced stochastic background challenging. It has long been recognized that LISA could directly observe gravitational waves from thousands of galactic binaries, while also having to contend with a stochastic background from unresolvable galactic and extragalactic binaries~\cite{1987ApJ...323..129E,Schutz:1997bw,0264-9381-14-6-008}.

Having a mass model for the Milky Way helps to predict the distribution of close compact binary stars~\cite{doi:10.1046/j.1365-8711.1998.01282.x,doi:10.1111/j.1365-2966.2011.18564.x}. This can then be used to predict the gravitational waves from these binary systems, including their distribution in the sky for LISA observations~\cite{0264-9381-19-7-306}. Knowing the distribution of galactic gravitational wave sources on the sky could help LISA to remove this signal and get to a cosmologically produced stochastic background, similar to what is done with observations of the CMB, namely the effort to remove the contamination by the galaxy or other foreground sources~\cite{doi:10.1111/j.1365-2966.2010.17624.x,Planck_xii}. LISA will certainly be able to produce a sky map of the galactic binaries producing gravitational waves in its observational band~\cite{0264-9381-18-20-307}. 
Further knowledge about galactic binary systems, including white dwarf binaries, will be increasing rapidly with the observations by Gaia and its creation of a three-dimensional map of the Milky Way~\cite{gaia1,2018arXiv180409365G,doi:10.1093/mnrasl/sly110}.

The distribution of sources for gravitational waves from close compact binary stars can be seen in Fig.~\ref{fig:LISA}, along with the predicted sensitivity for LISA~\cite{2017arXiv170200786A}. There will be thousands of galactic binaries in the LISA observation band that will be individually observable via gravitational wave emission. The points in the figure above the LISA sensitivity curve reflect predictions for individual observations with marked with signal-to-noise ratio (SNR) $>$ 7. However there will be countless other binaries both in our galaxy and extragalactic that will contribute to an unresolvable gravitational background; this is also displayed in Fig.~\ref{fig:LISA}. It is predicted that in the LISA band, from 0.1 to 10 mHz, the gravitational wave background energy density from extragalactic binaries will be in the range $1 \times 10^{-12} < \Omega_{GW}(1 mHz) < 6 \times 10^{-12}$~\cite{Farmer:2003pa}.

Whether it is the gravitational wave signals from thousands of directly observable galactic binaries, or the unresolved gravitational wave background from galactic and extragalactic binaries, these gravitational wave signals will create a tremendous data analysis challenge for the attempt by LISA to observe a cosmologically produced stochastic background. Research progress has shown that the thousands of individually detectable gravitational wave signals from galactic binaries can be removed from the search for a cosmologically produced stochastic background~\cite{PhysRevD.72.022001,0264-9381-22-18-S04,0264-9381-24-19-S20,PhysRevD.80.064032,0264-9381-24-19-S17,PhysRevD.82.022002,PhysRevD.89.022001}. The unresolvable gravitational waves from close compact binary stars need to be removed in the search for the cosmologically produced stochastic background. Much progress has been made in addressing this problem ~\cite{PhysRevD.82.022002,PhysRevD.89.022001,0264-9381-34-24-244002}, but further confirmation will need to be made in the coming years through LISA mock data challenges~\cite{0264-9381-25-11-114037,1742-6596-840-1-012026}.

While the numerous gravitational wave signals from close compact binary stars will present a data analysis challenge, some of these binary systems will be especially valuable for the LISA mission. Many of these systems have already been observed and studied electromagnetically; for LISA these are referred to as the {\it Verification Binaries}. These binary systems will produce gravitational wave signals that will be observed and used to confirm the calibration and sensitivity of LISA~\cite{doi:10.1093/mnras/sty1545}. A comparison between the predicted and observed gravitational wave signals should provide significant confidence in the LISA observations and results. The verification binaries will also be used to test general relativity, including placing limits on the mass of the graviton~\cite{Gair2013}.

LISA will also gain important information on binary systems in our galaxy through the observation of gravitational waves from ultra-compact binaries in the galaxy~\cite{Nelemans:2013iq,1742-6596-610-1-012003}. These are binary systems consisting of two stars with an orbital period less than an hour. Of order 60 ultra-compact binaries have been identified via electromagnetic observations and these are thought to be composed of white dwarfs, neutron stars, and stellar mass black holes; the double white dwarf binary J0651 has already been observed to have an orbital decal that is consistent with general relativity and loss of energy via the emission of gravitational waves~\cite{1742-6596-610-1-012003}. The observation of these systems with gravitational waves will provide further tests of general relativity, and will also give information in helping to explain the formation and evolution of stellar binary systems~\cite{1742-6596-610-1-012003}.

The observation of gravitational waves from close compact binary stars are interesting in their own right. They provide a gravitational wave foreground and background containing much important astrophysical information.
Ultimately if the close compact binary stars can be addressed by LISA (such as subtracting signals from the data~\cite{PhysRevD.73.042001,PhysRevD.77.123010}, or accounting for them using Bayesian parameter estimation methods~\cite{PhysRevD.72.022001}) it could achieve a sensitivity of $\Omega_{GW} \sim 10^{-12}$ in a search for a cosmologically produced stochastic background of gravitational waves.

\subsection{Supernovae}
\label{subsec:SN}
Common and powerful astrophysical events throughout the history of the universe will contribute to the stochastic background. If a supernova has some asymmetry, then gravitational waves will be produced. The emission of gravitational waves from supernovae has been studied in many ways. Numerical simulations are providing some of the most comprehensive studies, but they are difficult and time consuming~\cite{Fryer2011,PhysRevD.90.044001,0264-9381-27-19-194005,PhysRevD.92.084040,PhysRevD.78.064056}.

There have been numerous studies trying to address the level of a stochastic background produced by supernovae in the universe. Population III stars \footnote{Population I are young and metal-rich stars and are often found in the arms or spiral galaxies, such as in the Milky Way. Population II stars are very old, metal-poor and tend to be found in the center of galaxies or in galactic halos~\cite{shapley2013galaxies}. The hypothesized Population III stars would have essentially no metals, only the material present after the Big Bang (hydrogen, helium, and trace amounts of lithium and beryllium). Population III stars would be the oldest population of stars~\cite{0004-637X-660-1-516}.} were formed in the early universe and had very large masses. Stars with high metallicity are more succeptible to mass loss via stellar winds~\cite{Stevenson:2017tfq}. Population III stars had very low metallicity (essentially zero), and as such, were able to live their stellar lives with minimal mass loss. Population II stars had low metallicity compared with present day Population I stars. In~\cite{doi:10.1111/j.1365-2966.2009.15120.x} the authors consider Population III stars in the mass range of $100 - 500 M_{\odot}$ and Population II stars in the mass range $8 - 40 M_{\odot}$. Using redshift dependent formation rates for these stars, the expected evolution of these stars once created, and then the stars' death through supernovae, the resulting stochastic background is predicted. Assumptions are made as to the amount of energy released in gravitational waves in these supernovae. This study predicts a stochastic background that peaks in the LIGO-Virgo band, with
$10^{-12} \leq \Omega_{GW} h^{2} \leq 7 \times 10^{-10}$ in the 387-850 Hz frequency band. This stochastic background is dominated by gravitational waves from the supernovae of Population II stars~\cite{doi:10.1111/j.1365-2966.2009.15120.x}.

Another study considers a stochastic background produced by the ring-down of black holes created via stellar core collapse~\cite{PhysRevD.92.063005}. Certainly this is only one of the different mechanisms for gravitational wave production in core collapse supernovae. Various models (including different star formation rates) predict a stochastic background of 
$10^{-10} \leq \Omega_{GW} \leq 5 \times 10^{-9}$ in the 50-1000 Hz frequency band.
It is interesting to note that most of the gravitational wave production for this background comes from regions having redshifts of 1 to 2. 
This post-supernova black hole ring down stochastic background is at a level that could be observed by the Advanced LIGO - Advanced Virgo network, or third generation detectors~\cite{PhysRevD.92.063005}. 
This level assumes that $10^{-6}$ to $10^{-4}$ of the rest mass of the black hole is converted into gravitational waves~\cite{PhysRevD.92.063005}. 
This efficiency assumption is probably quite optimistic.

Some of the members of the group who conducted the previous study extended their supernovae models to consider more general gravitational wave emission mechanisms~\cite{PhysRevD.95.063015}. The full supernova process and associated gravitaitonal wave emission is very difficult to calculate. In the new study two models are considered. One considers the form of the gravitational wave signals produced by two and three dimensional supernova simulations.
The form and frequency dependence of the gravitational wave emission from the core collapse supernova can be approximated~\cite{PhysRevD.73.104024,PhysRevD.72.084001}. This can then be combined with predictions for star formation and eventual supernovae over the history of the universe. This then provides a prediction for a core collapse supernova produced stochastic background. Based on reasonable assumptions for the parameters in this model the stochastic background is predicted to be possibly as large as $\Omega_{GW} \sim 10^{-9}$ around 300 Hz, while other parameter choices could reduce it to the $\Omega_{GW} \sim 10^{-12}$ level. The other model considered in this study concentrates on the low-frequency structure seen in the predicted gravitational wave emission from core-collapse supernova. This has been observed in the simulations from a number of groups, some of which suspect that it pertains to prompt convection. The most optimistic prediction for the stochastic background level for this model is $\Omega_{GW} \sim 10^{-10}$ in the 30 - 100 Hz band~\cite{PhysRevD.95.063015}, and might be observable with third generation gravitational wave detectors~\cite{0264-9381-27-19-194002,0264-9381-34-4-044001}.

Since the gravitational wave production from supernovae is difficult to predict, the absence of a detection of a stochastic background can be used to constrain the average amount of gravitational wave emission from supernovae. Using the upper limits reported by initial LIGO and initial Virgo for the analysis of the scientific run S5 data 
it is possible to say that a supernova can only only produce an upper limit in the range $0.49 - 1.98 M_{\odot} c^2$ of energy in gravitational waves~\cite{Zhu:2010af}. Future results on the upper limit of the gravitational wave energy density will provide further constraints. Note that since this study~\cite{Zhu:2010af} was conducted the constraints imposed by LIGO on the stochastic background have improved by a factor of $\sim 50$~\cite{PhysRevLett.118.121101}. The constraints on the energy emitted in gravitational waves from supernovae will improve accordingly.  

\subsection{Pulsars and Magnetars}
\label{subsec:pulsars}
Non-axisymmetric spinning neutron stars are expected to be a detectable source of gravitational waves~\cite{doi:10.1146/annurev.aa.10.090172.002003,Riles:2012yw}. The radio observations from pulsars indicate that neutron stars are rotating with periods that can be as rapid as milli-seconds. Gravitational waves would be emitted if the neutron star is not perfectly spherical, namely if there is an asymmetry in its shape; such a deformation might be created by having toroidal magnetic fields within the neutron star~\cite{Zimmermann:1978mk,Riles:2012yw}.
Another path for the production of gravitational waves would be the presence of a slight mountain on the neutron star surface. Such an effect could happen due to cracking of the crust through thermal effects~\cite{1976ApJ...208..550P,Riles:2012yw}.
With such asymmetries or defects gravitational waves would be emitted at twice the rotation frequency of the neutron star.

The excitation of internal mechanical oscillation modes is another way for the symmetry of the neutron star to be broken, and for gravitational waves to be produced. There can be an interplay between the viscocity of the material within the neutron star and the emission of gravitational waves~\cite{PhysRevLett.24.611,Friedman:1978hf}. It is also speculated that quadrupole mass currents can emit gravitational waves in such a way that the process actually amplifies the currents, leading to an unstable run-away process; these are associated with the so-called r-modes~\cite{Bildsten:1998ey,Andersson:1997xt,0004-637X-502-2-714,PhysRevD.58.084020,Riles:2012yw}.

Pulsars are numerous in our galaxy, and presumably in the universe. It was soon recognized that a stochastic background could be created by the sum of all neutron star produced gravitational waves in the universe. For example, one study~\cite{Ferrari:1998jf} considers newly created neutron stars that are spinning rapidly. The neutron star loses energy and spins down via gravitational wave emission. The r-mode instability~\cite{Bildsten:1998ey,Andersson:1997xt,0004-637X-502-2-714,PhysRevD.58.084020} is responsible for the gravitational wave emission.
The prediction for this study is an energy density of the stochastic background of 
$\Omega_{GW} h^{2} \sim (2.2-3.3) \times 10^{-8}$ in the 500-1700 Hz frequency band.
The results of this study are dependent on assumptions of the star formation rate, with the assumptions that this is peaking at a redshift of about $z \sim 1.3$~\cite{Ferrari:1998jf}. This study and results are similar (especially with respect to r-mode production of gravitational waves) to another~\cite{PhysRevD.58.084020}, with results that are slightly different due to different assumptions about the star formation rate and its redshift dependence~\cite{PhysRevD.58.084020}. The star formation rate in the study of Owen et al.~\cite{PhysRevD.58.084020} extends over the range $0 < z < 4$. The resulting predicted stochastic backgrund is $\Omega_{GW} h^{2} \sim 1.5 \times 10^{-8}$ at $\sim 300$ Hz, and diminishes for higher frequencies~\cite{PhysRevD.58.084020}. This corresponds to maximum gravitational wave production at a redshift of $z \sim 4$~\cite{Ferrari:1998jf}.

A recent study has continued this avenue of research and investigated the stochastic background created by newly formed magnetars~\cite{Cheng:2015rja,PhysRevD.95.083003}. 
A magnetar is a neutron star with an extraordinarily large magnetic field ($\sim 10^{14} - 10^{15}$) G~\cite{doi:10.1146/annurev-astro-081915-023329}.
Various equations of state for the neutron star matter are assumed, in addition to the merger rate for binary neutron star systems. Very strong magnetic fields for the newly formed magnetars are also assumed ($10^{15}$ G to even $10^{17}$ G). The most optimistic results produced predictions of $\Omega_{GW} \sim 10^{-10}$ at $\sim 100$ Hz, $\Omega_{GW} \sim 10^{-9}$ at $\sim 300$ Hz, and $\Omega_{GW} \sim 10^{-8}$ at $\sim 1000$ Hz~\cite{Cheng:2015rja,PhysRevD.95.083003}. 

It is also possible to calculate the gravitational wave production from all types of neutron stars, such as pulsars (typical magnetic fied strengths, $\sim 10^{8}$ T), magnetars (very large magnetic fields, $\sim 10^{10}$ T, potentially creating ellipticities that enhance gravitational wave production), and gravitars (low magnetic field strengths, $< 10^{8}$ T, thereby making gravitational wave emission the dominant source of rotational energy loss)~\cite{PhysRevD.86.104007}. Different assumptions are made on the distribution of spins for the neutron stars. If the assumption is (admittedly optimistic) that all rotating neutron stars are gravitars, then the predicted gravitational wave emission is quite large, reaching $\Omega_{GW} \sim 10^{-7}$ at 1 kHz, or $\Omega_{GW} \sim 10^{-8}$ at 100 Hz. If on the other hand, the assumption is that neutron stars are essentially pulsars then the estimated stochastic background level is more pessimistic, with $\Omega_{GW} \sim 10^{-10}$ at 1 kHz, or $\Omega_{GW} \sim 10^{-13}$ at 100 Hz. For magnetars, and assuming their distribution is as described in~\cite{MNR:MNR17861}, the prediction is that the resulting stochastic background would be $\Omega_{GW} \sim 10^{-8}$ at 1 kHz and $\Omega_{GW} \sim 10^{-10}$ at 100 Hz. The conclusion is that for realistic assumptions it will be difficult to detect this stochastic background, although with third generation detectors~\cite{0264-9381-27-19-194002,0264-9381-34-4-044001} it might be possible~\cite{PhysRevD.86.104007}.

The large number of neutron stars in the Milky Way, plus the fact that these neutron stars are relatively close, provides a means to constrain the average neutron star ellipticity based on the limits set on the stochastic background~\cite{PhysRevD.89.123008}. It is assumed that there are $10^{8}$ to $10^{9}$ neutron stars in our galaxy~\cite{refId0_Sartore}. Of these, it is predicted that of order $\sim 5 \times 10^{4}$ have rotation periods less than 200 ms, in which case they could produce gravitational waves in the observable band of LIGO and Virgo, $f > 10$ Hz. The Advanced LIGO - Advanced Virgo network should be able to constrain the 1-sigma sensitivity to neutron star ellipticity to be $\sim 2 \times 10^{-7}$, which is also the limit derived from the two co-located initial LIGO detectors~\cite{PhysRevD.91.022003}. Third generation gravitational wave detectors~\cite{0264-9381-27-19-194002,0264-9381-34-4-044001} may be able to constrain ellipticities to $\sim 6 \times 10^{-10}$~\cite{PhysRevD.89.123008}. Theoretical studies predict that the largest possible ellipticity for a neutron star is $\sim 10^{-5}$~\cite{PhysRevLett.102.191102,PhysRevD.88.044004}.

The recent observation of the binary neutron star inspiral gravitational wave signal GW170817~\cite{PhysRevLett.119.161101} generated much interest as to the post-merger remnant. The total mass of the system was $2.74^{+0.04}_{-0.01} M_{\odot}$. 
The merger of the two neutron stars could have formed a black hole directly, in which case the black hole ringdown gravitational wave signal would be above 6 kHz. Another possibility is that a hypermassive neutron star could be formed, and it would survive for timescales of up to thousands of seconds before collapsing into a black hole. This hypermassive neutron star would survive through thermal gradients and differential rotation~\cite{1538-4357-528-1-L29}. Another possibility is that a stable hypermassive neutron star is formed.
In the short time after the merger the remanant will likely be excited, and emit gravitational waves in the 1 kHz to 4 kHz regime~\cite{2017PhRvD..96f3011M,2000PhRvD..61f4001S,2013PhRvD..88d4026H}.
LIGO and Virgo conducted a search for a post-merger gravitational wave signal associated with GW170817~\cite{2041-8205-851-1-L16}.
A recent study considers a stochastic background created by such a post-merger remnant~\cite{Miao:2017qot}. This study also discusses how future gravitational wave detectors could be designed and constructed at higher frequencies (1 - 4 kHz) to search for post-merger remnant signals, either for direct observation of an individual event or a stochastic background from these types of sources. The study claims that the combination of the binary neutron star inspiral signals plus the post merger ringdown signals will contribute to a stochastic background of level $\Omega_{GW} \sim 10^{-9}$ from 1 to 3 kHz~\cite{Miao:2017qot}.

\section{Summary of methods to observe or constrain a stochastic gravitational wave background}
\label{sec:detection}
The search for a stochastic gravitational wave background is arguably one of the most important projects in cosmology and astrophysics. In contrast to the electromagnetic spectrum, gravitational waves will potentially provide a window to the earliest moments in the universe. In this section we review the methods by which one can attempt to observe the stochastic background. An extremely comprehensive review of the observational methods used and proposed to detect gravitational waves is given by Romano and Cornish~\cite{Romano2017}. 

\subsection{LIGO-Virgo}
\label{subsec:LIGO-Virgo}
The ground based gravitational wave detectors, LIGO and Virgo, have been attempting to measure the stochastic gravitational wave background since 2004~\cite{PhysRevD.69.122004,PhysRevLett.95.221101,0004-637X-659-2-918,Abadie:2011fx,Abbott:2011rs,PhysRevLett.113.231101,PhysRevD.91.022003,PhysRevLett.118.121101}. The magnitude of gravitational waves associated with the stochastic background will be random, so it will appear like noise in an individual detector. However, it will be coherent in two detectors (completely coherent for two co-located detectors, with the coherence falling off with distance if the detectors are displaced from one another). The way to extract the stochastic background signal from two detectors is essentially outlined in Eq.~\ref{eq:coherence}. 
The correlation between the data from two gravitational wave detectors is more complicated due to their physical separation and misalignment. While this makes the calculation somewhat more involved, it is nonetheless straightforward to account for the presence of the stochastic background in both detectors~\cite{Christensen:1992wi}. 

The LIGO-Virgo data analysis method follows the prescription of Allen and Romano~\cite{PhysRevD.59.102001}. Instead of working in the time domain, as is the case with Eqs.~\ref{eq:define} and \ref{eq:coherence}, one works in the frequency domain, using the Fourier transform of the signals, $\tilde{s}_{1}(f)$ and $\tilde{s}_{2}(f)$. An optimal filter is used to maximize the signal to noise ratio, but in order to do this, there must be assumptions made on the frequency dependence of the signal. The search is described in terms of the energy density of the stochastic gravitational wave with respect to the closure density of the universe, as described by Eqs.~\ref{eq:Omega1} and \ref{eq:Omega2}. Next, the frequency dependence of the energy density of the stochastic background is assumed to have the form
\begin{equation}
\label{eq:Omega3}
\Omega_{GW}(f) = \Omega_{\alpha} \Big(\frac{f}{f_{ref}}\Big)^{\alpha} ~ ,
\end{equation}
where $f_{ref}$ is an arbitrary reference frequency.
The search uses an estimator~\cite{PhysRevD.59.102001,PhysRevLett.118.121101}
\begin{equation}
\hat{Y}_\alpha=\int_{-\infty}^{\infty}df\,\int_{-\infty}^{\infty}df'\,\delta_T(f-f'){\tilde s}^{*}_{1}(f){\tilde s}_{2}(f'){\tilde Q}_\alpha(f')
\label{eq:cc_estimator}
\end{equation}
and variance
\begin{equation}
\sigma^2_Y{\approx}\frac{T}{2}\int_{0}^{\infty}df\,P_1(f)P_2(f)|{{\tilde Q}_\alpha(f)}|^2,
\label{eq:theor_sigma}
\end{equation}
where $\delta_T(f-f')$ is a finite-time Dirac delta function, $T$ is the observation time, $P_{1,2}$ are the one-sided power spectral densities for the detectors, and $\tilde Q_\alpha(f)$ is a filter function to optimize the search \footnote{The Hubble constant appears explicitly, rather than being absorbed into $\lambda_\alpha$, to emphasize that the estimator for $\Omega_{\rm GW}$ depends on the measured value of $H_0$.},
\begin{equation}
{\tilde Q}_\alpha(f)={\lambda_\alpha}\frac{{\gamma}(f)H_0^2}{f^{3}P_1(f)P_2(f)}\left(\frac{f}{f_{\rm ref}}\right)^\alpha.
\label{eq:optimal_filter}
\end{equation}
The $\gamma(f)$ term is what is known as the overlap reduction function~\cite{Christensen:1992wi,PhysRevD.55.448}; this accounts for the reduction in sensitivity due to
separation and relative misalignment between the two detectors used in the stochastic search. ${\gamma(f)} = 1$ if the detectors are co-located and co-aligned, and diminishes otherwise. Note that it is actually the magnitude, $|{\gamma(f)}|$, that is the most important; a rotation of a detector by 90$^o$ will not affect the sensitivity of the search for the stochastic background.

\subsection{Results from Advanced LIGO Observing Run O1}
\label{sec:O1_results}
\begin{figure}
\begin{center}
\includegraphics[width=5.5in]{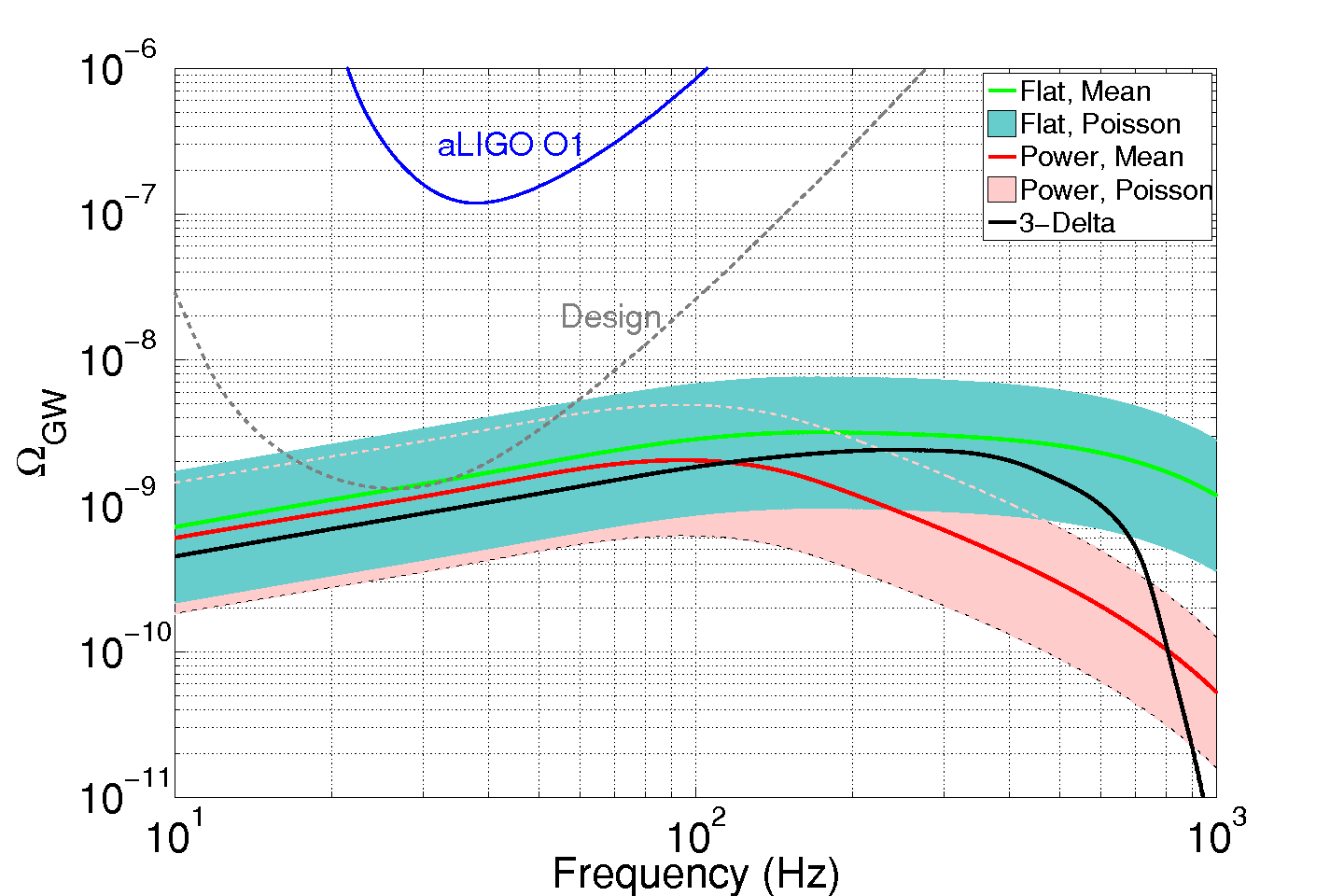}
\caption{The range of potential spectra for a binary black hole background assuming 
the flat-log, power-law, and three-delta mass distribution models
described in \protect \cite{2041-8205-818-2-L22,PhysRevX.6.041015},
with the local rate derived from the O1 observations \protect \cite{PhysRevX.6.041015}. Also displayed is the O1 sensitivity and the projected ultimate design sensitivity for Advanced LIGO and Advanced Virgo. Figure from \protect \cite{PhysRevLett.118.121101}.}
\label{fig:BBH_background}
\end{center}
\end{figure}

Advanced LIGO's first observing run went from September 2015 to January 2016. The data from the two Advanced LIGO detectors, LIGO Hanford and LIGO Livingston, were used for the search for a stochastic background. Data quality cuts removed problematic times and frequencies from the analysis. In total, 29.85 days of coincident data were analyzed. No stochastic background was detected. 
The dramatic improvement in the upper limit on the stochastic background energy density was important, but not the most important stochastic background outcome of observing run O1. The observation of the gravitational waves from stellar mass binary black hole mergers~\cite{PhysRevLett.116.061102,PhysRevLett.116.241103,PhysRevX.6.041015} implies that these events are far more numerous in the universe than previously expected. In fact, it is likely that the stochastic background produced from these type of events will be at the level of $\Omega_{GW} \sim 10^{-9}$ in the observing band of Advanced LIGO and Advanced Virgo~\cite{PhysRevLett.116.131102}. See Figure~\ref{fig:BBH_background}.

\subsubsection{O1 Isotropic Results}
\label{subsubsec:O1_isotropic}
\begin{figure}
\begin{center}
\includegraphics[width=6.5in]{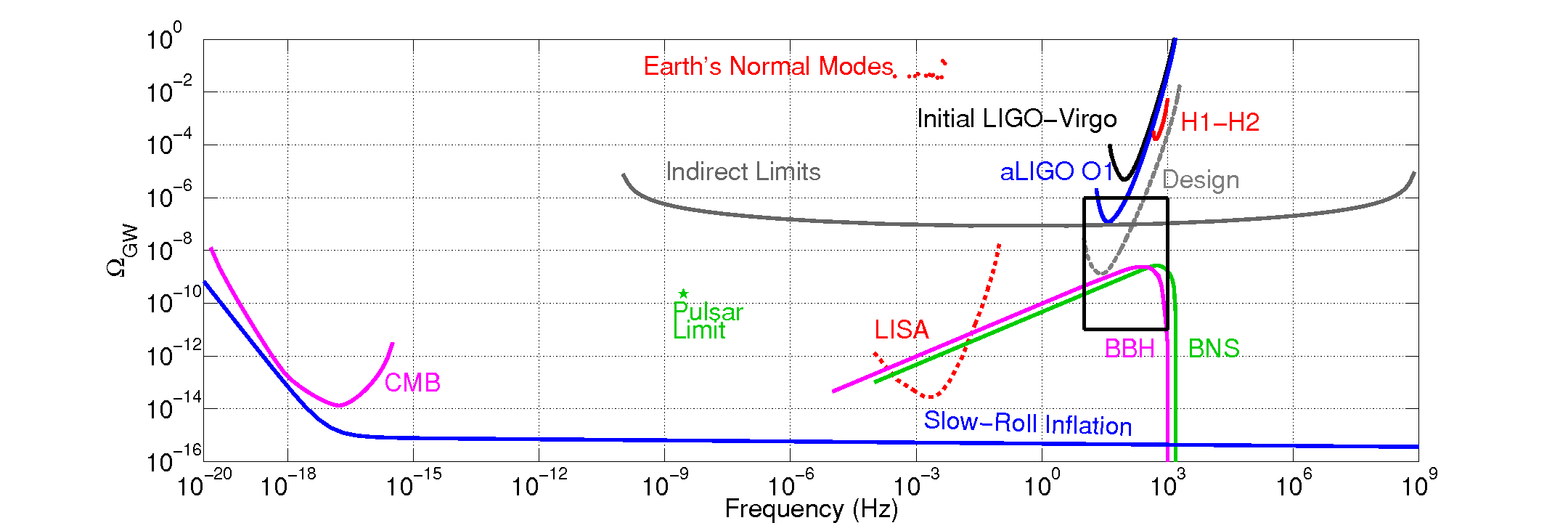}
\caption{Constraints on the stochastic background, as well as various predictions, across over 29 decades in frequency. Displayed are the limits from the final science run of initial LIGO-Virgo, the co-located detectors at Hanford (H1-H2) during run S5, Advanced LIGO for O1, and the expected design sensitivity of the Advanced LIGO - Advanced Virgo detector network assuming two years of coincident data. Also shown are the constraints on the energy density of the stochastic background from other observations: CMB measurements \protect \cite{refId0}, indirect limits from the Cosmic Microwave Background (CMB) and Big-Bang Nucleosynthesis \protect \cite{Pagano2016823,PhysRevX.6.011035}, pulsar timing \protect \cite{PhysRevX.6.011035}, and from the Earth's normal modes \protect \cite{PhysRevD.90.042005}. The predicted stochastic background from binary black holes (BBH) \protect \cite{PhysRevLett.116.131102} and binary neutron stars (BNS) \protect \cite{PhysRevD.92.063002} are displayed.
Also given is the predicted sensitivity for the proposed space-based detector LISA \protect \cite{2017arXiv170200786A}.
Displayed in Figure \protect\ref{fig:BBH_background} is the region in the black box in more detail. 
Finally, the stochastic gravitational-wave background
predicted from slow-roll inflation is displayed; this result is consistent with the Planck results~\cite{Planck2015} and for this plan a tensor-to-scalar-ratio
of $ r = 0.11$ is used.
Figure from \protect \cite{PhysRevLett.118.121101}.}
\label{fig:Landscape}
\end{center}
\end{figure}

Assuming that the frequency dependence of the energy density of the stochastic background is flat, namely $\alpha = 0$, the constraint on the energy density is $\Omega(f) < 1.7 \times 10^{-7}$ with 95\% confidence within the 20 Hz - 86 Hz frequency band~\cite{PhysRevLett.118.121101}. This is a factor of 33 better than the upper limit set by initial LIGO and initial Virgo~\cite{PhysRevLett.113.231101}. Assuming a spectral index of $\alpha = 2/3$ the constraint on the energy density is $\Omega(f) < 1.3 \times 10^{-7}$ with 95\% confidence within the 20 Hz - 98 Hz frequency band, while for $\alpha = 3$ it is $\Omega(f) < 1.7 \times 10^{-8}$ in the 20 Hz - 300 Hz band~\cite{PhysRevLett.118.121101} (the reference frequency is $f_{ref} = 25$ Hz when $\alpha \neq 0$). Figure~\ref{fig:Landscape} provides the O1 stochastic background results, as well as constraints from 
from previous analyses, theoretical predictions, the expected sensitivity at design sensitivity for Advanced LIGO and Advanced Virgo, and the projected sensitivity of the proposed Laser Interferometer Space Antenna (LISA)~\cite{2017arXiv170200786A}. 
The O1 results will be used to limit cosmic string parameters, similar to what was done with initial LIGO and initial Virgo~\cite{Abbott:2011rs,Aasi:2013vna}.

\subsubsection{O1 Anisotropic Results}
\label{subsubsec:O1_directional}
Within the LIGO-Virgo observational band it is expected that the stochastic background will be essentially isotropic. However, LIGO and Virgo have decided to look for a stochastic background that would be anisotropic. Such an anisotropic background could provide even more information about the early universe, or the astrophysical environment in our region of the universe. Using the recent O1 data there have been three different types of searches for an anisotropic background~\cite{PhysRevLett.118.121102}. To look for extended sources, LIGO and Virgo use what is known as the spherical harmonic decomposition~\cite{Abbott:2011rr}. In order to search for point sources, a broadband radiometer analysis is used~\cite{0264-9381-23-8-S23,S4radiom}. Finally, LIGO and Virgo employed a narrowband radiometer search to look for gravitational waves in the direction of interesting objects in the sky, such as the galactic center, Scorpius X-1 and SN 1987A.

An anisotropic stochastic background was not observed with the Advanced LIGO O1 data, but important upper limits were set~\cite{PhysRevLett.118.121102}.
For broadband point sources, the gravitational wave energy flux per unit frequency was constrained to be  
$F_{\alpha,\Theta}  < (0.1 - 56)\times 10^{-8}$ erg cm$^{-2}$ s$^{-1}$ Hz$^{-1} (f/25$ Hz)$^{\alpha-1}$
depending on the sky location 
$\Theta$
and the spectral power index $\alpha$.  
For extended sources, the upper limits on the fractional gravitational wave energy density required to close the Universe are 
$\Omega(f,\Theta) < (0.39 - 7.6)\times 10^{-8}$ sr$^{-1} (f/25 $ Hz$)^{\alpha}$,
again depending on 
$\Theta$
and $\alpha$.
The directed searches for narrowband gravitational waves from Scorpius X-1, Supernova 1987~A, and the Galactic Center had median frequency-dependent limits on strain amplitude of 
$h_0 < (6.7,\, 5.5,$ and $ 7.0)\times10^{-25}$
respectively, for the most sensitive detector frequencies 
130 - 175 Hz. See \cite{PhysRevLett.118.121102} for further details.

\subsubsection{Tests of General Relativity with the Stochastic Gravitational-Wave Background}
\label{subsubsec:O1_NON-GR}
LIGO and Virgo have used the recent observation of gravitational waves from binary black hole and binary neutron star inspirals to test general relativity~\cite{PhysRevLett.116.221101,PhysRevX.6.041015,2041-8205-848-2-L13}. The LIGO-Virgo stochastic background search has also been extended in order to test general relativity. Assuming that general relativity is the correct description of gravitation, there is no reason to expect extra polarizations of gravitational waves, nor extra polarizations in the stochastic background; however, LIGO and Virgo have the ability to search for these modes, and will do so. With general relativity there are only two possible polarizations for gravitational waves, namely the two tensor modes. Alternative theories of gravity can also generate gravitational waves with scalar or vector polarizations~\cite{Will2014}. The observation of the gravitational waves from the binary black hole merger by the three detectors of the Advanced LIGO - Advanced Virgo network, GW170814, allowed for the first direct test as to whether the polarizations of gravitational waves obey the predictions of general relativity; from this observation, the tensor-only polarizations of general relativity are preferred~\cite{PhysRevLett.119.141101}.

\begin{figure}
\begin{center}
\includegraphics[width=6.5in]{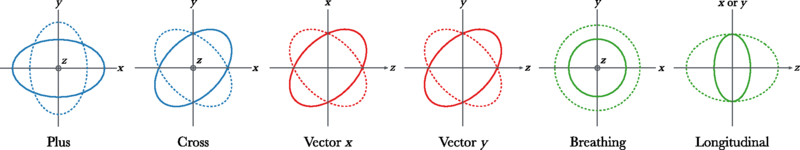}
\caption{The effect of different possible polarizations of gravitational waves on a ring of freely falling test particles. The six gravitational-wave polarizations are allowed with general metric theories of gravitation. The gravitational waves are assumed to be propagating in the z direction (out of the page for the plus, cross, and breathing modes; to the right for the vector- x, vector- y, and longitudinal modes). While general relativity allows only for two tensor polarizations (plus and cross), other theories allow for two vector (x and y) and/or two scalar (breathing and longitudinal) polarizations.
Figure from \protect \cite{Callister:2017ocg}.}
\label{fig:polarizations}
\end{center}
\end{figure}

Since there are six possible polarization modes (see Fig.~\ref{fig:polarizations}), Advanced LIGO (with only two detectors, that are essentially co-aligned with respect to each other) cannot identify the polarization of short duration gravitational wave signals~\cite{PhysRevX.6.041015,Romano2017,Will2014}, such as those that have been recently observed~\cite{PhysRevLett.116.061102,PhysRevLett.116.241103,PhysRevX.6.041015}.
A minimum of six detectors would be necessary to resolve the polarization content (scalar, vector and tensor)  of
a short duration gravitational wave~\cite{Will2014}.
A search for long duration gravitational waves, such as those from rotating neutron stars or the stochastic background by the two Advanced LIGO detectors,
can directly measure the polarizations of the gravitational waves~\cite{Romano2017,0264-9381-32-24-243001,Isi:2017equ,PhysRevLett.120.031104,Callister:2017ocg}.
A detection of a stochastic background by Advanced LIGO and Advanced Virgo would allow for a
verification of general relativity that is not possible with short duration gravitational wave signals.

The LIGO-Virgo search for a stochastic background has now been expanded to a search for 6 polarizations: two tensor modes, two vector modes, and two scalar modes~\cite{Callister:2017ocg,Abbott:2018utx}. This has been applied to Advanced LIGO Observing Run O1 data~\cite{Abbott:2018utx}.
In future observing runs, the addition of Advanced Virgo to the network will not improve detection prospects (because of its longer distance displacement from the LIGO detectors), however it will improve the ability to estimate
the parameters of a stochastic background of mixed polarizations. The eventual inclusion of KAGRA~\cite{Aso:2013eba} and LIGO-India~\cite{doi:10.1142/S0218271813410101} will further expand the ability to resolve different polarizations of the stochastic background, and further test general relativity. Bayesian parameter estimation techniques have been developed in order to search for tensor, vector and scalar polarizations in the LIGO-Virgo data~\cite{Callister:2017ocg}.

For the Advanced LIGO O1 data, there has been a search for tensorial gravitational waves, vector gravitational waves, and scalar gravitational waves~\cite{Abbott:2018utx}. While no signal was detected, upper limits have been place on the energy density of each of these stochastic backgrounds. This search assumed log-uniform priors~\footnote{This is a uniform prior between $log(\Omega_{min})$ and $log(\Omega_{max})$. For the analysis of the Advanced LIGO O1 data $\Omega_{min} = 10^{-13}$ and $\Omega_{max} = 10^{-5}$~\cite{Abbott:2018utx}.} for the energy density in each polarization; note that in the O1 Advanced LIGO results reported in~\cite{PhysRevLett.118.121101} it was assumed that the prior on the energy density was uniform in a particular band.
With 95\% credibility, the
limit for the energy density of the tensor modes is $\Omega_{GW}^{T} < 5.6 \times 10^{-8}$, for the vector modes $\Omega_{GW}^{V} < 6.4 \times 10^{-8}$, and scalar modes $\Omega_{GW}^{S} < 1.1 \times 10^{-7}$; for these limits the reference frequency is 25 Hz~\cite{Abbott:2018utx}.

\subsection{LIGO Co-Located Detectors}
\label{subsubsec:S5_H1H2}
In principle the best chance to detect a stochastic background would be with two co-located and co-aligned detectors. In this case the overlap reduction function $\gamma(f)$~\cite{Christensen:1992wi,PhysRevD.55.448}, would be equal to 1 for all of the frequencies in the search. For the first five scientific runs of initial LIGO, S1-S5, there were two interferometers operating at the LIGO Hanford site. H1 was the 4 km interferometer, while H2 was the 2 km interferometer. These two detectors were co-aligned and co-located, and operated within the same vacuum system. Using the LIGO H1 and H2 S5 data a search was conducted for a stochastic background~\cite{PhysRevD.91.022003}.

In reality this search proved to be very difficult. Common noise was coherent in both detectors. As such, the correlation that was done between the gravitational wave data from H1 and H2 was corrupted by the presence of coherent noise. This was especially true at low frequencies, $f < 460 ~ \rm{Hz}$. However at higher frequencies it was possible to conduct the search.  For the band of 460 - 1000 Hz, a 95\% confidence-level upper limit on the gravitational-wave energy density was found to be $\Omega_{GW}(f) < 7.7 \times 10^{-4} (f/900 ~ \rm{Hz})^{3}$~\cite{PhysRevD.91.022003}. These continue to be the best upper limits in this frequency band~\cite{PhysRevLett.118.121101}.

\subsection{Correlated magnetic noise in global networks of gravitational-wave detectors}
\label{subsec:Schumann}
A search for the stochastic background uses a cross-correlation between the data from two detectors. Inherent in such an analysis is the assumption that the noise in one detector is statistically independent from the noise in the other detector. Correlated noise would introduce an inherent bias in the analysis. It is for this reason that the data from two separated detectors is used. 
See Sec.~\ref{subsubsec:S5_H1H2} for the discussion of co-located detector measurement~\cite{PhysRevD.91.022003}.

The LIGO and Virgo detectors' sites are thousands of kilometers from one another, and the simple assumption is that the noise in the detectors at these sites is independent from one another. However, this assumption has been demonstrated to be false for magnetic noise. The Earth's surface and the ionosphere act like mirrors and form a spherical cavity for extremely low frequency electromagnetic waves. The Schumann resonances are a result of this spherical cavity, and resonances are observed at  8, 14, 20, 26, ... Hz~\cite{Sentman}. Most of these frequencies fall in the important stochastic background detection band (10 Hz to 100 Hz) for Advanced LIGO and Advanced Virgo. The resonances are driven by the 100 or so lightning strikes per second around the world.
The resonances result in magnetic fields of order 0.5 - 1.0 pT Hz$^{-1/2}$ on the Earth's surface~\cite{Sentman}. 
In the time domain, 10 pT bursts appear
above a 1 pT background  at a rate of $\approx$ 0.5 Hz ~\cite{FULLEKRUG1995479}.

This magnetic field noise correlation has been observed between magnetometers at the LIGO and Virgo sites~\cite{Thrane:2013npa}. Magnetic fields can couple into the gravitational wave detectors and create noise in the detectors' output strain channels. It has been determined that the correlated magnetic field noise did not affect the stochastic background upper limits measured by initial LIGO and Virgo, but it is possible that they could contaminate the future results of Advanced LIGO and Advanced Virgo~\cite{wiener}. If that is the case, then methods must be taken to try to monitor the magnetic fields and subtract their effects. This could be done, for example, via Wiener filtering~\cite{wiener,0264-9381-33-22-224003,Cirone:2018guh}. Low noise magnetometers are now installed at the LIGO and Virgo sites in order to monitor this correlated magnetic noise.
The data from these magnetometers will be used for Wiener filtering if it is necessary for the stochastic background searches. In addition to long term magnetic noise correlations, short duration magnetic transients, produced from lightning strikes around the world, are seen to be coincidently visible at the detector sites and could affect the search for short duration gravitational wave events~\cite{0264-9381-34-7-074002}.

\subsection{Future Observing Runs for LIGO and Virgo}
\label{subsec:AdLV_future}
Advanced LIGO has completed its first observing run, and the results of the search for a stochastic background have been published~\cite{PhysRevLett.118.121101,PhysRevLett.118.121102}. At the time of this writing Advanced LIGO has completed its second observing run, with Advanced Virgo joining for the last month. Over the next few years further observing runs will happen as Advanced LIGO and Advanced Virgo approach their target sensitivities~\cite{Abbott2016}. At their target sensitivities LIGO and Virgo should be able to constrain the energy density of the stochastic background to approximately $\Omega_{GW} \sim 1 \times 10^{-9}$ (in the 10 Hz to 100 Hz band) with a year of coincident data, while 3 years of data will give a limit of $\Omega_{GW} \sim 6 \times 10^{-10}$\footnote{Note that the predicted evolution of the LIGO-Virgo sensitivity for the stochastic background search, from O1 to reaching design sensitivity, is displayed in Fig.1 of \cite{Abbott:2017xzg}.}. 
At this point it is likely that LIGO and Virgo could observe a stochastic background produced by binary black holes and binary neutron stars~\cite{PhysRevLett.118.121101,PhysRevLett.116.131102,Abbott:2017xzg}. Various cosmological models~\cite{starobinksii,bar-kana,buonanno,mandic} or cosmic strings~\cite{kibble,Damour:2004kw,olmez1,olmez2} might produce a detectable stochastic background at this level as well.
Similar sensitivity advances will also be made with the directional searches as Advanced LIGO and Advanced Virgo reach their target sensitivities. In fact, the addition of Advanced Virgo to the network, with its long distance displacement from the LIGO sites, will make a further important contribution to the directional searches and their ability to map the sky~\cite{PhysRevLett.118.121102}. One can expect to see many important results pertaining to the search for a stochastic background from LIGO and Virgo in the coming years.

\subsection{Laser Interferometer Space Antenna - LISA}
\label{subsec:LISA}
A way to avoid the many deleterious noise sources found on the Earth is to put a gravitational wave detector in space. This is the idea behind the Laser Interferometer Space Antenna (LISA)~\cite{eLISA:2014,2017arXiv170200786A}. The LISA mission has been accepted by the European Space Agency (ESA), with the National Aeronautics and Space Administration (NASA) participating as a junior partner. The current plan is for a 2034 launch, with a mission lasting 4 years, with a possibility for an extension to 10 years of total observation time.

LISA will consist of three satellites in an equilateral triangle configuration, separated from one another by $2.5 \times 10^{6}$ km. This will allow for three gravitational wave {\it interferometers}. Strictly speaking, these will not be interferometers of the kind used by LIGO and Virgo. Of order of $\sim 1$ W of laser light will be emitted from one satellite, while only pico-Watts will be received by the other. As such, the phase of the incoming beam will be measuered, and the re-emitted light will have its phase set accordingly~\cite{2017arXiv170200786A}. At low frequencies only two of the interferometers' data streams will be independent~\cite{PhysRevD.73.102001}.

LISA Pathfinder has demonstrated that much of the technology required for the LISA mission can meet the requirements for its success~\cite{PhysRevLett.116.231101,PhysRevLett.120.061101}. For example, with LISA Pathfinder the relative acceleration noise of two test masses was measured to be 
$(1.74 \pm 0.05) ~ \rm{fm ~ s}^{-2}/\sqrt{\rm{Hz}}$ above 2 mHz and $(6 \pm 1) \times 10 ~ \rm{fm ~ s}^{-2}/\sqrt{\rm{Hz}}$ at 20 $~\mu$Hz.
This level of relative acceleration noise meets the requirements for the LISA mission.

LISA will be able to observe gravitational waves from any direction in the sky. It will also be generally sensitive to both polarizations of gravitational waves from any direction. The operating band for LISA will extend from frequencies smaller than
$10^{-4}$ Hz to those greater than $10^{-1}$ Hz. This will be an important observing band for observations, with many interesting signals predicted~\cite{2017arXiv170200786A,Colpi:2016fup}. 

One of the important signal sources for LISA will be the stochastic background. Certainly all of the compact galactic binaries will produce a stochastic background that will be significant for LISA; so significant, that it could mask other more interesting signals. Various methods have been suggested for accounting for galactic binary signals within the LISA data~\cite{PhysRevD.72.022001,0264-9381-22-18-S04,0264-9381-24-19-S20,0264-9381-24-19-S17,PhysRevD.80.064032,0264-9381-27-8-084009}.

Other important sources for a stochastic background will include binary black hole systems throughout the universe.
The detections by LIGO and Virgo of gravitational waves from binary black hole inspirals implies that there will be a stochastic background from these systems from throughout the history of the universe~\cite{PhysRevLett.116.131102}. This stochastic background will also be potentially observable by LISA. This background will have its energy density vary as $\Omega_{GW}(f) \propto f^{2/3}$. 
The predicted stochastic background was $\Omega_{GW}(f) = 1.1^{+2.7}_{-0.9} \times 10^{-9} ~ (f/25 ~ \rm{Hz})^{2/3}$. An assumption of the worst case scenario gives a background at the $\Omega_{GW} \sim 10^{-10} ~ (f/25 ~ \rm{Hz})^{2/3}$ level.
If LISA observes a stochastic background it will be important for it to also be able to measure its spectral variation. A goal of the LISA mission is to make measurement of this stochastic background in two bands, 0.8 mHz $< f <$ 4 mHz and, 4 mHz $< f <$ 20 mHz, each with a signal to noise ratio of 10 assuming 4 years of integration time. This should be achievable by LISA~\cite{2017arXiv170200786A}.

Certainly a cosmologically produced stochastic background would be the most interesting as it would give direct evidence about the universe at its earliest moments. For example, a first-order phase
transition in the energy range from hundreds of GeV to one TeV would produce gravitational waves that would fall within LISA's observing band~\cite{1475-7516-2016-04-001,1705.01783}. 
By measuring the spectral shape it will be possible to begin to decipher the source of the background. For example, LISA hopes to detect stochastic backgrounds produced by inflation~\cite{1475-7516-2016-12-026}, first order phase transitions~\cite{1475-7516-2016-04-001}, and cosmic strings~\cite{Blanco-Pillado:2017rnf}. In order to have sufficient sensitivity to make statements about the spectral characteristics of the stochastic background, LISA is being designed so that its sensitivity to is sufficient to achieve 
measurements of
$\Omega_{GW} = 1.3 \times 10^{-11} (f /10^{-4} ~ \rm{Hz})^{-1}$ for  0.1 mHz $< f <$ 20 mHz, 
and $\Omega_{GW} = 4.5 \times 10^{-12} (f /10^{-2} ~ \rm{Hz})^{3}$ for 2 mHz $< f <$ 200 mHz. Again this assumes
4 years of observation~\cite{2017arXiv170200786A}.
Because there are three detectors, each sharing an arm and laser beam with its neighbor, there will be correlations in the signals and the noise. It will be helpful to this search that a {\it null-stream} can be created; namely an output channel where there is no signal (to some approximation). However, it is possible that correlated noise could affect the data; this would be especially problematic for a search for a stochastic background. 
LISA will take the data from the three interferometers and recombine them to create three different channels using {\it Time Delay Interferometry}~\cite{Tinto2014}, a way to minimize laser noise when the arm lengths for the interferometers are unequal. Nominally the noise and signals will then be uncorrelated. 
Between correlated noise and the galactic binaries it will be a challenge for LISA to achieve the $\Omega_{GW} \sim 10^{-12}$ level, but certainly not impossible either. Much research is already underway in order to achieve the LISA goals for measuring or setting limits on a stochastic background~\cite{2017arXiv170200786A}.

\begin{figure}
\begin{center}
\includegraphics[width=5.5in]{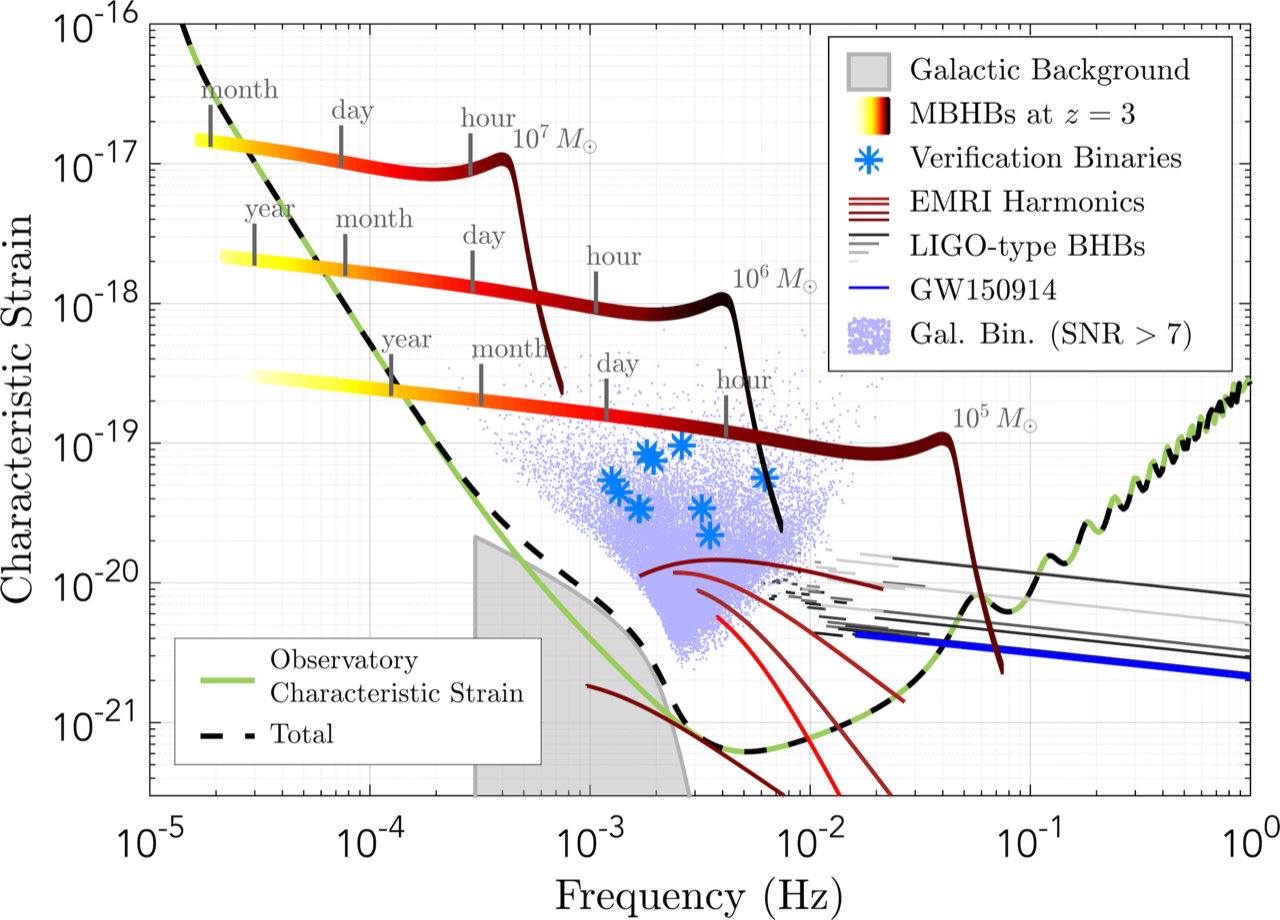}
\caption{The sensitivity (in terms of characteristic strain~\cite{0264-9381-32-1-015014,moore26gravitational}) of the proposed LISA 3-detector system with $2.5 \times 10^{6}$ km arms~\cite{2017arXiv170200786A}. Numerous sources that are expected to be observed by LISA are displayed. Especially important for the search for a stochastic background will be the galactic background (see Sec.~\ref{subsec:wdb}). Thousands of galactic binaries in LISA will produce signals with SNR $>$ 7, and will be individually resolvable. Some of these systems are well known and have already been studied; these will be the so-called {\it Verification Binaries}, that will produce gravitational wave signals that will help to confirm the sensitivity and calibration of LISA. However, countless other binary systems will contribute to a gravitational wave background that will complicate the LISA search for a cosmologically produced stochastic background~\cite{PhysRevD.82.022002,PhysRevD.89.022001}. This background is also displayed in this figure. Many other predicted signals for LISA are displayed, including massive black hole binaries (MBHBs, including GW150914), black hole binary systems that can be observed with Advanced LIGO and Advanced Virgo (LIGO-type BHBs), and extreme mass ratio inspirals (EMRIs). See \cite{2017arXiv170200786A} for more details on these signal sources. Figure from \protect \cite{2017arXiv170200786A}.}
\label{fig:LISA}
\end{center}
\end{figure}

The sensitivity of the proposed LISA 3-detector system with $2.5 \times 10^{6}$ km arms is presented in Fig. ~\ref{fig:LISA}~\cite{2017arXiv170200786A}. The signal sources that are expected to be observed by LISA are also presented. It is important to note the presence of the close compact binaries, as described in Sec.~\ref{subsec:wdb}. Those binaries producing gravitational waves above the LISA sensitivity (marked with SNR $>$ 7) will be individually resolvable, and in principle can be removed from contaminating the LISA stochastic background search~\cite{PhysRevD.72.022001,0264-9381-22-18-S04,0264-9381-24-19-S20,PhysRevD.80.064032,0264-9381-24-19-S17,PhysRevD.82.022002,PhysRevD.89.022001}. However, the sum of all other binaries will produce a gravitational wave background that must be addressed in a search for a cosmologically produced stochastic background~\cite{PhysRevD.89.022001}.

\subsection{DECi-hertz Interferometer Gravitational wave Observatory - DECIGO}
\label{subsec:DECIGO}
Another proposed space based gravitational wave detector is the DECi-hertz Interferometer Gravitational wave Observatory, a Japanese project~\cite{1742-6596-122-1-012006,1742-6596-840-1-012010}. Similar to LISA, DECIGO will consist of 3 satellites, in an equilateral triangle configuration, but with a distance separation of 1000 km. It will also be in a heliocentric orbit. The light traveling between each spacecraft will be within a Fabry-Perot cavity, similar to what is done in the arms of LIGO and Virgo. The proposal is for four DECIGO clusters (with a DECIGO cluster consisting of the 3 satellites in a 1000 km equilateral triangle configuration). Two of the DECIGO clusters will be overlapping, with the two equilateral triangles displaced from one another by a rotation of 30$^o$. This close proximity should make DECIGO especially sensitive to a stochastic background.

DECIGO's operating frequency band will be 0.1 Hz to 10 Hz. This will form an important bridge in frequency space between LISA, and the LIGO-Virgo-KAGRA ground based network. This frequency band is particularly advantageous in that the signals' {\it contamination} from white dwarf binaries will be extremely low, giving a window for a search for a cosmologically produced stochastic background~\cite{Farmer:2003pa}. Because of this reduced white dwarf binary foreground, and the sensitivity of DECIGO, it could be possible to achieve a detection limit for a stochastic background search of $\Omega_{GW} \sim 2 \times 10^{-16}$ with three years of observations. This impressive sensitivity could provide a direct observation of gravitational waves produced during inflation~\cite{PhysRevD.48.3513}. In addition, with DECIGO it could also be possible to measure the Stokes V parameter, namely a measure of the circular polarization~\cite{PhysRevD.75.061302}. A measured asymmetry in right-handed and left-handed polarizations of a stochastic background could indicate parity violation in the early universe. An adjustment in the positions of the DECIGO clusters will allow DECIGO to be sensitive to an asymmetry in the right-handed and left-handed gravitational waves, as quantified by the Stokes V parameter ~\cite{PhysRevD.75.061302}.
Initial LIGO data has been used to search for a parity violation, but with no detected stochastic background the results are consistent with $\Pi = 0$, 
with $\Pi = \pm 1$ representing fully right- or left-handed gravitational waves polarizations~\cite{CROWDER201366}. Since DECIGO could in principle measure a cosmologically produced stochastic background, it could then subsequently search for these signatures of parity violation.

The current planning for the mission estimates that DECIGO will be launched in the 2030s~\cite{1742-6596-840-1-012010}. In preparation for this ambitious mission a smaller version of DECIGO is being planned for launch in the late 2020s, called B-DECIGO. This will consist of three satellites, but with a separation of 100 km, and orbiting the Earth. B-DECIGO is intended to demonstrate the technology needed for the full DECIGO mission, but it could detect gravitational waves in its own right~\cite{1742-6596-840-1-012010}.

\subsubsection{Big Bang Observer and Other Space Mission Proposals} 
\label{subsubsec:BBO}

A project similar to DECIGO is the Big Bang Observer (BBO)~\cite{PhysRevD.80.104009}. Like DECIGO, it would have a triangular configuration, but with arm lengths of $5 \times 10^{4}$ km. With two overlapping triangular clusters a cross-correlation can be made between independent detector data sets~\cite{0264-9381-23-7-014}. BBO is designed to look for a cosmologically produced stochastic background, with a sensitivity of $\Omega_{GW} \sim 10^{-17}$ in the 0.03 Hz to 3 Hz frequency band~\cite{PhysRevD.80.104009,0264-9381-23-7-014}. This important frequency band should be free of astrophysical contamination~\cite{0264-9381-23-7-014}.

Cornish and collaborators have explored various modifications to the LISA-DECIGO-BBO designs, especially the concept of two overlapping triangular clusters~\cite{0264-9381-18-17-308}. The cross-correlation of the data from the two overlapping (but independent) detectors creates an opportunity to achieve a sensitivity whereby gravitational waves from inflation could be detected. A major goal would be to search for a stochastic background around a $\mu$Hz, thereby operating in a regime with minimal contamination from astrophysical sources~\cite{0264-9381-18-17-308}. The proposal would be for a successor for LISA, namely a LISA II with arm lengths of $\sqrt{3}$ AU. LISA II is proposed to be a system of 6 spacecraft in a configuration of two equilateral triangles, so essentially two overlapping LISA systems~\cite{0264-9381-18-17-308}. The $\sqrt{3}$ AU large arm lengths require an orbit farther out, which results in reduced thermal effects because of the diminished solar heating. In addition, a relative acceleration noise for the proof masses is assumed to be at the level of $\delta a \sim 3 \times 10^{-16}$ m s$^{-2}$. And while the recent observations of the relative acceleration of the proof masses for LISA Pathfinder were impressive~\cite{PhysRevLett.116.231101,PhysRevLett.120.061101}, an improvement will still be necessary, especially at this low frequency of a $\mu$Hz. Given the assumptions for the detectors' performance it is speculated that the LISA II design could observe a stochastic background at the level of $\Omega_{GW} \sim 4 \times 10^{-13}$. Even more ambitious would be the LISA III design, with arm lengths of 35 AU. In this case the sensitivity to a stochastic background could reach $\Omega_{GW} \sim 2 \times 10^{-18}$ at $10^{-8}$ Hz. This would certainly be sufficient to observe gravitational waves from inflation~\cite{PhysRevD.48.3513}.

\subsection{Fermilab Holometer}
\label{subsec:Holometer}
The Fermilab Holometer consists of two Michelson interferometers that are nearly overlapping (a separation of 0.635 m), with arm lengths of 39.2 m~\cite{0264-9381-34-6-065005}. The Holometer was constructed with the goal to attempt to observe correlations in space-time variations. It was speculated~\cite{PhysRevD.18.1747,HAWKING1980283,PhysRevLett.69.237} that this could be a consequence of quantum gravity. However the two co-aligned and co-situated interferometers also provide a unique means to try to measure a stochastic background at MHz frequencies~\cite{PhysRevD.95.063002}.

The Holometer has recently demonstrated that its strain sensitivity $h(f) = \sqrt{S_{h}}$ (see Eq.~\ref{Eq:specden}) is better than $10^{-21}$ Hz$^{-1/2}$ in the 1 MHz to 13 MHz band. With 130 hours of coincident data between the two interferometers a $3 \sigma$ limit on the energy density of the stochastic background was made, $\Omega_{GW} < 5.6 \times 10^{12}$ at 1 MHz, and $\Omega_{GW} < 8.4 \times 10^{15}$ at 13 MHz ~\cite{PhysRevD.95.063002}. These are the best limits to date in this high frequency band for a direct measurement, although Big Bang Nucleosynthesis~\cite{CYBURT2005313}, CMB observations~\cite{PhysRevD.85.123002,0264-9381-32-4-045003} and indirect limits~\cite{PhysRevX.6.011035} do place much better constraints at these frequencies.

\subsection{Pulsar Timing}
\label{subsec:PT}
Pulsars are like clocks in space. These are rapidly rotating neutron stars with large magnetic fields. Presumably there is a misalignment between the magentic field dipole axis and the rotation axis. As such, the sweeping magnetic field creates a regularly arriving radio pulse. These pulses were first detected on Earth in 1967 and reported in {\it Observation of a Rapidly Pulsating Radio Source}~\cite{1968Natur.217..709H}. It was quickly deciphered that these radio signals were coming from rapidly rotating neutron stars, namely {\it pulsars}~\cite{1968Natur.218..731G,1969Natur.221...25G}.

Sazhin~\cite{1978SvA....22...36S} and Detweiler~\cite{1979ApJ...234.1100D} were the first to recognize that the regularity of the signals received from pulsars could be used to search for gravitational waves.
For the detection of gravitational waves, one can consider a pulsar and an observer on Earth to be analogous to the two ends of a single interferometer arm. 
For the long gravitational wave periods ($T \sim 1$ yr) the energy density of the stochastic background can be expressed as
\begin{equation}
\label{eq:Omega_pulsar}
\Omega_{GW}(f) = \frac{2 \pi^{2}}{3 H_{0}^2} A^{2}_{GW} f^{2}_{yr}  \Big( \frac{f}{f_{yr}} \Big)^{n_{t}} ~,
\end{equation}
where $A_{GW}$ is the characteristic strain amplitude at the reference frequency $f_{yr} = 1/\rm{year}$, and $n_{t}$ is the spectral index; see Eq. 6 of~\cite{PhysRevX.6.011035}.
In addition, see~\cite{Romano2017} for a comprehensive description of how one can extract a gravitational wave signal from the pulsar timing data. 

Needless to say, while the signal from a pulsar can be regular, numerous effects can modify the phase of the arriving signal. Typically pulsars lose energy and their rotation frequency decreases. If pulsar signals are to be used to try to detect gravitational waves, then the physical effects of the pulsars themselves must be well understood. One must account for dispersion of the signal by the interstellar medium, and also account for fluctuations in the dispersion. The period derivative of the pulsar, caused by the loss of rotational energy via the emission of gravitational waves, must be included. The exact location of the pulsar in the sky, along with its proper motion, must be known to high precision~\cite{RAWLEY761}.

After the discovery of the first pulsar, and subsequent detections of others, it was observed that some pulsars, such as PSR 1937+21, could be as stable as atomic clocks~\cite{RAWLEY761}. For this pulsar the frequency stability was observed to be $\Delta f/f \sim 6 \times 10^{-14}$ when averaged over times longer than 4 months. With the observations of this pulsar, and through the observed frequency stability, it was possible in 1987 to set a limit on the energy density of the stochastic background to be $\Omega_{GW}(f) ~ h^{2} < 4 \times 10^{-7}$ at a frequency of $7 \times 10^{-7}$ Hz~\cite{RAWLEY761}. During this early period in pulsar observations many quickly used their observations to also constrain the stochastic background. 
For example, Hellings and Downs~\cite{1983ApJ...265L..39H} used the observation of four pulsars to contrain the stochastic background to $\Omega_{GW}(f) ~ h^{2} < 1.4 \times 10^{-4}$ at a frequencies around $10^{-8}$ Hz.

Using pulsar timing to try to observe gravitational waves is currently a very active research area, involving numerous collaborations around the world~\cite{0264-9381-30-22-220301,0264-9381-30-22-224007,0264-9381-30-22-224008,0264-9381-30-22-224009,0264-9381-30-22-224010,0264-9381-30-22-224011}.
The current observations are concentrating on signals with frequencies in the range of $10^{-9}$ to $10^{-7}$ Hz~\cite{Romano2017}. 
A stochastic background is the most likely signal source for the current pulsar timing experiments, namely the background produced by all of the inspiral and mergers of super massive black hole binaries over the history of the universe~\cite{Romano2017,0004-637X-583-2-616}.

The European Pulsar Timing Array has recently reported limits on the stochastic background based on the observation of 6 pulsars over 18 years. Their upper limits on the energy density of the stochastic background is $\Omega_{GW} h^{2} < 1.1 \times 10^{-9}$ at 2.8 nHz. This limit places stringent constraints on the super-massive binary black hole population in the universe. 
This analysis also constrains the string tension to 
$G \mu < 1.3 \times 10^{-7}$
for a Nambu-Goto field theory cosmic string network~\cite{doi:10.1093/mnras/stv1538}.

NANOGrav~\cite{0004-637X-821-1-13} has reported the results from an examination of nine years of pulsar data involving 37 pulsars~\cite{0004-637X-821-1-13}.
The upper limit on the energy density of the stochastic background was reported to be $\Omega_{GW} h^{2} < 4.2 \times 10^{-10}$ at
frequency $3.3 \times 10^{-9}$ Hz~\cite{0004-637X-821-1-13}. These results were then improved with the goal to constrain cosmic string parameters~\cite{Blanco-Pillado:2017rnf}. Using a new analysis of the NANOGrav results, a constraint has been made on the cosmic string tension to be $G\mu < 1.5 \times 10^{-11}$~\cite{Blanco-Pillado:2017rnf}.

The Parkes Pulsar Timing Array (PPTA)~\cite{0264-9381-30-22-224007,2013PASA...30...17M} uses the Parkes 64-m radio telescope to observe 24 pulsars. With this data they have constrained the energy density of the stochastic background to be $\Omega_{GW} < 2.3 \times 10^{-10}$ at 6.3 nHz for a spectral index of $n_{t} = 0.5$~\cite{Shannon1522}. The limit for a spectral index of $n_{t} = 0$ is the same to two decimal places~\cite{PhysRevX.6.011035}. It is expected that with five subsequent years of data the PPTA could achieve a limit of $\Omega_{GW} < 5 \times 10^{-11}$, but that will be even further improved by combining the results from the different pulsar timing collaborations~\cite{PhysRevX.6.011035} as part of the International Pulsar Timing Array~\cite{0264-9381-30-22-224010}.

\subsection{Doppler Tracking Limits}
\label{subsec:DTL}
The same techniques that are applied to radio signals from pulsars for the detection of gravitational waves can be applied to signals transmitted from spacecraft traveling through our solar system~\cite{PhysRevD.23.832,Armstrong2006}. In fact, the Doppler tracking of spacecraft was considered and analyzed before pulsar timing~\cite{1975GReGr...6..439E}. Originally intended to look for gravitational waves emitted from pulsars, the Doppler tracking technique is also applicable to searches for a stochastic backgorund of gravitational waves. The Earth and the spacecraft are considered as free masses. A limit can be placed on the energy density of the stochastic background in the frequency range of $10^{-6}$ to $10^{-2}$ Hz~\cite{Thorne_300yrs}. Signals from many different spacecraft have been used, including the Viking~\cite{1979ApJ...230..570A}, Voyager~\cite{PhysRevD.23.844}, Pioneer 10~\cite{1984Natur.308..158A}, Pioneer 11~\cite{1987ApJ...318..536A}, and Cassini~\cite{2003SPIE.4856...90A,2003Natur.425..374B,0004-637X-599-2-806}. The best upper limit on the energy density of the stochastic background comes from the analysis of the Cassini data, giving $\Omega_{GW} < 0.025$ at a frequency of $1.2 \times 10^{-6}$ Hz and assuming a value for the Hubble constant of $75 ~\rm{km/s/Mpc}$~\cite{0004-637X-599-2-806,Armstrong2006}. Using the currently accepted $H_{0} = 67.74 ~\rm{km/s/Mpc}$~\cite{Planck2015} this reduces the limit to $\Omega_{GW} < 0.03$.

\subsection{Cosmic Microwave Background Anisotropy Limits}
\label{subsec:CMB}
The near isotropy of the comsic microwave background (CMB) can be used to constrain the energy density of the stochastic background at very low frequencies. There are two ways in which gravitational waves will disturb the CMB. Gravitational waves today with wavelengths on the order of the horizon size will produce a quadrupole anisotropy, while gravitational waves at the time of recombination will cause fluctuations on smaller angular scales that can be oberved today~\cite{1967ApJ...147...73S}. In Fig.~\ref{fig:Landscape} the curve labeled CMB corresponds to the limits on $\Omega_{GW}(f)$ from the CMB measurements of the Planck satellite~\cite{Planck_CMB_2011}. An energy of gravitational waves above this level would have changed the observations made on the CMB~\cite{PhysRevD.50.3713,PhysRevD.52.1902,Allen:1996vm}, such as those made by Planck~\cite{Planck_CMB_2011}.

\subsection{Indirect Limits}
\label{subsec:indirect}
The production of deuterium, helium and lithium in the early universe can be used to constrain the energy density of the stochastic background. This Big Bang Nucleosynthesis (BBN) limit provides an important constraint on the stochastic background. If the energy density of the gravitational waves is too large when these light nuclei are produced, the abundances today will be different from what is actually observed~\cite{1980A&A....89....6C}. Too much gravitational wave energy will speed up the universe's expansion rate, thereby reducing the amount of helium formed from deuterium, altering the observed ratios.

The baryon density in the universe today is in the range of $\rho_{b} = (3.9 - 4.6) \times 10^{-31}$ g cm$^{-3}$. This then translates into a relationship with the critical density of the universe, namely $\Omega_{b} = \rho_{b}/\rho_{crit} = 0.046 - 0.053$. The majority of the baryon mass of the universe is made up of neutral hydrogen. The primordial mass fraction of helium $^{4}$He is $Y_{p} = \rho(^{4}$He$)/\rho_{b} \approx 0.25$. The primordial mass fraction for deuterium D and helium $^{3}$He are of the order $10^{-5}$, while for Lithium $^{7}$Li it is at the $10^{-10}$ level~\cite{Patrignani:2016xqp}. The observations of the mass ratios for primordial nucleosynthesis limit the energy density of gravitational waves to $\Omega_{GW} < 1.8 \times 10^{-5}$
for frequencies in excess of $10^{-10}$ Hz~\cite{0264-9381-32-4-045003,Maggiore:1999vm,CYBURT2005313}.

Observations of the CMB, BBN, and baryon acoustic oscillations (BAO)~\cite{doi:10.1093/mnras/stt2206} can be combined to provide a limit on the energy density of the stochastic background~\cite{0264-9381-32-4-045003,Pagano2016823,PhysRevX.6.011035}. It can be shown that an upper limit on the energy density of the stochastic background for frequencies above $10^{-15}$ Hz can be made with
\begin{equation}
\label{eq:Neff}
\Omega_{GW} \leq \frac{7}{8} \Big(\frac{4}{11}\Big)^{4/3} (N_{\rm{eff}} - 3.046) ~ \Omega_{\gamma}
\end{equation}
where the energy density of the CMB is $\Omega_{\gamma} = 2.473 \times 10^{-5} /h^2$~\cite{0264-9381-32-4-045003}. The term $N_{\rm{eff}}$ is the effective number of neutrinos, and the measurements of the $Z$ boson width~\cite{Patrignani:2016xqp}  
limit its value. Studies considering the behavior of the three neutrino families in the early universe give a value of $N_{\rm{eff}} \approx 3.046$~\cite{2005NuPhB.729..221M}. The presence of a large energy density of gravitational waves would alter the value of $N_{\rm{eff}}$ observed via cosmological observations today. Combining Eq.~\ref{eq:Neff} with the value of $\Omega_{\gamma}$ implies
\begin{equation}
\label{eq:indirect}
\Omega_{GW} h^{2} \leq 5.6 \times 10^{-6} (N_{\rm{eff}} - 3.046) ,
\end{equation}                     
which can then be used to limit $\Omega_{GW}$ based on BBN, CMB and BAO observations~\cite{0264-9381-32-4-045003}. Recent observations place a limit of $\Omega_{GW} \leq 3.8 \times 10^{-6}$~\cite{Pagano2016823,PhysRevX.6.011035}.

\subsection{B-modes in the Cosmic Microwave Background}
\label{subsec:Bmodes}
The CMB holds much information pertaining to a stochastic background produced at the earliest moments of the universe. For example, the gravitational waves produced during inflation should leave their imprint on the CMB when it was produced
 $3.8 \times 10^{5}$ years after the Big Bang; this is the recombination time when the temperature of the universe was  $\sim 3 \times 10^{3}$ K. 

As described above, quantum fluctuations during inflation will create a stochastic background of gravitational waves. Density fluctuations will also be created. Both of these can affect the polarization content of the CMB. However, they can be differentiated from one another, namely by breaking down the composition of the CMB polarization into a curl-free component (an E mode), and a curl component (a B mode)~\cite{doi:10.1146/annurev-astro-081915-023433}. The presence of gravitational waves produced during inflation would be responsible for introducing B modes into the CMB polarization at the time of recombination. Gravitational waves can also induce fluctuations in the temperature of the CMB. An excellent summary of all aspects of B modes is presented in~\cite{doi:10.1146/annurev-astro-081915-023433}.

Gravitational waves affect the metric of spacetime, which can then consequently affect a photon's energy. At the time of recombination the gravitational waves and the photons were traveling within the cosmic fluid of material present at the time, mostly protons, electrons and neutrinos. Of course there is also a change in the energy of the photons due to the expansion of the universe. The presence of gravitational waves alone does not affect the polarization of the photons, only their energy. Similarly, density fluctuations in the cosmic fluid will induce a gravitational redshift in the photons, but not affect their polarization. However, as photons Thomson scatter off of the electrons present, a net polarization can be induced.


The measure for the amount of gravitational waves produced during inflation is typically expressed in terms of the tensor to scalar ratio,
\begin{equation}
r = \frac{\Delta_{h}^{2}}{\Delta_{R}^{2}} ~ ,
\end{equation}
where $\Delta_{h}^{2}$ is the gravitational wave power spectrum and $\Delta_{R}^{2}$ is the curvature power spectrum. The $r$ value can also be directly related to the potential of the inflaton, $\phi$, during inflation, namely $V(\phi)$; see~\cite{doi:10.1146/annurev-astro-081915-023433} for details.

Unfortunately gravitational waves are not the only means to create B modes in the polarization of the CMB. Gravitational lensing of the CMB can also produce B modes. This would be caused by massive objects between us (as observors) and the surface of last scattering of the CMB~\cite{PhysRevD.58.023003,LEWIS20061,doi:10.1146/annurev-astro-081915-023433}. This effect has been observed~\cite{PhysRevLett.111.141301,0004-637X-808-1-7,2015ApJ...807..151K}. However, with the present knowledge of the parameters describing our universe, $\Lambda$CDM, it is possible to accurately predict the amount of B modes in the CMB polarization created by lensing. The influence of gravitation waves on the B modes will be prominent in the spherical harmonic range from $\sim l = 10$ to $\sim l = 100$, or roughly an angular scale of $~\sim 0.1^{o}$ to $~\sim 1^{o}$~\cite{doi:10.1146/annurev-astro-081915-023433}. 

The most serious obstacle to directly observing the effects of gravitational waves on the CMB is the presence of the material in and about our galaxy. Synchrotron emission in the galaxy is a foreground which will contaminate CMB polarization studies for photon frequencies under 100 GHz~\cite{Adam:2015tpy,doi:10.1146/annurev-astro-081915-023433}. Dust grains tend to align themselves with the galactic magnetic field; the thermal emission from these grains tends to be polarized~\cite{doi:10.1146/annurev-astro-082214-122414}. The presence of the material makes the search for B modes in the galactic plane impractical, and hence observations need to take place at high galactic latitudes~\cite{doi:10.1146/annurev-astro-081915-023433}.

When observations are made of the polarization of the CMB across a patch of sky, a decomposition can be made of the E modes and B modes. The polarization power as a function of angular scale (or exactly, spherical harmonic number $l$) is measured and plotted. From that the tensor to scalar ratio, $r$, can be extracted~\cite{doi:10.1146/annurev-astro-081915-023433}. Numerous observation teams are currently attempting to find the B modes produced by gravitational waves. In 2014 the BICEP2 Collabroration claimed an observation of B modes in the range $30 < l < 150$, or roughly $0.3^{o}$ to $1.5^{o}$~\cite{PhysRevLett.112.241101}. However, the results were quickly challenged~\cite{Flauger:2014qra}, and subsequent analyses showed that the observed B modes were actually due to galactic dust, and reported an upper limit of $r < 0.12$ at 95\% confidence~\cite{PhysRevLett.114.101301}. Subsequent observations by BICEP2 and the KECK Array have further reduced this limit to $r < 0.09$ at 95\% confidence; combining the results with Planck CMB temperature data and baryon acoustic oscillation results further constrains the ratio to $r < 0.07$ at 95\% confidence~\cite{PhysRevLett.116.031302}. There are other attempts by other groups to observe or constrain the B modes due to gravitational waves~\cite{Kusaka:2018yzq,2014ApJ...794..171P,2014SPIE.9153E..11M,Manzotti:2017net}.

\subsection{Normal Modes of the Earth, Moon and Sun}
\label{subsec:ENM}
The measurement of the normal modes of the Sun, Earth and the Moon have been used to limit the energy density of the stochastic gravitational wave background. The idea of using the Earth itself as a gravitational wave detector goes back to 1969 with a proposal from Freeman Dyson~\cite{1969ApJ...156..529D}. The use of applying actual data pertaining to motions of the Sun and Earth started as early as 1984 when Boughn and Kuhn~\cite{1984ApJ...286..387B} analyzed the process by which a gravitational wave background drives the normal modes of a spherical body. Using data of the observed line of sight velocity of the surface of the Sun they were able to constrain $\Omega_{GW}(f)$ to be less that 100 at a frequency of $4 \times 10^{-4}$ Hz. The Earth's cross section to the background of gravitational waves is smaller than the Sun's because the Earth is much smaller. However, the data on the seismic activity is much better for the Earth. The limit achieved from the Earth data was also $\Omega_{GW}(f) < 100$  at frequencies of $2 \times 10^{-3}$ Hz and $2 \times 10^{-2}$ Hz. 

Much progress has subsequently been made with these types of studies. Recent observations of the Sun have used helioseismology. A stochastic background of gravitational waves would excite stars like our Sun, causing them to oscillate. For the Sun, high precision radial velocity data is used to monitor the motion. Specifically, limits on the the high frequency quadrupolar $g$ modes~\cite{2017A&A...604A..40F} are used to constrain the stochastic background. A model of the sun has been used where it is assumed to be a spherical body with a negligible shear modulus. The best constraint with this method is $\Omega_{GW} < 4.0 \times 10^{5}$ at 0.171 mHz~\cite{0004-637X-784-2-88}.

The method of Dyson~\cite{1969ApJ...156..529D} using the Earth to attempt to measure gravitational waves was implemented using seismometer data~\cite{PhysRevLett.112.101102}. Correlations were made between pairs of seismometers. The seismometers used in this study were located around the world. The surface of the Earth was considered to be a free and flat surface in its response to gravitational waves. The limit derived was $\Omega_{GW} < 1.2 \times 10^{8}$ in the $0.05 - 1$ Hz band using one year of data~\cite{PhysRevLett.112.101102}. 

This study was then extended to take into account the internal structure of the Earth~\cite{PhysRevD.90.042005}. This allowed for lower frequencies to be addressed since below 50 mHz there is global coherence in the seismic motion. The new study used both the data from gravimeters, and a model of the response of the Earth's modes to gravitational waves. Ten years of data from the superconducting gravimeters for the Global Geodynamics Project~\cite{10.1007/978-3-642-10634-7_83} were analyzed. For frequencies between 0.3 mHz and 5 mHz limits were placed on the energy density of the stochastic background, $\Omega_{GW}$, with the limits ranging between 0.035 and 0.15~\cite{PhysRevD.90.042005}. 

Seismic arrays on the moon have also been used to limit the energy density of the stochastic background~\cite{PhysRevD.90.102001}. Seismometers were placed on the moon between 1969 and 1972 as part of the Apollo 12, 14, 15 and 16 missions. Data was acquired until 1977. The seismic noise on the moon is less than that on Earth. From the lunar seismometer data the integrated energy density of the stochastic background from 0.1 to 1 Hz can be constrained to $\Omega_{GW} < 1.2 \times 10^{5}$~\cite{PhysRevD.90.102001}. This is currently the best limit in this frequency band.

\section{Conclusions}
\label{sec:conclusions}
The observations of gravitational waves by Advanced LIGO~\cite{0264-9381-27-8-084006,0264-9381-32-7-074001} and Advanced Virgo~\cite{0264-9381-32-2-024001} have created tremendous excitement in the world of physics~\cite{PhysRevLett.116.061102,PhysRevLett.116.241103,PhysRevX.6.041015,PhysRevLett.118.221101,GW170608,PhysRevLett.119.141101,PhysRevLett.119.161101}. In addition to signals from the coalescence of binary black hole and binary neutron star systems, numerous other types of signals are expected~\cite{Abbott2016}. One of those is a stochastic background of gravitational waves. The observations of Advanced LIGO and Advanced Virgo predict that these instruments, in the coming few years, should detect a stochastic background created by all binary black hole and binary neutron star mergers throughout the history of the universe~\cite{PhysRevLett.116.131102,PhysRevLett.118.121101,Abbott:2017xzg}. It is also possible that in the coming years LIGO and Virgo could detect a stochastic background from other sources, for example from cosmic strings~\cite{Abbott:2017mem}. The observations by the Advanced LIGO - Advanced Virgo network will likely be made in the 20 to 100 Hz band.

In the coming years it is likely that pulsar timing could make an observation, and most likely that from a stochastic background. This would be the stochastic background produced by all of the inspiral and mergers of super massive black hole binaries over the history of the universe~\cite{Romano2017,0004-637X-583-2-616}. The frequency band for these observations would be $10^{-9}$ to $10^{-7}$ Hz. Numerous collaborations around the world are attempting to detect gravitational waves, and especially the stochastic background~\cite{0264-9381-30-22-220301,0264-9381-30-22-224007,0264-9381-30-22-224008,0264-9381-30-22-224009,0264-9381-30-22-224010,0264-9381-30-22-224011}.

It is also probable that in the coming years the imprint made into the polarization of the CMB by graviational waves created by quantum fluctuations during inflation will be measured.
Observations by BICEP2 and the KECK Array have set a limit on $r$, the tensor to scalar ratio, of $r < 0.09$ at 95\% confidence; when combining the results with Planck CMB temperature data and baryon acoustic oscillation results the constraint is narrowed to $r < 0.07$ at 95\% confidence~\cite{PhysRevLett.116.031302}. Galactic dust is a continual problem in the quest to observe the effect of primordial gravitational waves~\cite{Flauger:2014qra}, however observing the CMB at multiple frequencies may allow the effectts of the dust to be disentangled if $r$ is not inordinately small.. Many groups are trying to observe or constrain the B modes due to gravitational waves~\cite{Kusaka:2018yzq,2014ApJ...794..171P,2014SPIE.9153E..11M,Manzotti:2017net}.

Future gravitational wave detectors will offer exciting prospects for observing the stochastic background. Third generation ground-based gravitational wave detectors, such as the Einstein Telescope~\cite{0264-9381-27-19-194002} or the Cosmic Explorer~\cite{0264-9381-34-4-044001}, will have a factor of $\approx 10$ better sensitivity than the target sensitivity of Advanced LIGO or Advanced Virgo. An exciting prospect for these detectors is that they
should be able to directly observe almost every stellar mass binary black hole merger in the observable universe. 
This can allow them to directly detect and remove from the stochastic search the astrophysical foreground.
By removing this foreground the third generation detection detectors could be sensitive to a cosmologically produced background at the $\Omega_{GW} \sim 10^{-13}$ level with 5 years of observations ~\cite{PhysRevLett.118.151105}. This will then bring the third generation detectors into a sensitivity regime for important cosmological observations.

The LISA mission has been accepted by ESA, with contributions to be made by NASA~\cite{eLISA:2014,2017arXiv170200786A}. The current plan is for a 2034 launch, with a mission lasting 4 years, with a possibility for an extension to 10 years of total observation time. While a major goal of LISA will be to observe a cosmologically produced stochastic background, there will be a significant astrophysically produced foreground that will make this task difficult. For example,  galactic binaries will mask other more interesting signals, and different techniques have been proposed for addressing the galactic binary signals~\cite{PhysRevD.72.022001,0264-9381-22-18-S04,0264-9381-24-19-S20,0264-9381-24-19-S17,PhysRevD.80.064032,0264-9381-27-8-084009}.
The detections of gravitational waves from binary black hole inspirals implies that there will be a stochastic background from these systems~\cite{PhysRevLett.116.131102}, and this stochastic background will also be observable by LISA. If the astrophysical foreground can be addressed LISA could potentially have a sensitivity to a stochastic background at the $\Omega_{GW}(f) \approx 10^{-12}$ level in the $10^{-4}$ Hz to $10^{-1}$ Hz band. This sensitivity could allow LISA to observe the consequences, for example, of a first order electroweak phase transition~\cite{1475-7516-2016-04-001,1705.01783}, or of the presence of cosmic strings~\cite{Blanco-Pillado:2017rnf}.

The recent detection of gravitational waves is the start of a new era. The stochastic gravitational wave background will hold information on some of the most important events in the history of the universe. In the coming years we can expect this background to be observed, and stunning revelations about the universe should be discovered.

\section{Acknowledgements}
NC is supported by National Science Foundation (NSF) grants PHY-1505373 and PHY-1806990 to Carleton College. Thanks to Joe Romano and Philip Charlton for comments on the manuscript. This article has been assigned LIGO Document Number P1800071.

LIGO and Virgo gratefully acknowledge the support of the United States
National Science Foundation (NSF) for the construction and operation of the
LIGO Laboratory and Advanced LIGO as well as the Science and Technology Facilities Council (STFC) of the
United Kingdom, the Max-Planck-Society (MPS), and the State of
Niedersachsen/Germany for support of the construction of Advanced LIGO 
and construction and operation of the GEO600 detector. 
Additional support for Advanced LIGO was provided by the Australian Research Council.
The LIGO and Virgo gratefully acknowledge the Italian Istituto Nazionale di Fisica Nucleare (INFN),  
the French Centre National de la Recherche Scientifique (CNRS) and
the Foundation for Fundamental Research on Matter supported by the Netherlands Organisation for Scientific Research, 
for the construction and operation of the Virgo detector
and the creation and support  of the EGO consortium. 
The LIGO and Virgo also gratefully acknowledge research support from these agencies as well as by 
the Council of Scientific and Industrial Research of India, 
the Department of Science and Technology, India,
the Science \& Engineering Research Board (SERB), India,
the Ministry of Human Resource Development, India,
the Spanish  Agencia Estatal de Investigaci\'on,
the  Vicepresid\`encia i Conselleria d'Innovaci\'o, Recerca i Turisme and the Conselleria d'Educaci\'o i Universitat del Govern de les Illes Balears,
the Conselleria d'Educaci\'o, Investigaci\'o, Cultura i Esport de la Generalitat Valenciana,
the National Science Centre of Poland,
the Swiss National Science Foundation (SNSF),
the Russian Foundation for Basic Research, 
the Russian Science Foundation,
the European Commission,
the European Regional Development Funds (ERDF),
the Royal Society, 
the Scottish Funding Council, 
the Scottish Universities Physics Alliance, 
the Hungarian Scientific Research Fund (OTKA),
the Lyon Institute of Origins (LIO),
the National Research, Development and Innovation Office Hungary (NKFI), 
the National Research Foundation of Korea,
Industry Canada and the Province of Ontario through the Ministry of Economic Development and Innovation, 
the Natural Science and Engineering Research Council Canada,
the Canadian Institute for Advanced Research,
the Brazilian Ministry of Science, Technology, Innovations, and Communications,
the International Center for Theoretical Physics South American Institute for Fundamental Research (ICTP-SAIFR), 
the Research Grants Council of Hong Kong,
the National Natural Science Foundation of China (NSFC),
the Leverhulme Trust, 
the Research Corporation, 
the Ministry of Science and Technology (MOST), Taiwan
and
the Kavli Foundation.
The LIGO and Virgo gratefully acknowledge the support of the NSF, STFC, MPS, INFN, CNRS and the
State of Niedersachsen/Germany for provision of computational resources.

\section*{References}
\bibliography{report}

\providecommand{\newblock}{}
\begin{thebibliography}{100}
\expandafter\ifx\csname url\endcsname\relax
  \def\url#1{{\tt #1}}\fi
\expandafter\ifx\csname urlprefix\endcsname\relax\def\urlprefix{URL }\fi
\providecommand{\eprint}[2][]{\url{#2}}

\bibitem{Einstein:1916b}
Einstein A 1916 {\em Preuss. Akad. Wiss. Berlin\/} {\bf 1} 688

\bibitem{Einstein:1918}
Einstein A 1918 {\em Preuss. Akad. Wiss. Berlin\/} {\bf 1} 154

\bibitem{Einstein:1916a}
Einstein A 1916 {\em Annalen der Physik\/} {\bf 49} 769

\bibitem{PhysRevLett.116.061102}
Abbott B~P {\em et~al.\/} (LIGO Scientific Collaboration and Virgo
  Collaboration) 2016 {\em Phys. Rev. Lett.\/} {\bf 116}(6) 061102

\bibitem{0264-9381-32-7-074001}
Aasi J {\em et~al.\/} (LIGO Scientific Collaboration and Virgo Collaboration)
  2015 {\em Classical and Quantum Gravity\/} {\bf 32} 074001

\bibitem{0264-9381-27-8-084006}
Harry G~M {\em et~al.\/} (LIGO Scientific Collaboration) 2010 {\em Classical
  and Quantum Gravity\/} {\bf 27} 084006

\bibitem{1982ApJ...253..908T}
{Taylor} J~H and {Weisberg} J~M 1982 {\em Astrophys.~J.\/} {\bf 253} 908--920

\bibitem{PhysRevLett.116.241103}
Abbott B~P {\em et~al.\/} (LIGO Scientific Collaboration and Virgo
  Collaboration) 2016 {\em Phys. Rev. Lett.\/} {\bf 116}(24) 241103

\bibitem{PhysRevX.6.041015}
Abbott B~P {\em et~al.\/} (LIGO Scientific Collaboration and Virgo
  Collaboration) 2016 {\em Phys. Rev. X\/} {\bf 6}(4) 041015

\bibitem{PhysRevLett.118.221101}
Abbott B~P {\em et~al.\/} (LIGO Scientific and Virgo Collaboration) 2017 {\em
  Phys. Rev. Lett.\/} {\bf 118}(22) 221101

\bibitem{GW170608}
Abbott B~P {\em et~al.\/} 2017 {\em The Astrophysical Journal Letters\/} {\bf
  851} L35

\bibitem{0264-9381-32-2-024001}
Acernese F {\em et~al.\/} 2015 {\em Classical and Quantum Gravity\/} {\bf 32}
  024001

\bibitem{PhysRevLett.119.141101}
Abbott B~P {\em et~al.\/} (LIGO Scientific Collaboration and Virgo
  Collaboration) 2017 {\em Phys. Rev. Lett.\/} {\bf 119}(14) 141101

\bibitem{PhysRevLett.119.161101}
Abbott B~P {\em et~al.\/} (LIGO Scientific Collaboration and Virgo
  Collaboration) 2017 {\em Phys. Rev. Lett.\/} {\bf 119}(16) 161101

\bibitem{2041-8205-848-2-L12}
Abbott B~P {\em et~al.\/} 2017 {\em The Astrophysical Journal Letters\/} {\bf
  848} L12

\bibitem{1989ARA&A..27..629A}
{Arnett} W~D, {Bahcall} J~N, {Kirshner} R~P and {Woosley} S~E 1989 {\em Annual
  Review of Astronomy and Astrophysics\/} {\bf 27} 629

\bibitem{PhysRevLett.116.131102}
Abbott B~P {\em et~al.\/} (LIGO Scientific Collaboration and Virgo
  Collaboration) 2016 {\em Phys. Rev. Lett.\/} {\bf 116}(13) 131102

\bibitem{Abbott:2017xzg}
Abbott B~P {\em et~al.\/} (LIGO Scientific Collaboration and Virgo
  Collaboration) 2018 {\em Phys. Rev. Lett.\/} {\bf 120}(9) 091101

\bibitem{PhysRevD.23.347}
Guth A~H 1981 {\em Phys. Rev. D\/} {\bf 23}(2) 347--356

\bibitem{PhysRevD.48.3513}
Grishchuk L~P 1993 {\em Phys. Rev. D\/} {\bf 48}(8) 3513--3516

\bibitem{maggiore2007gravitational}
Maggiore M 2007 {\em Gravitational Waves: Volume 1: Theory and Experiments\/}
  Gravitational Waves (OUP Oxford) ISBN 9780198570745
  \urlprefix\url{https://books.google.com/books?id=AqVpQgAACAAJ}

\bibitem{creighton2012gravitational}
Creighton J and Anderson W 2012 {\em Gravitational-Wave Physics and Astronomy:
  An Introduction to Theory, Experiment and Data Analysis\/} Wiley series in
  cosmology (Wiley) ISBN 9783527636051
  \urlprefix\url{https://books.google.com/books?id=n75zEK98WDsC}

\bibitem{Saulson:2013dga}
Saulson P~R 2013 {\em C.~R.~Physique\/} {\bf 14} 288--305

\bibitem{Romano2017}
Romano J~D and Cornish N~J 2017 {\em Living Reviews in Relativity\/} {\bf 20} 2
  ISSN 1433-8351

\bibitem{0004-637X-829-1-55}
Weisberg J~M and Huang Y 2016 {\em The Astrophysical Journal\/} {\bf 829} 55

\bibitem{0264-9381-33-21-215004}
Usman S~A, Nitz A~H, Harry I~W, Biwer C~M, Brown D~A, Cabero M, Capano C~D,
  Canton T~D, Dent T, Fairhurst S, Kehl M~S, Keppel D, Krishnan B, Lenon A,
  Lundgren A, Nielsen A~B, Pekowsky L~P, Pfeiffer H~P, Saulson P~R, West M and
  Willis J~L 2016 {\em Classical and Quantum Gravity\/} {\bf 33} 215004

\bibitem{PhysRevD.95.042001}
Messick C, Blackburn K, Brady P, Brockill P, Cannon K, Cariou R, Caudill S,
  Chamberlin S~J, Creighton J~D~E, Everett R, Hanna C, Keppel D, Lang R~N, Li
  T~G~F, Meacher D, Nielsen A, Pankow C, Privitera S, Qi H, Sachdev S,
  Sadeghian L, Singer L, Thomas E~G, Wade L, Wade M, Weinstein A and Wiesner K
  2017 {\em Phys. Rev. D\/} {\bf 95}(4) 042001

\bibitem{PhysRevD.89.024003}
Privitera S, Mohapatra S~R~P, Ajith P, Cannon K, Fotopoulos N, Frei M~A, Hanna
  C, Weinstein A~J and Whelan J~T 2014 {\em Phys. Rev. D\/} {\bf 89}(2) 024003

\bibitem{PhysRevX.6.041014}
Abbott B~P {\em et~al.\/} (LIGO Scientific Collaboration and Virgo
  Collaboration) 2016 {\em Phys. Rev. X\/} {\bf 6}(4) 041014

\bibitem{PhysRevLett.116.231102}
Sesana A 2016 {\em Phys. Rev. Lett.\/} {\bf 116}(23) 231102

\bibitem{eLISA:2014}
Vitale S 2014 {\em General Relativity and Gravitation\/} {\bf 46} 1730 ISSN
  0001-7701

\bibitem{2017arXiv170200786A}
{Amaro-Seoane} P {\em et~al.\/} 2017 {\em ArXiv e-prints\/} (\textit{Preprint}
  \eprint{1702.00786})

\bibitem{0264-9381-30-22-224005}
Cornish N~J and Sesana A 2013 {\em Classical and Quantum Gravity\/} {\bf 30}
  224005

\bibitem{Fryer2011}
Fryer C~L and New K~C~B 2011 {\em Living Reviews in Relativity\/} {\bf 14} 1
  ISSN 1433-8351

\bibitem{PhysRevD.95.042003}
Abbott B~P {\em et~al.\/} (LIGO Scientific Collaboration and Virgo
  Collaboration) 2017 {\em Phys. Rev. D\/} {\bf 95}(4) 042003

\bibitem{PhysRevD.93.042005}
Abbott B~P {\em et~al.\/} (The LIGO Scientific Collaboration and the Virgo
  Collaboration) 2016 {\em Phys. Rev. D\/} {\bf 93}(4) 042005

\bibitem{0264-9381-35-6-065009}
Abbott B~P {\em et~al.\/} 2018 {\em Classical and Quantum Gravity\/} {\bf 35}
  065009

\bibitem{Aasi:2013vna}
Aasi J, Abadie J, Abbott B, Abbott R, Abbott T {\em et~al.\/} 2014 {\em
  Phys.~Rev.~Lett.\/} {\bf 112} 131101 (\textit{Preprint} \eprint{1310.2384})

\bibitem{0305-4470-9-8-029}
Kibble T~W~B 1976 {\em Journal of Physics A: Mathematical and General\/} {\bf
  9} 1387

\bibitem{vilenkin:1994}
Vilenkin A and Shellard E~P~S 1994 {\em Cosmic Strings and Other Topological
  Defects\/} (Cambridge: Cambridge University Press)

\bibitem{Hessels1901}
Hessels J~W~T, Ransom S~M, Stairs I~H, Freire P~C~C, Kaspi V~M and Camilo F
  2006 {\em Science\/} {\bf 311} 1901--1904 ISSN 0036-8075 (\textit{Preprint}
  \eprint{http://science.sciencemag.org/content/311/5769/1901.full.pdf})

\bibitem{PhysRevD.94.102002}
Abbott B~P and other (LIGO Scientific Collaboration and Virgo Collaboration)
  2016 {\em Phys. Rev. D\/} {\bf 94}(10) 102002

\bibitem{PhysRevD.85.042003}
Wette K 2012 {\em Phys. Rev. D\/} {\bf 85}(4) 042003

\bibitem{Planck2015}
{Planck Collaboration}, {Ade, P A R} {\em et~al.\/} 2016 {\em Astronomy \&
  Astrophysics\/} {\bf 594} A13

\bibitem{PhysRevX.6.011035}
Lasky P~D, Mingarelli C~M~F, Smith T~L, Giblin J~T, Thrane E, Reardon D~J,
  Caldwell R, Bailes M, Bhat N~D~R, Burke-Spolaor S, Dai S, Dempsey J, Hobbs G,
  Kerr M, Levin Y, Manchester R~N, Os\l{}owski S, Ravi V, Rosado P~A, Shannon
  R~M, Spiewak R, van Straten W, Toomey L, Wang J, Wen L, You X and Zhu X 2016
  {\em Phys. Rev. X\/} {\bf 6}(1) 011035

\bibitem{2041-8205-848-2-L13}
Abbott B~P {\em et~al.\/} 2017 {\em The Astrophysical Journal Letters\/} {\bf
  848} L13

\bibitem{2041-8205-848-2-L14}
Goldstein A, Veres P, Burns E, Briggs M~S, Hamburg R, Kocevski D, Wilson-Hodge
  C~A, Preece R~D, Poolakkil S, Roberts O~J, Hui C~M, Connaughton V, Racusin J,
  von Kienlin A, Canton T~D, Christensen N, Littenberg T, Siellez K, Blackburn
  L, Broida J, Bissaldi E, Cleveland W~H, Gibby M~H, Giles M~M, Kippen R~M,
  McBreen S, McEnery J, Meegan C~A, Paciesas W~S and Stanbro M 2017 {\em The
  Astrophysical Journal Letters\/} {\bf 848} L14

\bibitem{2041-8205-848-2-L15}
Savchenko V, Ferrigno C, Kuulkers E, Bazzano A, Bozzo E, Brandt S, Chenevez J,
  Courvoisier T~J~L, Diehl R, Domingo A, Hanlon L, Jourdain E, von Kienlin A,
  Laurent P, Lebrun F, Lutovinov A, Martin-Carrillo A, Mereghetti S, Natalucci
  L, Rodi J, Roques J~P, Sunyaev R and Ubertini P 2017 {\em The Astrophysical
  Journal Letters\/} {\bf 848} L15

\bibitem{NLCthesis}
Christensen N 1990 {\em On measuring the stochastic gravitational radiation
  background with laser interferometric antennas\/} Ph.D. thesis Massachusetts
  Institute of Technology. \urlprefix\url{http://hdl.handle.net/1721.1/13619}

\bibitem{Christensen:1992wi}
Christensen N 1992 {\em Phys.~Rev.\/} {\bf D46} 5250--5266

\bibitem{PhysRevLett.118.121101}
Abbott B~P {\em et~al.\/} (LIGO Scientific Collaboration and Virgo
  Collaboration) 2017 {\em Phys. Rev. Lett.\/} {\bf 118}(12) 121101

\bibitem{0067-0049-208-2-20}
Bennett C~L {\em et~al.\/} 2013 {\em The Astrophysical Journal Supplement
  Series\/} {\bf 208} 20

\bibitem{0067-0049-208-2-19}
Hinshaw G {\em et~al.\/} 2013 {\em The Astrophysical Journal Supplement
  Series\/} {\bf 208} 19

\bibitem{PhysRevLett.118.121102}
Abbott B~P {\em et~al.\/} (LIGO Scientific Collaboration and Virgo
  Collaboration) 2017 {\em Phys. Rev. Lett.\/} {\bf 118}(12) 121102

\bibitem{PhysRevD.91.022003}
Aasi J {\em et~al.\/} (LIGO Scientific Collaboration and Virgo Collaboration)
  2015 {\em Phys. Rev. D\/} {\bf 91}(2) 022003

\bibitem{Thrane:2013npa}
Thrane E, Christensen N and Schofield R 2013 {\em Phys.~Rev.\/} {\bf D87}
  123009 (\textit{Preprint} \eprint{1303.2613})

\bibitem{0264-9381-34-7-074002}
Kowalska-Leszczynska I, Bizouard M~A, Bulik T, Christensen N, Coughlin M,
  Gołkowski M, Kubisz J, Kulak A, Mlynarczyk J, Robinet F and Rohde M 2017
  {\em Classical and Quantum Gravity\/} {\bf 34} 074002

\bibitem{0264-9381-33-22-224003}
Coughlin M~W, Christensen N~L, Rosa R~D, Fiori I, Golkowski M, Guidry M, Harms
  J, Kubisz J, Kulak A, Mlynarczyk J, Paoletti F and Thrane E 2016 {\em
  Classical and Quantum Gravity\/} {\bf 33} 224003

\bibitem{1959ZNatA..14..767B}
{Bendat} J~S 1959 {\em Zeitschrift Naturforschung Teil A\/} {\bf 14} 767

\bibitem{PhysRevLett.102.231301}
Takahashi T and Soda J 2009 {\em Phys. Rev. Lett.\/} {\bf 102}(23) 231301

\bibitem{refId0}
{Planck Collaboration}, {Ade, P A R} {\em et~al.\/} 2014 {\em Astronomy \&
  Astrophysics\/} {\bf 571} A16

\bibitem{PhysRevD.50.1157}
Bar-Kana R 1994 {\em Phys. Rev. D\/} {\bf 50}(2) 1157--1160

\bibitem{Starobinski:1979aa}
Starobinski A 1979 {\em JETP~Lett.\/} {\bf 30} 682--686

\bibitem{PhysRevLett.99.221301}
Easther R, Giblin J~T and Lim E~A 2007 {\em Phys. Rev. Lett.\/} {\bf 99}(22)
  221301

\bibitem{PhysRevD.85.023525}
Barnaby N, Pajer E and Peloso M 2012 {\em Phys. Rev. D\/} {\bf 85}(2) 023525

\bibitem{PhysRevD.85.023534}
Cook J~L and Sorbo L 2012 {\em Phys. Rev. D\/} {\bf 85}(2) 023534

\bibitem{1475-7516-2015-01-037}
Lopez A and Freese K 2015 {\em Journal of Cosmology and Astroparticle
  Physics\/} {\bf 2015} 037

\bibitem{PhysRevD.55.R435}
Turner M~S 1997 {\em Phys. Rev. D\/} {\bf 55}(2) R435--R439

\bibitem{2006JCAP...04..010E}
{Easther} R and {Lim} E~A 2006 {\em JCAP\/} {\bf 4} 010

\bibitem{GASPERINI1993317}
Gasperini M and Veneziano G 1993 {\em Astroparticle Physics\/} {\bf 1} 317 --
  339 ISSN 0927-6505

\bibitem{doi:10.1142/S0217732393003433}
Gasperini M and Veneziano G 1993 {\em Modern Physics Letters A\/} {\bf 08}
  3701--3713 (\textit{Preprint}
  \eprint{http://www.worldscientific.com/doi/pdf/10.1142/S0217732393003433})

\bibitem{PhysRevD.73.063008}
Mandic V and Buonanno A 2006 {\em Phys. Rev. D\/} {\bf 73}(6) 063008

\bibitem{1475-7516-2016-12-010}
Gasperini M 2016 {\em Journal of Cosmology and Astroparticle Physics\/} {\bf
  2016} 010

\bibitem{Sarangi2002185}
Sarangi S and Tye S~H 2002 {\em Physics Letters B\/} {\bf 536} 185 -- 192 ISSN
  0370-2693

\bibitem{PhysRevLett.98.111101}
Siemens X, Mandic V and Creighton J 2007 {\em Phys. Rev. Lett.\/} {\bf 98}(11)
  111101

\bibitem{Damour:2004kw}
Damour T and Vilenkin A 2005 {\em Phys.~Rev.\/} {\bf D71} 063510
  (\textit{Preprint} \eprint{hep-th/0410222})

\bibitem{PhysRevD.86.023503}
Kuroyanagi S, Miyamoto K, Sekiguchi T, Takahashi K and Silk J 2012 {\em Phys.
  Rev. D\/} {\bf 86}(2) 023503

\bibitem{PhysRevLett.69.2026}
Kosowsky A, Turner M~S and Watkins R 1992 {\em Phys. Rev. Lett.\/} {\bf 69}(14)
  2026--2029

\bibitem{PhysRevD.49.2837}
Kamionkowski M, Kosowsky A and Turner M~S 1994 {\em Phys. Rev. D\/} {\bf 49}(6)
  2837--2851

\bibitem{PhysRevD.90.107502}
Giblin J~T and Thrane E 2014 {\em Phys. Rev. D\/} {\bf 90}(10) 107502

\bibitem{doi:10.1093/mnras/stt207}
Zhu X~J, Howell E~J, Blair D~G and Zhu Z~H 2013 {\em Monthly Notices of the
  Royal Astronomical Society\/} {\bf 431} 882 (\textit{Preprint}
  \eprint{/oup/backfile/Content_public/Journal/mnras/431/1/10.1093/mnras/stt207/3/stt207.pdf})

\bibitem{PhysRevD.85.104024}
Wu C, Mandic V and Regimbau T 2012 {\em Phys. Rev. D\/} {\bf 85}(10) 104024

\bibitem{PhysRevD.84.084004}
Rosado P~A 2011 {\em Phys. Rev. D\/} {\bf 84}(8) 084004

\bibitem{PhysRevD.72.084001}
Buonanno A, Sigl G, Raffelt G~G, Janka H~T and M\"uller E 2005 {\em Phys. Rev.
  D\/} {\bf 72}(8) 084001

\bibitem{Zhu:2010af}
Zhu X~J, Howell E and Blair D 2010 {\em Mon. Not. Roy. Astron. Soc.\/} {\bf
  409} L132--L136 (\textit{Preprint} \eprint{1008.0472})

\bibitem{PhysRevD.86.104007}
Rosado P~A 2012 {\em Phys. Rev. D\/} {\bf 86}(10) 104007

\bibitem{PhysRevD.87.063004}
Lasky P~D, Bennett M~F and Melatos A 2013 {\em Phys. Rev. D\/} {\bf 87}(6)
  063004

\bibitem{PhysRevD.95.083003}
Cheng Q, Zhang S~N and Zheng X~P 2017 {\em Phys. Rev. D\/} {\bf 95}(8) 083003

\bibitem{PhysRevD.59.102001}
Allen B and Romano J~D 1999 {\em Phys. Rev. D\/} {\bf 59}(10) 102001

\bibitem{PhysRevD.69.122004}
Abbott B {\em et~al.\/} (LIGO Scientific Collaboration) 2004 {\em Phys. Rev.
  D\/} {\bf 69}(12) 122004

\bibitem{PhysRevLett.95.221101}
Abbott B~o (LIGO Scientific Collaboration) 2005 {\em Phys. Rev. Lett.\/} {\bf
  95}(22) 221101

\bibitem{0004-637X-659-2-918}
Abbott B {\em et~al.\/} 2007 {\em The Astrophysical Journal\/} {\bf 659} 918

\bibitem{Abadie:2011fx}
Abadie J {\em et~al.\/} (The LIGO Scientific Collaboration and the Virgo
  Collaboration) 2012 {\em Phys.~Rev.\/} {\bf {D85}} 122001 (\textit{Preprint}
  \eprint{1112.5004})

\bibitem{Abbott:2011rs}
Abbott B {\em et~al.\/} 2009 {\em Nature\/} {\bf 460} 990--994

\bibitem{PhysRevLett.113.231101}
Aasi J {\em et~al.\/} ((LIGO and Virgo Collaboration)) 2014 {\em Phys. Rev.
  Lett.\/} {\bf 113}(23) 231101

\bibitem{Pagano2016823}
Pagano L, Salvati L and Melchiorri A 2016 {\em Physics Letters B\/} {\bf 760}
  823 -- 825 ISSN 0370-2693

\bibitem{PhysRevD.90.042005}
Coughlin M and Harms J 2014 {\em Phys. Rev. D\/} {\bf 90}(4) 042005

\bibitem{PhysRevD.93.021101}
Chaibi W, Geiger R, Canuel B, Bertoldi A, Landragin A and Bouyer P 2016 {\em
  Phys. Rev. D\/} {\bf 93}(2) 021101

\bibitem{1742-6596-122-1-012006}
Kawamura S {\em et~al.\/} 2008 {\em Journal of Physics: Conference Series\/}
  {\bf 122} 012006

\bibitem{1742-6596-840-1-012010}
Sato S {\em et~al.\/} 2017 {\em Journal of Physics: Conference Series\/} {\bf
  840} 012010

\bibitem{2011RAA....11..369R}
{Regimbau} T 2011 {\em Research in Astronomy and Astrophysics\/} {\bf 11}
  369--390

\bibitem{1538-4357-563-2-L95}
Knox L, Christensen N and Skordis C 2001 {\em The Astrophysical Journal
  Letters\/} {\bf 563} L95

\bibitem{1965ApJ...142..419P}
Penzias A and Wilson R 1965 {\em Astrophys. J.\/} {\bf 142} 419--421

\bibitem{1965ApJ...142..414D}
Dicke R, Peebles P, Roll P and Wilkinson D 1965 {\em Astrophys. J.\/} {\bf 142}
  414--419

\bibitem{1948Natur.162..774A}
Alpher R and Herman R 1948 {\em Nature\/} {\bf 162} 774--775

\bibitem{1994ApJ...420..439M}
Mather J {\em et~al.\/} 1994 {\em Astrophys. J.\/} {\bf 420} 439--444

\bibitem{0004-637X-707-2-916}
Fixsen D~J 2009 {\em Astrophys. J.\/} {\bf 707} 916

\bibitem{1992ApJ...396L...1S}
Smoot G {\em et~al.\/} 1992 {\em Astrophys. J. Lett.\/} {\bf 396} L1--L5

\bibitem{0264-9381-18-14-306}
Christensen N, Meyer R, Knox L and Luey B 2001 {\em Classical and Quantum
  Gravity\/} {\bf 18} 2677

\bibitem{Linde82}
Linde A 1982 {\em Phys. Lett. B\/} {\bf 108} 389--393

\bibitem{Binetruy_2015}
Binetruy P 2015 {\em ArXiv e-prints\/} (\textit{Preprint} \eprint{1504.07050})

\bibitem{starobinsky1979jetp}
Starobinsky A 1979 {\em Pisma Zh. Eksp. Teor. Fiz\/} {\bf 30} 719

\bibitem{starobinksii}
Starobinskii A~A 1979 {\em JETP Lett.\/} {\bf 30} 682

\bibitem{kolbturner}
{E W Kolb \& M S Turner} 1994 {\em The Early Universe\/} (Westview Press)

\bibitem{bar-kana}
Bar-Kana R 1994 {\em Phys. Rev. D\/} {\bf 50} 1157

\bibitem{PhysRevD.68.103514}
Jeannerot R, Rocher J and Sakellariadou M 2003 {\em Phys. Rev. D\/} {\bf
  68}(10) 103514

\bibitem{SAKELLARIADOU200968}
Sakellariadou M 2009 {\em Nuclear Physics B - Proceedings Supplements\/} {\bf
  192} 68 -- 90 ISSN 0920-5632

\bibitem{Sakellariadou2881}
Sakellariadou M 2008 {\em Philosophical Transactions of the Royal Society of
  London A: Mathematical, Physical and Engineering Sciences\/} {\bf 366}
  2881--2894 ISSN 1364-503X (\textit{Preprint}
  \eprint{http://rsta.royalsocietypublishing.org/content/366/1877/2881.full.pdf})

\bibitem{SHELLARD1987624}
Shellard E 1987 {\em Nuclear Physics B\/} {\bf 283} 624 -- 656 ISSN 0550-3213

\bibitem{PhysRevD.41.1751}
Laguna P and Matzner R~A 1990 {\em Phys. Rev. D\/} {\bf 41}(6) 1751--1763

\bibitem{1475-7516-2005-04-003}
Sakellariadou M 2005 {\em Journal of Cosmology and Astroparticle Physics\/}
  {\bf 2005} 003

\bibitem{Jackson:2004zg}
Jackson M~G, Jones N~T and Polchinski J 2005 {\em JHEP\/} {\bf 0510} 013
  (\textit{Preprint} \eprint{hep-th/0405229})

\bibitem{PhysRevD.50.2496}
Allen B and Casper P 1994 {\em Phys. Rev. D\/} {\bf 50}(4) 2496--2518

\bibitem{PhysRevD.43.3173}
Allen B and Caldwell R~R 1991 {\em Phys. Rev. D\/} {\bf 43}(10) 3173--3187

\bibitem{PhysRevD.95.023519}
Wachter J~M and Olum K~D 2017 {\em Phys. Rev. D\/} {\bf 95}(2) 023519

\bibitem{Damour:2001bk}
Damour T and Vilenkin A 2001 {\em Phys.~Rev.\/} {\bf D64} 064008
  (\textit{Preprint} \eprint{gr-qc/0104026})

\bibitem{Damour:2000wa}
Damour T and Vilenkin A 2000 {\em Phys.~Rev.~Lett.\/} {\bf 85} 3761--3764
  (\textit{Preprint} \eprint{gr-qc/0004075})

\bibitem{0264-9381-32-4-045003}
Henrot-Versill\'e S, Robinet F, Leroy N, Plaszczynski S, Arnaud N, Bizouard
  M~A, Cavalier F, Christensen N, Couchot F, Franco S, Hello P, Huet D,
  Kasprzack M, Perdereau O, Spinelli M and Tristram M 2015 {\em Classical and
  Quantum Gravity\/} {\bf 32} 045003

\bibitem{0004-637X-821-1-13}
Arzoumanian Z {\em et~al.\/} 2016 {\em The Astrophysical Journal\/} {\bf 821}
  13

\bibitem{Ade:2013xla}
Ade P {\em et~al.\/} (Planck) 2014 {\em Astron. Astrophys.\/} {\bf 571} A25
  (\textit{Preprint} \eprint{1303.5085})

\bibitem{Lizarraga:2016onn}
Lizarraga J, Urrestilla J, Daverio D, Hindmarsh M and Kunz M 2016 {\em JCAP\/}
  {\bf 1610} 042 (\textit{Preprint} \eprint{1609.03386})

\bibitem{Lazanu:2014eya}
Lazanu A and Shellard P 2015 {\em JCAP\/} {\bf 1502} 024 (\textit{Preprint}
  \eprint{1410.5046})

\bibitem{PhysRevD.42.354}
Sakellariadou M 1990 {\em Phys. Rev. D\/} {\bf 42}(2) 354--360

\bibitem{Hogan87}
Hogan C 1987 {\em Nature\/} {\bf 326} 853--855

\bibitem{refId0A15}
Ade P~A~R {\em et~al.\/} 2014 {\em A\&A\/} {\bf 571} A15

\bibitem{2041-8205-749-1-L9}
Reichardt C~L, de~Putter R, Zahn O and Hou Z 2012 {\em The Astrophysical
  Journal Letters\/} {\bf 749} L9

\bibitem{1475-7516-2014-04-014}
Das S, Louis T, Nolta M~R, Addison G~E, Battistelli E~S, Bond J~R, Calabrese E,
  Crichton D, Devlin M~J, Dicker S, Dunkley J, Dünner R, Fowler J~W, Gralla M,
  Hajian A, Halpern M, Hasselfield M, Hilton M, Hincks A~D, Hlozek R,
  Huffenberger K~M, Hughes J~P, Irwin K~D, Kosowsky A, Lupton R~H, Marriage
  T~A, Marsden D, Menanteau F, Moodley K, Niemack M~D, Page L~A, Partridge B,
  Reese E~D, Schmitt B~L, Sehgal N, Sherwin B~D, Sievers J~L, Spergel D~N,
  Staggs S~T, Swetz D~S, Switzer E~R, Thornton R, Trac H and Wollack E 2014
  {\em Journal of Cosmology and Astroparticle Physics\/} {\bf 2014} 014

\bibitem{doi:10.1111/j.1365-2966.2011.19250.x}
Beutler F, Blake C, Colless M, Jones D~H, Staveley-Smith L, Campbell L, Parker
  Q, Saunders W and Watson F 2011 {\em Monthly Notices of the Royal
  Astronomical Society\/} {\bf 416} 3017 (\textit{Preprint}
  \eprint{/oup/backfile/content_public/journal/mnras/416/4/10.1111/j.1365-2966.2011.19250.x/2/mnras0416-3017.pdf})

\bibitem{doi:10.1111/j.1365-2966.2012.21888.x}
Padmanabhan N, Xu X, Eisenstein D~J, Scalzo R, Cuesta A~J, Mehta K~T and Kazin
  E 2012 {\em Monthly Notices of the Royal Astronomical Society\/} {\bf 427}
  2132 (\textit{Preprint}
  \eprint{/oup/backfile/content_public/journal/mnras/427/3/10.1111/j.1365-2966.2012.21888.x/2/427-3-2132.pdf})

\bibitem{doi:10.1111/j.1365-2966.2012.22066.x}
Anderson L {\em et~al.\/} 2012 {\em Monthly Notices of the Royal Astronomical
  Society\/} {\bf 427} 3435 (\textit{Preprint}
  \eprint{/oup/backfile/content_public/journal/mnras/427/4/10.1111/j.1365-2966.2012.22066.x/2/427-4-3435.pdf})

\bibitem{refId0A17}
Ade P~A~R {\em et~al.\/} 2014 {\em A\&A\/} {\bf 571} A17

\bibitem{Abbott:2017mem}
Abbott B {\em et~al.\/} (Virgo, LIGO Scientific) 2018 {\em Phys. Rev.\/} {\bf
  D97} 102002 (\textit{Preprint} \eprint{1712.01168})

\bibitem{vilenkin2000cosmic}
Vilenkin A and Shellard E 2000 {\em Cosmic Strings and Other Topological
  Defects\/} Cambridge Monographs on Mathematical Physics (Cambridge University
  Press) ISBN 9780521654760
  \urlprefix\url{https://books.google.fr/books?id=eW4bB\_LAthEC}

\bibitem{PhysRevD.73.105001}
Siemens X, Creighton J, Maor I, Majumder S~R, Cannon K and Read J 2006 {\em
  Phys. Rev. D\/} {\bf 73}(10) 105001

\bibitem{PhysRevD.89.023512}
Blanco-Pillado J~J, Olum K~D and Shlaer B 2014 {\em Phys. Rev. D\/} {\bf 89}(2)
  023512

\bibitem{1475-7516-2007-02-023}
Ringeval C, Sakellariadou M and Bouchet F~R 2007 {\em Journal of Cosmology and
  Astroparticle Physics\/} {\bf 2007} 023

\bibitem{1475-7516-2010-10-003}
Lorenz L, Ringeval C and Sakellariadou M 2010 {\em Journal of Cosmology and
  Astroparticle Physics\/} {\bf 2010} 003

\bibitem{Papon:1339609}
Papon P, Leblond J and Meijer P~H 2006 {\em {The Physics of Phase Transitions:
  Concepts and Applications}\/} (Berlin, Heidelberg: Springer)
  \urlprefix\url{https://cds.cern.ch/record/1339609}

\bibitem{PhysRevLett.119.141301}
Iso S, Serpico P~D and Shimada K 2017 {\em Phys. Rev. Lett.\/} {\bf 119}(14)
  141301

\bibitem{PhysRevD.93.025003}
D'Onofrio M and Rummukainen K 2016 {\em Phys. Rev. D\/} {\bf 93}(2) 025003

\bibitem{1367-2630-14-12-125003}
Morrissey D~E and Ramsey-Musolf M~J 2012 {\em New Journal of Physics\/} {\bf
  14} 125003

\bibitem{0038-5670-34-5-A08}
Sakharov A~D 1991 {\em Soviet Physics Uspekhi\/} {\bf 34} 392

\bibitem{1705.01783}
Weir D~J 2018 {\em Phil. Trans. Roy. Soc. Lond.\/} {\bf A376} 20170126
  (\textit{Preprint} \eprint{1705.01783})

\bibitem{1475-7516-2016-04-001}
Caprini C, Hindmarsh M, Huber S, Konstandin T, Kozaczuk J, Nardini G, No J~M,
  Petiteau A, Schwaller P, Servant G and Weir D~J 2016 {\em Journal of
  Cosmology and Astroparticle Physics\/} {\bf 2016} 001

\bibitem{PhysRevD.45.4514}
Kosowsky A, Turner M~S and Watkins R 1992 {\em Phys. Rev. D\/} {\bf 45}(12)
  4514--4535

\bibitem{PhysRevD.77.124015}
Caprini C, Durrer R and Servant G 2008 {\em Phys. Rev. D\/} {\bf 77}(12) 124015

\bibitem{1475-7516-2008-09-022}
Huber S~J and Konstandin T 2008 {\em Journal of Cosmology and Astroparticle
  Physics\/} {\bf 2008} 022

\bibitem{PhysRevLett.112.041301}
Hindmarsh M, Huber S~J, Rummukainen K and Weir D~J 2014 {\em Phys. Rev.
  Lett.\/} {\bf 112}(4) 041301

\bibitem{PhysRevD.92.123009}
Hindmarsh M, Huber S~J, Rummukainen K and Weir D~J 2015 {\em Phys. Rev. D\/}
  {\bf 92}(12) 123009

\bibitem{1475-7516-2009-12-024}
Caprini C, Durrer R and Servant G 2009 {\em Journal of Cosmology and
  Astroparticle Physics\/} {\bf 2009} 024

\bibitem{PhysRevD.74.063521}
Caprini C and Durrer R 2006 {\em Phys. Rev. D\/} {\bf 74}(6) 063521

\bibitem{PhysRevD.92.043006}
Kisslinger L and Kahniashvili T 2015 {\em Phys. Rev. D\/} {\bf 92}(4) 043006

\bibitem{1475-7516-2016-12-026}
Bartolo N, Caprini C, Domcke V, Figueroa D~G, Garcia-Bellido J, Guzzetti M~C,
  Liguori M, Matarrese S, Peloso M, Petiteau A, Ricciardone A, Sakellariadou M,
  Sorbo L and Tasinato G 2016 {\em Journal of Cosmology and Astroparticle
  Physics\/} {\bf 2016} 026

\bibitem{GASPERINI20031}
Gasperini M and Veneziano G 2003 {\em Physics Reports\/} {\bf 373} 1 -- 212
  ISSN 0370-1573

\bibitem{PhysRevLett.72.3305}
Borde A and Vilenkin A 1994 {\em Phys. Rev. Lett.\/} {\bf 72}(21) 3305--3308

\bibitem{PhysRevLett.90.151301}
Borde A, Guth A~H and Vilenkin A 2003 {\em Phys. Rev. Lett.\/} {\bf 90}(15)
  151301

\bibitem{Buonanno:1996xc}
Buonanno A, Maggiore M and Ungarelli C 1997 {\em Phys.~Rev.\/} {\bf D55}
  3330--3336 (\textit{Preprint} \eprint{gr-qc/9605072})

\bibitem{1475-7516-2017-06-017}
Gasperini M 2017 {\em Journal of Cosmology and Astroparticle Physics\/} {\bf
  2017} 017

\bibitem{Shannon1522}
Shannon R~M, Ravi V, Lentati L~T, Lasky P~D, Hobbs G, Kerr M, Manchester R~N,
  Coles W~A, Levin Y, Bailes M, Bhat N~D~R, Burke-Spolaor S, Dai S, Keith M~J,
  Os{\l}owski S, Reardon D~J, van Straten W, Toomey L, Wang J~B, Wen L, Wyithe
  J~S~B and Zhu X~J 2015 {\em Science\/} {\bf 349} 1522--1525 ISSN 0036-8075
  (\textit{Preprint}
  \eprint{http://science.sciencemag.org/content/349/6255/1522.full.pdf})

\bibitem{2041-8205-833-1-L1}
Abbott B~P {\em et~al.\/} 2016 {\em The Astrophysical Journal Letters\/} {\bf
  833} L1

\bibitem{0067-0049-227-2-14}
Abbott B~P {\em et~al.\/} 2016 {\em The Astrophysical Journal Supplement
  Series\/} {\bf 227} 14

\bibitem{2041-8205-832-2-L21}
Abbott B~P {\em et~al.\/} 2016 {\em The Astrophysical Journal Letters\/} {\bf
  832} L21

\bibitem{Kim2003}
Kim C, Kalogera V and Lorimer D 2003 {\em Astrophys. J.\/} {\bf 584} 985

\bibitem{Nakar:2007}
Nakar E 2007 {\em Phys.~Rep.\/} {\bf 442} 166--236 (\textit{Preprint}
  \eprint{astro-ph/0701748})

\bibitem{2011ApJ...739...86Z}
{Zhu} X~J, {Howell} E, {Regimbau} T, {Blair} D and {Zhu} Z~H 2011 {\em apj\/}
  {\bf 739} 86

\bibitem{2011PhRvD..84l4037M}
{Marassi} S, {Schneider} R, {Corvino} G, {Ferrari} V and {Zwart} S~P 2011 {\em
  prd\/} {\bf 84} 124037

\bibitem{2013PhRvD..87d2002W}
{Wu} C~J, {Mandic} V and {Regimbau} T 2013 {\em prd\/} {\bf 87} 042002

\bibitem{2015A&A...574A..58K}
{Kowalska-Leszczynska} I, {Regimbau} T, {Bulik} T, {Dominik} M and {Belczynski}
  K 2015 {\em Astronomy and Astrophysics\/} {\bf 574} A58

\bibitem{2041-8205-818-2-L22}
Abbott B~P {\em et~al.\/} 2016 {\em The Astrophysical Journal Letters\/} {\bf
  818} L22

\bibitem{Belczynski:2016obo}
Belczynski K, Holz D~E, Bulik T and O'Shaughnessy R 2016 {\em Nature\/} {\bf
  534} 512 (\textit{Preprint} \eprint{1602.04531})

\bibitem{lrr-2006-6}
Postnov K and Yungelson L 2007 {\em Living~Rev.~Relativity\/} {\bf 9} 6

\bibitem{Rodriguez:2016kxx}
Rodriguez C~L, Chatterjee S and Rasio F~A 2016 {\em Phys. Rev.\/} {\bf D93}
  084029 (\textit{Preprint} \eprint{1602.02444})

\bibitem{2013LRR....16....4B}
{Benacquista} M~J and {Downing} J~M~B 2013 {\em Living Reviews in Relativity\/}
  {\bf 16} 4 (\textit{Preprint} \eprint{1110.4423})

\bibitem{1967SvA....10..602Z}
{Zel'dovich} Y~B and {Novikov} I~D 1967 {\em Soviet Astronomy\/} {\bf 10} 602

\bibitem{1971MNRAS.152...75H}
{Hawking} S 1971 {\em Mon. Not. R. Ast. Soc.\/} {\bf 152} 75

\bibitem{1974MNRAS.168..399C}
{Carr} B~J and {Hawking} S~W 1974 {\em Mon. Not. R. Ast. Soc.\/} {\bf 168}
  399--416

\bibitem{1974A&A....37..225M}
{Meszaros} P 1974 {\em Astron. Astrophys.\/} {\bf 37} 225--228

\bibitem{1975Natur.253..251C}
{Chapline} G~F 1975 {\em Nature\/} {\bf 253} 251

\bibitem{PhysRevD.94.083504}
Carr B, K\"uhnel F and Sandstad M 2016 {\em Phys. Rev. D\/} {\bf 94}(8) 083504

\bibitem{PhysRevLett.116.201301}
Bird S, Cholis I, Mu\~noz J~B, Ali-Ha\"{\i}moud Y, Kamionkowski M, Kovetz E~D,
  Raccanelli A and Riess A~G 2016 {\em Phys. Rev. Lett.\/} {\bf 116}(20) 201301

\bibitem{PhysRevLett.117.201102}
Mandic V, Bird S and Cholis I 2016 {\em Phys. Rev. Lett.\/} {\bf 117}(20)
  201102

\bibitem{PhysRevD.95.043511}
Nakama T, Silk J and Kamionkowski M 2017 {\em Phys. Rev. D\/} {\bf 95}(4)
  043511

\bibitem{PhysRevD.92.063002}
Meacher D, Coughlin M, Morris S, Regimbau T, Christensen N, Kandhasamy S,
  Mandic V, Romano J~D and Thrane E 2015 {\em Phys. Rev. D\/} {\bf 92}(6)
  063002

\bibitem{0264-9381-27-19-194002}
Punturo M {\em et~al.\/} 2010 {\em Classical and Quantum Gravity\/} {\bf 27}
  194002

\bibitem{0264-9381-34-4-044001}
others B~P~A 2017 {\em Classical and Quantum Gravity\/} {\bf 34} 044001

\bibitem{PhysRevLett.118.151105}
Regimbau T, Evans M, Christensen N, Katsavounidis E, Sathyaprakash B and Vitale
  S 2017 {\em Phys. Rev. Lett.\/} {\bf 118}(15) 151105

\bibitem{Caprini:2018mtu}
Caprini C and Figueroa D~G 2018 {\em Class. Quant. Grav.\/} {\bf 35} 163001
  (\textit{Preprint} \eprint{1801.04268})

\bibitem{PhysRevD.80.104009}
Cutler C and Holz D~E 2009 {\em Phys. Rev. D\/} {\bf 80}(10) 104009

\bibitem{PhysRevD.72.022001}
Umst\"atter R, Christensen N, Hendry M, Meyer R, Simha V, Veitch J, Vigeland S
  and Woan G 2005 {\em Phys. Rev. D\/} {\bf 72}(2) 022001

\bibitem{0264-9381-22-18-S04}
Umst\"atter R, Christensen N, Hendry M, Meyer R, Simha V, Veitch J, Vigeland S
  and Woan G 2005 {\em Classical and Quantum Gravity\/} {\bf 22} S901

\bibitem{PhysRevD.89.022001}
Adams M~R and Cornish N~J 2014 {\em Phys. Rev. D\/} {\bf 89}(2) 022001

\bibitem{PhysRevD.73.042001}
Cutler C and Harms J 2006 {\em Phys. Rev. D\/} {\bf 73}(4) 042001

\bibitem{PhysRevD.77.123010}
Harms J, Mahrdt C, Otto M and Prie\ss{} M 2008 {\em Phys. Rev. D\/} {\bf
  77}(12) 123010

\bibitem{Abbott:2017xzu}
Abbott B~P {\em et~al.\/} (LIGO Scientific, VINROUGE, Las Cumbres Observatory,
  DES, DLT40, Virgo, 1M2H, Dark Energy Camera GW-E, MASTER) 2017 {\em Nature\/}
  {\bf 551} 85--88 (\textit{Preprint} \eprint{1710.05835})

\bibitem{Aasi:2013wya}
Aasi J {\em et~al.\/} 2018 {\em Living Reviews in Relativity\/} {\bf 21} 3

\bibitem{Nelemans:2001hp}
Nelemans G, Yungelson L~R and Portegies~Zwart S~F 2001 {\em Astron.
  Astrophys.\/} {\bf 375} 890--898 (\textit{Preprint}
  \eprint{astro-ph/0105221})

\bibitem{Farmer:2003pa}
Farmer A~J and Phinney E~S 2003 {\em Mon. Not. Roy. Astron. Soc.\/} {\bf 346}
  1197 (\textit{Preprint} \eprint{astro-ph/0304393})

\bibitem{1987ApJ...323..129E}
{Evans} C~R, {Iben} Jr I and {Smarr} L 1987 {\em Astrophys. J.\/} {\bf 323}
  129--139

\bibitem{Schutz:1997bw}
Schutz B~F 1997 {\em ESA Spec. Publ.\/} {\bf 420} 229 (\textit{Preprint}
  \eprint{gr-qc/9710079})

\bibitem{0264-9381-14-6-008}
Bender P~L and Hils D 1997 {\em Classical and Quantum Gravity\/} {\bf 14} 1439

\bibitem{doi:10.1046/j.1365-8711.1998.01282.x}
Walter D and James B 1998 {\em Monthly Notices of the Royal Astronomical
  Society\/} {\bf 294} 429--438

\bibitem{doi:10.1111/j.1365-2966.2011.18564.x}
McMillan P~J 2011 {\em Monthly Notices of the Royal Astronomical Society\/}
  {\bf 414} 2446--2457 (\textit{Preprint}
  \eprint{/oup/backfile/content_public/journal/mnras/414/3/10.1111/j.1365-2966.2011.18564.x/2/mnras0414-2446.pdf})

\bibitem{0264-9381-19-7-306}
Cornish N~J 2002 {\em Classical and Quantum Gravity\/} {\bf 19} 1279

\bibitem{doi:10.1111/j.1365-2966.2010.17624.x}
Remazeilles M, Delabrouille J and Cardoso J~F 2011 {\em Monthly Notices of the
  Royal Astronomical Society\/} {\bf 410} 2481--2487 (\textit{Preprint}
  \eprint{/oup/backfile/content_public/journal/mnras/410/4/10.1111_j.1365-2966.2010.17624.x/1/mnras0410-2481.pdf})

\bibitem{Planck_xii}
{Planck Collaboration}, {Ade, P A R} {\em et~al.\/} 2014 {\em A\&A\/} {\bf 571}
  A12

\bibitem{0264-9381-18-20-307}
Cornish N~J 2001 {\em Classical and Quantum Gravity\/} {\bf 18} 4277

\bibitem{gaia1}
{Gaia Collaboration}, {Brown, A G A} {\em et~al.\/} 2016 {\em A\&A\/} {\bf 595}
  A2

\bibitem{2018arXiv180409365G}
{Gaia Collaboration}, {Brown, A G A} {\em et~al.\/} 2018 {\em A\&A\/} {\bf 616}
  A1

\bibitem{doi:10.1093/mnrasl/sly110}
Kilic M, Hambly N~C, Bergeron P, Genest-Beaulieu C and Rowell N 2018 {\em
  Monthly Notices of the Royal Astronomical Society: Letters\/} {\bf 479}
  L113--L117

\bibitem{0264-9381-24-19-S20}
Crowder J and Cornish N~J 2007 {\em Classical and Quantum Gravity\/} {\bf 24}
  S575

\bibitem{PhysRevD.80.064032}
Stroeer A and Veitch J 2009 {\em Phys. Rev. D\/} {\bf 80}(6) 064032

\bibitem{0264-9381-24-19-S17}
Stroeer A, Veitch J, R\"over C, Bloomer E, Clark J, Christensen N, Hendry M,
  Messenger C, Meyer R, Pitkin M, Toher J, Umstätter R, Vecchio A and Woan G
  2007 {\em Classical and Quantum Gravity\/} {\bf 24} S541

\bibitem{PhysRevD.82.022002}
Adams M~R and Cornish N~J 2010 {\em Phys. Rev. D\/} {\bf 82}(2) 022002

\bibitem{0264-9381-34-24-244002}
Robson T and Cornish N 2017 {\em Classical and Quantum Gravity\/} {\bf 34}
  244002

\bibitem{0264-9381-25-11-114037}
Babak S {\em et~al.\/} 2008 {\em Classical and Quantum Gravity\/} {\bf 25}
  114037

\bibitem{1742-6596-840-1-012026}
Babak S 2017 {\em Journal of Physics: Conference Series\/} {\bf 840} 012026

\bibitem{doi:10.1093/mnras/sty1545}
Kupfer T, Korol V, Shah S, Nelemans G, Marsh T~R, Ramsay G, Groot P~J, Steeghs
  D~T~H and Rossi E~M 2018 {\em Monthly Notices of the Royal Astronomical
  Society\/} {\bf 480} 302--309

\bibitem{Gair2013}
Gair J~R, Vallisneri M, Larson S~L and Baker J~G 2013 {\em Living Reviews in
  Relativity\/} {\bf 16} 7 ISSN 1433-8351

\bibitem{Nelemans:2013iq}
Nelemans G and van Haaften L 2013 {\em ASP Conf. Ser.\/} {\bf 470} 153
  (\textit{Preprint} \eprint{1302.0878})

\bibitem{1742-6596-610-1-012003}
Shah S, Larson S~L and Brown W 2015 {\em Journal of Physics: Conference
  Series\/} {\bf 610} 012003

\bibitem{PhysRevD.90.044001}
Abdikamalov E, Gossan S, DeMaio A~M and Ott C~D 2014 {\em Phys. Rev. D\/} {\bf
  90}(4) 044001

\bibitem{0264-9381-27-19-194005}
Yakunin K~N, Marronetti P, Mezzacappa A, Bruenn S~W, Lee C~T, Chertkow M~A, Hix
  W~R, Blondin J~M, Lentz E~J, Messer O~E~B and Yoshida S 2010 {\em Classical
  and Quantum Gravity\/} {\bf 27} 194005

\bibitem{PhysRevD.92.084040}
Yakunin K~N, Mezzacappa A, Marronetti P, Yoshida S, Bruenn S~W, Hix W~R, Lentz
  E~J, Bronson~Messer O~E, Harris J~A, Endeve E, Blondin J~M and Lingerfelt E~J
  2015 {\em Phys. Rev. D\/} {\bf 92}(8) 084040

\bibitem{PhysRevD.78.064056}
Dimmelmeier H, Ott C~D, Marek A and Janka H~T 2008 {\em Phys. Rev. D\/} {\bf
  78}(6) 064056

\bibitem{shapley2013galaxies}
Shapley H 2013 {\em Galaxies\/} Harvard Books on Astronomy Series (Harvard
  University Press) ISBN 9780674421790

\bibitem{0004-637X-660-1-516}
Tominaga N, Umeda H and Nomoto K 2007 {\em The Astrophysical Journal\/} {\bf
  660} 516

\bibitem{Stevenson:2017tfq}
Stevenson S, Vigna-Gómez A, Mandel I, Barrett J~W, Neijssel C~J, Perkins D and
  de~Mink S~E 2017  [Nature Commun.8,14906(2017)] (\textit{Preprint}
  \eprint{1704.01352})

\bibitem{doi:10.1111/j.1365-2966.2009.15120.x}
Marassi S, Schneider R and Ferrari V 2009 {\em Monthly Notices of the Royal
  Astronomical Society\/} {\bf 398} 293--302 (\textit{Preprint}
  \eprint{/oup/backfile/content_public/journal/mnras/398/1/10.1111/j.1365-2966.2009.15120.x/2/mnras0398-0293.pdf})

\bibitem{PhysRevD.92.063005}
Crocker K, Mandic V, Regimbau T, Belczynski K, Gladysz W, Olive K, Prestegard T
  and Vangioni E 2015 {\em Phys. Rev. D\/} {\bf 92}(6) 063005

\bibitem{PhysRevD.95.063015}
Crocker K, Prestegard T, Mandic V, Regimbau T, Olive K and Vangioni E 2017 {\em
  Phys. Rev. D\/} {\bf 95}(6) 063015

\bibitem{PhysRevD.73.104024}
Sandick P, Olive K~A, Daigne F and Vangioni E 2006 {\em Phys. Rev. D\/} {\bf
  73}(10) 104024

\bibitem{doi:10.1146/annurev.aa.10.090172.002003}
Press W~H and Thorne K~S 1972 {\em Annual Review of Astronomy and
  Astrophysics\/} {\bf 10} 335--374 (\textit{Preprint}
  \eprint{http://dx.doi.org/10.1146/annurev.aa.10.090172.002003})

\bibitem{Riles:2012yw}
Riles K 2013 {\em Prog. Part. Nucl. Phys.\/} {\bf 68} 1--54 (\textit{Preprint}
  \eprint{1209.0667})

\bibitem{Zimmermann:1978mk}
Zimmermann M 1978 {\em Nature\/} {\bf 271} 524--525

\bibitem{1976ApJ...208..550P}
{Pandharipande} V~R, {Pines} D and {Smith} R~A 1976 {\em "Astrophys. J."\/}
  {\bf 208} 550--566

\bibitem{PhysRevLett.24.611}
Chandrasekhar S 1970 {\em Phys. Rev. Lett.\/} {\bf 24}(11) 611--615

\bibitem{Friedman:1978hf}
Friedman J and Schutz B~F 1978 {\em Astrophys.~J.\/} {\bf 222} 281

\bibitem{Bildsten:1998ey}
Bildsten L 1998 {\em Astrophys.~J.\/} {\bf 501} L89 (\textit{Preprint}
  \eprint{astro-ph/9804325})

\bibitem{Andersson:1997xt}
Andersson N 1998 {\em Astrophys.~J.\/} {\bf 502} 708--713 (\textit{Preprint}
  \eprint{gr-qc/9706075})

\bibitem{0004-637X-502-2-714}
Friedman J~L and Morsink S~M 1998 {\em The Astrophysical Journal\/} {\bf 502}
  714

\bibitem{PhysRevD.58.084020}
Owen B~J, Lindblom L, Cutler C, Schutz B~F, Vecchio A and Andersson N 1998 {\em
  Phys. Rev. D\/} {\bf 58}(8) 084020

\bibitem{Ferrari:1998jf}
Ferrari V, Matarrese S and Schneider R 1999 {\em Mon. Not. Roy. Astron. Soc.\/}
  {\bf 303} 258 (\textit{Preprint} \eprint{astro-ph/9806357})

\bibitem{Cheng:2015rja}
Cheng Q, Yu Y~W and Zheng X~P 2015 {\em Mon. Not. Roy. Astron. Soc.\/} {\bf
  454} 2299--2304 (\textit{Preprint} \eprint{1509.07651})

\bibitem{doi:10.1146/annurev-astro-081915-023329}
Kaspi V~M and Beloborodov A~M 2017 {\em Annual Review of Astronomy and
  Astrophysics\/} {\bf 55} 261--301

\bibitem{MNR:MNR17861}
Marassi S, Ciolfi R, Schneider R, Stella L and Ferrari V 2011 {\em Monthly
  Notices of the Royal Astronomical Society\/} {\bf 411} 2549--2557 ISSN
  1365-2966

\bibitem{PhysRevD.89.123008}
Talukder D, Thrane E, Bose S and Regimbau T 2014 {\em Phys. Rev. D\/} {\bf
  89}(12) 123008

\bibitem{refId0_Sartore}
{Sartore, N}, {Ripamonti, E}, {Treves, A} and {Turolla, R} 2010 {\em Astron.
  Astrophys.\/} {\bf 510} A23

\bibitem{PhysRevLett.102.191102}
Horowitz C~J and Kadau K 2009 {\em Phys. Rev. Lett.\/} {\bf 102}(19) 191102

\bibitem{PhysRevD.88.044004}
Johnson-McDaniel N~K and Owen B~J 2013 {\em Phys. Rev. D\/} {\bf 88}(4) 044004

\bibitem{1538-4357-528-1-L29}
Baumgarte T~W, Shapiro S~L and Shibata M 2000 {\em The Astrophysical Journal
  Letters\/} {\bf 528} L29

\bibitem{2017PhRvD..96f3011M}
{Maione} F, {De Pietri} R, {Feo} A and {L{\"o}ffler} F 2017 {\em Phys. Rev.
  D\/} {\bf 96} 063011 (\textit{Preprint} \eprint{1707.03368})

\bibitem{2000PhRvD..61f4001S}
{Shibata} M and {Ury{\= u}} K~{\= o} 2000 {\em "Phys. Rev. D"\/} {\bf 61}
  064001 (\textit{Preprint} \eprint{gr-qc/9911058})

\bibitem{2013PhRvD..88d4026H}
{Hotokezaka} K, {Kiuchi} K, {Kyutoku} K, {Muranushi} T, {Sekiguchi} Y~i,
  {Shibata} M and {Taniguchi} K 2013 {\em "Phys. Rev. D"\/} {\bf 88} 044026
  (\textit{Preprint} \eprint{1307.5888})

\bibitem{2041-8205-851-1-L16}
Abbott B~P {\em et~al.\/} 2017 {\em The Astrophysical Journal Letters\/} {\bf
  851} L16

\bibitem{Miao:2017qot}
Miao H, Yang H and Martynov D 2018 {\em Phys. Rev.\/} {\bf D98} 044044
  (\textit{Preprint} \eprint{1712.07345})

\bibitem{PhysRevD.55.448}
Christensen N 1997 {\em Phys. Rev. D\/} {\bf 55}(2) 448--454

\bibitem{Abbott:2011rr}
Abbott B {\em et~al.\/} (The LIGO Scientific Collaboration and the Virgo
  Collaboration) 2011 {\em Phys.~Rev.~Lett.\/} {\bf 107}(27) 271102

\bibitem{0264-9381-23-8-S23}
Ballmer S~W 2006 {\em Classical and Quantum Gravity\/} {\bf 23} S179

\bibitem{S4radiom}
Abbott B {\em et~al.\/} 2007 {\em Phys. Rev. D\/} {\bf 76} 082003

\bibitem{PhysRevLett.116.221101}
Abbott B~P {\em et~al.\/} (LIGO Scientific and Virgo Collaborations) 2016 {\em
  Phys. Rev. Lett.\/} {\bf 116}(22) 221101

\bibitem{Will2014}
Will C~M 2014 {\em Living Reviews in Relativity\/} {\bf 17} 4 ISSN 1433-8351

\bibitem{0264-9381-32-24-243001}
Berti E {\em et~al.\/} 2015 {\em Classical and Quantum Gravity\/} {\bf 32}
  243001

\bibitem{Isi:2017equ}
Isi M, Pitkin M and Weinstein A~J 2017 {\em Phys. Rev. D\/} {\bf 96}(4) 042001

\bibitem{PhysRevLett.120.031104}
Abbott B~P {\em et~al.\/} (LIGO Scientific Collaboration and Virgo
  Collaboration) 2018 {\em Phys. Rev. Lett.\/} {\bf 120}(3) 031104

\bibitem{Callister:2017ocg}
Callister T, Biscoveanu A~S, Christensen N, Isi M, Matas A, Minazzoli O,
  Regimbau T, Sakellariadou M, Tasson J and Thrane E 2017 {\em Phys. Rev. X\/}
  {\bf 7}(4) 041058

\bibitem{Abbott:2018utx}
Abbott B~P {\em et~al.\/} (Virgo, LIGO Scientific) 2018 {\em Phys. Rev.
  Lett.\/} {\bf 120} 201102 (\textit{Preprint} \eprint{1802.10194})

\bibitem{Aso:2013eba}
Aso Y {\em et~al.\/} (KAGRA) 2013 {\em Phys.~Rev.\/} {\bf D88} 043007
  (\textit{Preprint} \eprint{1306.6747})

\bibitem{doi:10.1142/S0218271813410101}
Unnikrishnan C~S 2013 {\em International Journal of Modern Physics D\/} {\bf
  22} 1341010 (\textit{Preprint}
  \eprint{http://www.worldscientific.com/doi/pdf/10.1142/S0218271813410101})

\bibitem{Sentman}
{DD Sentman} 1995 {\em Handbook of Atmospheric Electrodynamics\/} vol~1 (CRC
  Press, Boca Raton)

\bibitem{FULLEKRUG1995479}
F{\"u}llekrug M 1995 {\em Journal of Atmospheric and Terrestrial Physics\/}
  {\bf 57} 479 -- 484 ISSN 0021-9169

\bibitem{wiener}
Thrane E, Christensen N, Schofield R~M~S and Effler A 2014 {\em Phys. Rev. D\/}
  {\bf 90} 023013

\bibitem{Cirone:2018guh}
Coughlin M~W {\em et~al.\/} 2018 {\em Phys. Rev.\/} {\bf D97} 102007
  (\textit{Preprint} \eprint{1802.00885})

\bibitem{Abbott2016}
Abbott B~P {\em et~al.\/} 2016 {\em Living Reviews in Relativity\/} {\bf 19} 1
  ISSN 1433-8351

\bibitem{buonanno}
Buonanno A 1997 {\em Phys. Rev. D\/} {\bf 55} 3330

\bibitem{mandic}
{V Mandic and A Buonanno} 2006 {\em Phys. Rev. D\/} {\bf 73} 063008

\bibitem{kibble}
Kibble T~W~B 1976 {\em J. Phys.\/} {\bf A9} 1387

\bibitem{olmez1}
Olmez S, Mandic V and Siemens X 2010 {\em Phys. Rev. D\/} {\bf 81} 104028

\bibitem{olmez2}
Olmez S, Mandic V and Siemens X 2011 {\em J. Cosmol. Astropart. Phys.\/} {\bf
  2012} 009

\bibitem{PhysRevD.73.102001}
Romano J~D and Woan G 2006 {\em Phys. Rev. D\/} {\bf 73}(10) 102001

\bibitem{PhysRevLett.116.231101}
Armano M {\em et~al.\/} 2016 {\em Phys. Rev. Lett.\/} {\bf 116}(23) 231101

\bibitem{PhysRevLett.120.061101}
Armano M {\em et~al.\/} 2018 {\em Phys. Rev. Lett.\/} {\bf 120}(6) 061101

\bibitem{Colpi:2016fup}
Colpi M and Sesana A 2017 {Gravitational Wave Sources in the Era of Multi-Band
  Gravitational Wave Astronomy} {\em An Overview of Gravitational Waves:
  Theory, Sources and Detection\/} ed Auger G pp 43--140 (\textit{Preprint}
  \eprint{1610.05309})
  \urlprefix\url{https://inspirehep.net/record/1492489/files/arXiv:1610.05309.pdf}

\bibitem{0264-9381-27-8-084009}
Babak S, Baker J~G, Benacquista M~J, Cornish N~J, Larson S~L, Mandel I,
  McWilliams S~T, Petiteau A, Porter E~K, Robinson E~L, Vallisneri M, Vecchio
  A, the Mock LISA Data Challenge Task~Force, Adams M, Arnaud K~A, Błaut A,
  Bridges M, Cohen M, Cutler C, Feroz F, Gair J~R, Graff P, Hobson M, Key J~S,
  Królak A, Lasenby A, Prix R, Shang Y, Trias M, Veitch J, Whelan J~T and the
  Challenge 3~participants 2010 {\em Classical and Quantum Gravity\/} {\bf 27}
  084009

\bibitem{Blanco-Pillado:2017rnf}
Blanco-Pillado J~J, Olum K~D and Siemens X 2018 {\em Phys. Lett.\/} {\bf B778}
  392--396 (\textit{Preprint} \eprint{1709.02434})

\bibitem{Tinto2014}
Tinto M and Dhurandhar S~V 2014 {\em Living Reviews in Relativity\/} {\bf 17} 6

\bibitem{0264-9381-32-1-015014}
Moore C~J, Cole R~H and Berry C~P~L 2015 {\em Classical and Quantum Gravity\/}
  {\bf 32} 015014

\bibitem{moore26gravitational}
Moore C, Cole R and Berry C {\em {Gravitational Wave Sensitivity Curve Plotter,
  Project homepage, University of Cambridge.}\/}  http://gwplotter.com/

\bibitem{PhysRevD.75.061302}
Seto N 2007 {\em Phys. Rev. D\/} {\bf 75}(6) 061302

\bibitem{CROWDER201366}
Crowder S, Namba R, Mandic V, Mukohyama S and Peloso M 2013 {\em Physics
  Letters B\/} {\bf 726} 66 -- 71 ISSN 0370-2693

\bibitem{0264-9381-23-7-014}
Corbin V and Cornish N~J 2006 {\em Classical and Quantum Gravity\/} {\bf 23}
  2435

\bibitem{0264-9381-18-17-308}
Cornish N~J and Larson S~L 2001 {\em Classical and Quantum Gravity\/} {\bf 18}
  3473

\bibitem{0264-9381-34-6-065005}
Chou A, Glass H, Gustafson H~R, Hogan C, Kamai B~L, Kwon O, Lanza R, McCuller
  L, Meyer S~S, Richardson J, Stoughton C, Tomlin R and Weiss R 2017 {\em
  Classical and Quantum Gravity\/} {\bf 34} 065005

\bibitem{PhysRevD.18.1747}
Hawking S~W 1978 {\em Phys. Rev. D\/} {\bf 18}(6) 1747--1753

\bibitem{HAWKING1980283}
Hawking S, Page D and Pope C 1980 {\em Nuclear Physics B\/} {\bf 170} 283 --
  306 ISSN 0550-3213 volume B170 [FSI] No. 3 to follow in Approximately Two
  Months

\bibitem{PhysRevLett.69.237}
Ashtekar A, Rovelli C and Smolin L 1992 {\em Phys. Rev. Lett.\/} {\bf 69}(2)
  237--240

\bibitem{PhysRevD.95.063002}
Chou A~S, Gustafson R, Hogan C, Kamai B, Kwon O, Lanza R, Larson S~L, McCuller
  L, Meyer S~S, Richardson J, Stoughton C, Tomlin R and Weiss R (Holometer
  Collaboration) 2017 {\em Phys. Rev. D\/} {\bf 95}(6) 063002

\bibitem{CYBURT2005313}
Cyburt R~H, Fields B~D, Olive K~A and Skillman E 2005 {\em Astroparticle
  Physics\/} {\bf 23} 313 -- 323 ISSN 0927-6505

\bibitem{PhysRevD.85.123002}
Sendra I and Smith T 2012 {\em Phys. Rev. D\/} {\bf 85}(12) 123002

\bibitem{1968Natur.217..709H}
{Hewish} A, {Bell} S~J, {Pilkington} J~D~H, {Scott} P~F and {Collins} R~A 1968
  {\em Nature\/} {\bf 217} 709--713

\bibitem{1968Natur.218..731G}
{Gold} T 1968 {\em Nature\/} {\bf 218} 731--732

\bibitem{1969Natur.221...25G}
{Gold} T 1969 {\em Nature\/} {\bf 221} 25--27

\bibitem{1978SvA....22...36S}
{Sazhin} M~V 1978 {\em Sov. Astron.\/} {\bf 22} 36--38

\bibitem{1979ApJ...234.1100D}
{Detweiler} S 1979 {\em Astrophys. J.\/} {\bf 234} 1100--1104

\bibitem{RAWLEY761}
RAWLEY L~A, TAYLOR J~H, DAVIS M~M and ALLAN D~W 1987 {\em Science\/} {\bf 238}
  761--765 ISSN 0036-8075 (\textit{Preprint}
  \eprint{http://science.sciencemag.org/content/238/4828/761.full.pdf})

\bibitem{1983ApJ...265L..39H}
{Hellings} R~W and {Downs} G~S 1983 {\em Astrophysical Journal Letters\/} {\bf
  265} L39--L42

\bibitem{0264-9381-30-22-220301}
Bizouard M~A, Jenet F, Price R and Will C 2013 {\em Classical and Quantum
  Gravity\/} {\bf 30} 220301

\bibitem{0264-9381-30-22-224007}
Hobbs G 2013 {\em Classical and Quantum Gravity\/} {\bf 30} 224007

\bibitem{0264-9381-30-22-224008}
McLaughlin M~A 2013 {\em Classical and Quantum Gravity\/} {\bf 30} 224008

\bibitem{0264-9381-30-22-224009}
Kramer M and Champion D~J 2013 {\em Classical and Quantum Gravity\/} {\bf 30}
  224009

\bibitem{0264-9381-30-22-224010}
(for~the IPTA) R~N~M 2013 {\em Classical and Quantum Gravity\/} {\bf 30} 224010

\bibitem{0264-9381-30-22-224011}
Lazio T~J~W 2013 {\em Classical and Quantum Gravity\/} {\bf 30} 224011

\bibitem{0004-637X-583-2-616}
Jaffe A~H and Backer D~C 2003 {\em Astrophys. J.\/} {\bf 583} 616

\bibitem{doi:10.1093/mnras/stv1538}
Lentati L {\em et~al.\/} 2015 {\em Monthly Notices of the Royal Astronomical
  Society\/} {\bf 453} 2576--2598 (\textit{Preprint}
  \eprint{/oup/backfile/content_public/journal/mnras/453/3/10.1093/mnras/stv1538/2/stv1538.pdf})

\bibitem{2013PASA...30...17M}
{Manchester} R~N, {Hobbs} G, {Bailes} M, {Coles} W~A, {van Straten} W, {Keith}
  M~J, {Shannon} R~M, {Bhat} N~D~R, {Brown} A, {Burke-Spolaor} S~G, {Champion}
  D~J, {Chaudhary} A, {Edwards} R~T, {Hampson} G, {Hotan} A~W, {Jameson} A,
  {Jenet} F~A, {Kesteven} M~J, {Khoo} J, {Kocz} J, {Maciesiak} K, {Oslowski} S,
  {Ravi} V, {Reynolds} J~R, {Sarkissian} J~M, {Verbiest} J~P~W, {Wen} Z~L,
  {Wilson} W~E, {Yardley} D, {Yan} W~M and {You} X~P 2013 {\em Publications of
  the Astronomical Society of Australia\/} {\bf 30} e017 (\textit{Preprint}
  \eprint{1210.6130})

\bibitem{PhysRevD.23.832}
Hellings R~W 1981 {\em Phys. Rev. D\/} {\bf 23}(4) 832--843

\bibitem{Armstrong2006}
Armstrong J~W 2006 {\em Living Reviews in Relativity\/} {\bf 9} 1 ISSN
  1433-8351

\bibitem{1975GReGr...6..439E}
{Estabrook} F~B and {Wahlquist} H~D 1975 {\em General Relativity and
  Gravitation\/} {\bf 6} 439--447

\bibitem{Thorne_300yrs}
Thorne K 1987 {Gravitational radiation} {\em Three Hundred Years of
  Gravitation\/} ed Hawking S and Israel W (Cambridge U.K.: Cambridge
  University Press) pp 330--458

\bibitem{1979ApJ...230..570A}
{Armstrong} J~W, {Woo} R and {Estabrook} F~B 1979 {\em Astrophys. J.\/} {\bf
  230} 570--574

\bibitem{PhysRevD.23.844}
Hellings R~W, Callahan P~S, Anderson J~D and Moffet A~T 1981 {\em Phys. Rev.
  D\/} {\bf 23}(4) 844--851

\bibitem{1984Natur.308..158A}
{Anderson} J~D, {Armstrong} J~W, {Estabrook} F~B, {Hellings} R~W, {Lau} E~K and
  {Wahlquist} H~D 1984 {\em Nature\/} {\bf 308} 158--160

\bibitem{1987ApJ...318..536A}
{Armstrong} J~W, {Estabrook} F~B and {Wahlquist} H~D 1987 {\em Astrophys. J.\/}
  {\bf 318} 536--541

\bibitem{2003SPIE.4856...90A}
{Abbate} S~F, {Armstrong} J~W, {Asmar} S~W, {Barbinis} E, {Bertotti} B,
  {Fleischman} D~U, {Gatti} M~S, {Goltz} G~L, {Herrera} R~G, {Iess} L, {Lee}
  K~J, {Ray} T~L, {Tinto} M, {Tortora} P and {Wahlquist} H~D 2003 {The Cassini
  gravitational wave experiment} {\em Gravitational-Wave Detection\/} ({\em
  Proceedings of the SPIE"\/} vol 4856) ed {Cruise} M and {Saulson} P pp 90--97

\bibitem{2003Natur.425..374B}
{Bertotti} B, {Iess} L and {Tortora} P 2003 {\em Nature\/} {\bf 425} 374--376

\bibitem{0004-637X-599-2-806}
Armstrong J~W, Iess L, Tortora P and Bertotti B 2003 {\em The Astrophysical
  Journal\/} {\bf 599} 806

\bibitem{1967ApJ...147...73S}
{Sachs} R~K and {Wolfe} A~M 1967 {\em Astrophys. J.\/} {\bf 147} 73

\bibitem{Planck_CMB_2011}
{Planck Collaboration}, {Ade, P A R} {\em et~al.\/} 2011 {\em Astron.
  Astrophys.\/} {\bf 536} A1

\bibitem{PhysRevD.50.3713}
Allen B and Koranda S 1994 {\em Phys. Rev. D\/} {\bf 50}(6) 3713--3737

\bibitem{PhysRevD.52.1902}
Koranda S and Allen B 1995 {\em Phys. Rev. D\/} {\bf 52}(4) 1902--1919

\bibitem{Allen:1996vm}
Allen B 1996 {The Stochastic gravity wave background: Sources and detection}
  {\em {Relativistic gravitation and gravitational radiation. Proceedings,
  School of Physics, Les Houches, France, September 26-October 6, 1995}\/} pp
  373--417 (\textit{Preprint} \eprint{gr-qc/9604033})

\bibitem{1980A&A....89....6C}
{Carr} B~J 1980 {\em Astron. Astrophys.\/} {\bf 89} 6--21

\bibitem{Patrignani:2016xqp}
Patrignani C {\em et~al.\/} (Particle Data Group) 2016 {\em Chin. Phys.\/} {\bf
  C40} 100001

\bibitem{Maggiore:1999vm}
Maggiore M 2000 {\em Phys.~Rep.\/} {\bf 331} 283--367 (\textit{Preprint}
  \eprint{gr-qc/9909001})

\bibitem{doi:10.1093/mnras/stt2206}
Anderson L {\em et~al.\/} 2014 {\em Monthly Notices of the Royal Astronomical
  Society\/} {\bf 439} 83--101 (\textit{Preprint}
  \eprint{/oup/backfile/content_public/journal/mnras/439/1/10.1093/mnras/stt2206/2/stt2206.pdf})

\bibitem{2005NuPhB.729..221M}
{Mangano} G, {Miele} G, {Pastor} S, {Pinto} T, {Pisanti} O and {Serpico} P~D
  2005 {\em Nuclear Physics B\/} {\bf 729} 221--234 (\textit{Preprint}
  \eprint{hep-ph/0506164})

\bibitem{doi:10.1146/annurev-astro-081915-023433}
Kamionkowski M and Kovetz E~D 2016 {\em Annual Review of Astronomy and
  Astrophysics\/} {\bf 54} 227--269

\bibitem{PhysRevD.58.023003}
Zaldarriaga M and Seljak U~c~v 1998 {\em Phys. Rev. D\/} {\bf 58}(2) 023003

\bibitem{LEWIS20061}
Lewis A and Challinor A 2006 {\em Physics Reports\/} {\bf 429} 1 -- 65 ISSN
  0370-1573

\bibitem{PhysRevLett.111.141301}
Hanson D {\em et~al.\/} (SPTpol Collaboration) 2013 {\em Phys. Rev. Lett.\/}
  {\bf 111}(14) 141301

\bibitem{0004-637X-808-1-7}
van Engelen A {\em et~al.\/} 2015 {\em The Astrophysical Journal\/} {\bf 808} 7

\bibitem{2015ApJ...807..151K}
{Keisler} R {\em et~al.\/} 2015 {\em Astrophys. J.\/} {\bf 807} 151
  (\textit{Preprint} \eprint{1503.02315})

\bibitem{Adam:2015tpy}
Adam R {\em et~al.\/} (Planck) 2016 {\em Astron. Astrophys.\/} {\bf 594} A9
  (\textit{Preprint} \eprint{1502.05956})

\bibitem{doi:10.1146/annurev-astro-082214-122414}
Andersson B~G, Lazarian A and Vaillancourt J~E 2015 {\em Annual Review of
  Astronomy and Astrophysics\/} {\bf 53} 501--539 (\textit{Preprint}
  \eprint{https://doi.org/10.1146/annurev-astro-082214-122414})

\bibitem{PhysRevLett.112.241101}
Ade P~A~R {\em et~al.\/} ((BICEP2 Collaboration)) 2014 {\em Phys. Rev. Lett.\/}
  {\bf 112}(24) 241101

\bibitem{Flauger:2014qra}
Flauger R, Hill J~C and Spergel D~N 2014 {\em JCAP\/} {\bf 1408} 039
  (\textit{Preprint} \eprint{1405.7351})

\bibitem{PhysRevLett.114.101301}
Ade P~A~R {\em et~al.\/} (BICEP2/Keck and Planck Collaborations) 2015 {\em
  Phys. Rev. Lett.\/} {\bf 114}(10) 101301

\bibitem{PhysRevLett.116.031302}
Ade P~A~R {\em et~al.\/} (Keck Array and BICEP2 Collaborations) 2016 {\em Phys.
  Rev. Lett.\/} {\bf 116}(3) 031302

\bibitem{Kusaka:2018yzq}
Kusaka A {\em et~al.\/} 2018 {\em JCAP\/} {\bf 1809} 005 (\textit{Preprint}
  \eprint{1801.01218})

\bibitem{2014ApJ...794..171P}
{Polarbear Collaboration}, {P~A~R~Ade} {\em et~al.\/} 2014 {\em Astrophys.
  J.\/} {\bf 794} 171 (\textit{Preprint} \eprint{1403.2369})

\bibitem{2014SPIE.9153E..11M}
{MacDermid} K {\em et~al.\/} 2014 {The performance of the bolometer array and
  readout system during the 2012/2013 flight of the E and B experiment (EBEX)}
  {\em Millimeter, Submillimeter, and Far-Infrared Detectors and
  Instrumentation for Astronomy VII\/} ({\em Proc. SPIE\/} vol 9153) p 915311
  (\textit{Preprint} \eprint{1407.6894})

\bibitem{Manzotti:2017net}
Manzotti A {\em et~al.\/} (Herschel, SPT) 2017 {\em Astrophys. J.\/} {\bf 846}
  45 (\textit{Preprint} \eprint{1701.04396})

\bibitem{1969ApJ...156..529D}
{Dyson} F~J 1969 {\em Astrophys. J.\/} {\bf 156} 529

\bibitem{1984ApJ...286..387B}
{Boughn} S~P and {Kuhn} J~R 1984 {\em Astrophys. J.\/} {\bf 286} 387--391

\bibitem{2017A&A...604A..40F}
{Fossat} E, {Boumier} P, {Corbard} T, {Provost} J, {Salabert} D, {Schmider}
  F~X, {Gabriel} A~H, {Grec} G, {Renaud} C, {Robillot} J~M, {Roca-Cort{\'e}s}
  T, {Turck-Chi{\`e}ze} S, {Ulrich} R~K and {Lazrek} M 2017 {\em Astron.
  Astrophys.\/} {\bf 604} A40 (\textit{Preprint} \eprint{1708.00259})

\bibitem{0004-637X-784-2-88}
Siegel D~M and Roth M 2014 {\em The Astrophysical Journal\/} {\bf 784} 88

\bibitem{PhysRevLett.112.101102}
Coughlin M and Harms J 2014 {\em Phys. Rev. Lett.\/} {\bf 112}(10) 101102

\bibitem{10.1007/978-3-642-10634-7_83}
Crossley D and Hinderer J 2010 Ggp (global geodynamics project): An
  international network of superconducting gravimeters to study time-variable
  gravity {\em Gravity, Geoid and Earth Observation\/} ed Mertikas S~P (Berlin,
  Heidelberg: Springer Berlin Heidelberg) pp 627--635 ISBN 978-3-642-10634-7

\bibitem{PhysRevD.90.102001}
Coughlin M and Harms J 2014 {\em Phys. Rev. D\/} {\bf 90}(10) 102001

\end{thebibliography}
\end{document}